\documentclass[aps,prx,showpacs,preprintnumbers,twocolumn,superscriptaddress,nofootinbib]{revtex4-2}
\usepackage{amsmath,amssymb}
\usepackage{bm}
\usepackage{tipa}
\usepackage{upgreek}
\usepackage{comment}
\usepackage{mathrsfs}
\usepackage{graphicx}
\usepackage{lipsum}
\usepackage{braket}
\usepackage{enumitem}
\usepackage{mathbbol}
\usepackage{booktabs}
\usepackage{gensymb}
\usepackage[normalem]{ulem}
\usepackage{color}

\usepackage[colorlinks,bookmarks=true,citecolor=blue,linkcolor=red,urlcolor=blue]{hyperref}
\usepackage{hyperref}

\usepackage{pifont}

\makeatletter 
    
\renewcommand\onecolumngrid{
\do@columngrid{one}{\@ne}%
\def\set@footnotewidth{\onecolumngrid}
\def\footnoterule{\kern-6pt\hrule width 1.5in\kern6pt}%
}

\renewcommand\twocolumngrid{
        \def\footnoterule{
        \dimen@\skip\footins\divide\dimen@\thr@@
        \kern-\dimen@\hrule width.5in\kern\dimen@}
        \do@columngrid{mlt}{\tw@}
}%

\makeatother    

\begin{document}
	
\title{Chern-Textured Exciton Insulators with Valley Spiral Order in Moiré Materials} 

 \author{Ziwei Wang}\thanks{These authors contributed equally.} 
	\affiliation{Rudolf Peierls Centre for Theoretical Physics, Parks Road, Oxford, OX1 3PU, UK}

  \author{Yves H. Kwan}\thanks{These authors contributed equally.} 
	\affiliation{Princeton Center for Theoretical Science, Princeton University, Princeton NJ 08544, USA}

 	\author{Glenn Wagner}
	\affiliation{Department of Physics, University of Zurich, Winterthurerstrasse 190, 8057 Zurich, Switzerland}
	
	\author{Steven H. Simon}
	\affiliation{Rudolf Peierls Centre for Theoretical Physics, Parks Road, Oxford, OX1 3PU, UK}
\author{Nick Bultinck}
	\affiliation{Department of Physics, Ghent University, Krijgslaan 281, 9000 Gent, Belgium} 
	\author{S.A. Parameswaran}
	\affiliation{Rudolf Peierls Centre for Theoretical Physics, Parks Road, Oxford, OX1 3PU, UK}

\begin{abstract}
We explore the phase diagrams of moiré materials in search of a new class of intervalley-coherent correlated insulating state: the Chern texture insulator (CTI). This phase of matter, proposed in a companion paper~\cite{companion}, breaks valley $U(1)$ symmetry in a  nontrivial fashion wherein the valley order parameter is forced to texture in momentum space as a consequence of band topology. Using detailed Hartree-Fock studies, we establish that the CTI emerges as an energetically competitive intermediate-coupling ground state in several  moiré systems which lack a twofold rotation symmetry that forbids the single-particle topology essential to the formation of the CTI  valley texture.
\end{abstract}

\maketitle

\section{Introduction}\label{sec:intro}

Moir\'e materials ---  heterostructures of layered atomically-thin constituents stacked with a lattice or rotational mismatch --- have ushered in a new paradigm of correlated electronic matter with unprecedented tunability and diversity. Experiments in recent years have uncovered a rich variety of strongly-interacting phenomena including superconductivity~\cite{Cao2018b,Yankowitz_2019,Lu2019,Park2021TTGSC}, correlated insulators~\cite{Cao2018,Lu2019,Yankowitz_2019}, orbital ferromagnetism~\cite{Sharpe_2019,Serlin900}, (fractional) Chern insulators~\cite{Park2023,Cai2023y,Zeng2023,Xu2023,lu2023fractional,Xie2021fractional}, linear-$T$ resistivity~\cite{Jaoui2022,polshyn2019large}, generalized Wigner crystals~\cite{Regan2020}, and many others. The pioneering studies focused on magic-angle twisted bilayer graphene (TBG), but an ever-growing research effort has been invested into other systems involving larger numbers of layers~\cite{park2021magicangle,zhang2022promotion} and alternative material building blocks~\cite{Mak2022}. While different moir\'e platforms can vary significantly in the details of their single-particle models and hence interacting phenomenology, they often share several fundamental features. Firstly, the region in parameter space of greatest interest is typically where some subset of the bands are isolated and narrow, and electronic interactions can be expected to play a non-trivial role. Secondly, most moir\'e materials carry flavor degrees of freedom such as spin and valley, which can participate in symmetry-breaking phenomena. Finally, the low-energy moir\'e bands are often  topologically non-trivial. 

The confluence of these three features have motivated parallels to an older problem: namely, that of quantum Hall ferromagnetism (QHFM)~\cite{Sondhi1993}. This is usually exhibited by electrons with flavor degrees of freedom in exactly-flat bands characterized by a topological Chern number $C=1$: the celebrated Landau levels of electrons in a magnetic field. A direct analogy between the two problems is complicated by two facts. The first is that single-particle moiré bands intrinsically preserve time-reversal symmetry ($\hat{\mathcal{T}}$), whose breaking is a prerequisite for $C\neq 0$. This means that the bands either (i) do not have a readily definable Chern number (as is the case, for instance, when they are endowed with an additional $\hat{\mathcal{C}}_{2z}$ spatial rotation symmetry, leading to a distinct `Euler invariant' topological index); or else (ii)  come in time-reversed partners with equal and opposite Chern numbers. The second is that moiré bands are never exactly flat, and moreover interactions can feed back nontrivially on their dispersion, potentially modifying the balance of interaction and kinetic energy. Together, these open the possibility for new  phases of matter unanticipated from the strong-coupling, QHFM-like limit.

Pursuing this line of inquiry, in a companion work~\cite{companion} we introduced a new class of  broken-symmetry topological phase that we termed the ``textured exciton insulator'' (TEI), as a candidate ground state for  moiré materials whose single-particle bands are topological and carry a conserved $U(1)_V$ `valley' charge. Here the excitonic  order parameter introduces intervalley coherence, thereby spontaneously breaks $U(1)_V$, and is forced to `texture' across the Brillouin zone due to the mandates of band topology~\cite{Bultinck2019mechanism}. We  established several universal aspects of TEIs linked to their intertwining of symmetry-breaking and topology, including identifying  a topological obstruction to giving them a local-moment description~\cite{companion}. 

TEIs come in two varieties, depending on the two classes of time-reversal-invariant bandstructure structure discussed above. The first, the ``Euler texture insulator'' (ETI) is linked to $\hat{\mathcal{C}}_{2z}\hat{\mathcal{T}}$-invariant bands with a nonzero Euler index in each valley~\cite{Ahn2019,Song2019PHS}, whereas the second, the ``Chern texture insulator'' emerges in $\hat{\mathcal{C}}_{2z}$-broken systems with valley-contrasting Chern numbers. Using general arguments and  simplified toy models, we  demonstrated in Ref.~\cite{companion} that TEIs are  energetically competitive ground states at intermediate coupling at some partial (integer) filling of the bands. In the ETI case, we further identified  the incommensurate Kekulé spiral (IKS) phase predicted~\cite{kwan_kekule_2021}, and subsequently observed~\cite{nuckolls2023quantum, kim2023imaging}  in magic-angle bi/tri-layer~\cite{wang_kekule_2024}  graphene as an ETI. Given that this establishes the ETI as an experimentally-relevant concept, this naturally stimulates a search for its Chern-textured counterpart in other systems.

With this motivation, here we explore the phase diagrams of different moiré materials that break $\hat{\mathcal{C}}_{2z}$ either explicitly or spontaneously.  Using microscopically faithful self-consistent Hartree-Fock (HF) calculations, we demonstrate that CTIs are  energetically competitive ground states in many of these  materials. En route, we address  the various further complications of  realistic materials absent in simplified models, such as various competing orders,  the interaction-induced hybridization of single-particle bands within each valley, etc. Even after incorporating these features, we find  robust CTI order survives within a reasonable parameter regime across a large class of materials, as summarized in Tab.~\ref{tab:moire_materials}. In each of these cases, the order is linked to a moiré-scale modulation of intervalley coherence, that in several cases can potentially be directly detected through scanning probe measurements similar to those used to identify IKS order~\cite{nuckolls2023quantum, kim2023imaging}. Our work thus provides yet another illustration of how the combinatorial versatility of moiré systems provides fertile ground for seeking new and complex correlated states of matter with cutting-edge probes.

The remainder of this paper is organized as follows.
In Sec.~\ref{sec:methods}, briefly summarize the theory of the CTI,   summarize the single-particle continuum models of the moir\'e materials considered in this paper, and discuss details of the HF calculations. We present the results of our numerical studies in the form of HF phase diagrams  for each of the materials summarized in Tab.~\ref{tab:moire_materials} in Sec.~\ref{sec:results}, and close with a final discussion in Sec.~\ref{sec:discussion}. 

\begin{table}[t!]
\centering
\newcommand{\colskip}{\hskip 0.15in}
\renewcommand{\arraystretch}{1.15}
\begin{tabular}{
l @{\hskip 0.2in} 
c @{\colskip} 
c  
c
}\toprule[1.3pt]\addlinespace[0.3em]
System & 
$\hat{\mathcal{C}}_{2z}$-breaking & 
CTI$_n$ index  & Phase Diag.
\\  \midrule
TDBG (ABAB) & explicit & $2$ & Fig.~\ref{fig:abab123} \\
TDBG (ABBA) & explicit &  $0, 1,-1$ & Figs.~\ref{fig:abba123},\ref{fig:abba_angle}\\
TMBG & explicit &  0, $2$ & Fig.~\ref{fig:tmbg123}\\
HTG & explicit & 0, $1$ & Fig.~\ref{fig:htg}\\
$t$MoTe$_2$ & explicit & 0 & Fig.~\ref{fig:tMoTe2}b\\
TSTG & spontaneous & 0& Fig.~\ref{fig:TSTG}b \\
\bottomrule[1.3pt]
\end{tabular}
\caption{\textbf{Chern-textured exciton insulators (CTIs) in moir\'e materials studied in this work.} The IVC order parameter winding is $4\pi n$ in the CTI$_n$ phase. For $n=0$, the state is considered a `trivial IVC' phase. [$t$MoTe$_2$: twisted bilayer MoTe$_2$; TDBG: twisted double-bilayer graphene; TMBG: twisted mono-bilayer graphene; HTG: helical trilayer graphene; TSTG: twisted symmetric trilayer graphene; TBG: twisted bilayer graphene].\label{tab:moire_materials}}
\end{table}

\section{Background, Models, and Methods}\label{sec:methods}

\subsection{Chern Texture Insulators}
The simplest setting in which a CTI can arise consists of a pair of Chern bands with Chern number $C_\tau =  \tau n$,  distinguished by a time-reversal-odd `valley' index $\tau =\pm 1$. The interacting Hamiltonian has time-reversal symmetry (TRS) $\hat{\mathcal{T}}$, which relates the two valleys, and $U_{V}(1)$ symmetry, i.e.~the independent conservation of charge in each valley. In all the cases we study, the spin structure is such that we can choose implementation of time reversal to satisfy $\hat{\mathcal{T}}^2 = +1$. The absence of a valley-exchanging $\hat{\mathcal{C}}_{2z}$ symmetry, whether due to explicit single-particle symmetry-breaking effects or interaction driven spontaneous symmetry breaking, is therefore crucial for allowing a non-zero valley Chern number\footnote{This should be contrasted against the ETI setting, where single-valley $\hat{\mathcal{C}}_{2z}\hat{\mathcal{T}}$ symmetry is essential~\cite{companion}.}. In the limit where interactions dominate the kinetic energy, the ground state of the system at half-filling is expected to be a valley-polarized Chern insulator that breaks TRS~\cite{Bultinck2019mechanism,Zhang2019anomalous,Zhang2019}. In the intermediate-coupling regime where the kinetic energy is of similar magnitude as the interaction scale, we argued in Ref.~\onlinecite{companion} that a time-reversal-symmetric intervalley coherent insulator, which spontaneously breaks the $U_\text{v}(1)$ symmetry, is a generically competitive ground state. Due to the different Chern numbers in the two valleys, the intervalley coherence (IVC) acquires a $4\pi C$ winding around the Brillouin zone (BZ), which forces its vanishing at vortices within the Brillouin zone. In order to open an insulating gap, the valley pseudospin forms a complex texture in the BZ and fully valley polarizes at the cores of the  IVC vortices. The resulting state is a Chern texture insulator denoted CTI$_n$, where the additional index $n$  indicates the Chern number of the underlying valley-diagonal bands $C =\pm n$ (and hence  fixes the enforced winding of the IVC order in the BZ to be $4\pi n$). We demonstrated that the CTI exhibits a form of delicate topology in an obstruction to an atomically-localized limit. One further complication is that, in general, it is possible for IVC to develop between Bloch states with momentum $\bm{k}$ in valley $\tau = +$ and momentum $\bm{k} + \bm{q}$ in valley $\tau = -$, with $\bm{q}$ the wavevector of the resulting intervalley spiral order. The value of $\bm{q}$ realized in the ground state depends on the kinetic energy landscape of the system.
We note that there  exists a $\mathcal{T}$-breaking cousin of the CTI which also exhibits a winding IVC across the BZ, but where the IVC vortex cores have the {\it same sense} of VP. This `tilted valley polarized' (TVP) phase has nonzero Chern number, yet is distinct on grounds of both symmetry and topology from the fully-VP state, and is often found proximate to the CTI in the phase diagram of the toy models we have explored.
Readers interested in details of the CTI and TVP and their topological properties beyond this brief account are directed to the companion paper, Ref.~\cite{companion}.

\subsection{Continuum Models}
We perform self-consistent HF calculations on several moir\'e materials. To do so, we adopt the  approach of first modeling the non-interacting moir\'e band structure within the  `continuum model' approximation~\cite{Bistritzer2011} and then incorporating Coulomb interactions projected to the resulting  low-energy Hilbert space of the single-particle problem. In this subsection we describe the single-particle models for each of the different materials summarized in Tab.~\ref{tab:moire_materials}, before turning to technical details of our HF computations in the next subsection.

\subsubsection{Twisted double-bilayer graphene (ABAB)}\label{subsec:ABAB_model}

Twisted double-bilayer graphene (TDBG)~\cite{burg_correlated_2019, cao_tunable_2020, du_ferroelectricity_2024, he_symmetry_2021, liu_isospin_2022, liu_tunable_2020, rickhaus_correlated_2021, rubio-verdu_moire_2022, shen_correlated_2020, zhang_visualizing_2021, zhu_tunable_2022, he_symmetry-broken_2023} is a heterostructure consisting of two Bernal (i.e.~crystallographic AB or BA stacked) graphene bilayers placed on top of each other with  a twist angle $\theta$. Owing to the presence of Bernal bilayer stacks, the system does not possess $\hat{\mathcal{C}}_{2z}$ symmetry. The ABAB and ABBA stacking configurations are inequivalent;  we first consider the former  stacking order. In this case, the single-particle Hamiltonian for valley $K$ is given by $H_K = H_{K, 0} + H_V$, where the Hamiltonian in the absence of external displacement fields is~\cite{koshino_band_2019, liu_quantum_2019-1, chebrolu_flat_2019, Lee2019, haddadi_moire_2020}
\begin{equation}\label{eq:HK0_ABAB}
H_{K, 0} = \begin{pmatrix}
    H^A_0(\bm{k}_1) & g(\bm{k}_1) & & \\
    g^\dagger(\bm{k}_1) & H^B_0(\bm{k}_1) & U & \\
    & U^\dagger & H^A_0(\bm{k}_2) & g(\bm{k}_2) \\
    & & g^\dagger(\bm{k}_2) & H^B_0(\bm{k}_2)
\end{pmatrix},
\end{equation}
and we approximate the effect of an external displacement field with a linear interlayer potential\footnote{For all materials considered, we model the displacement field by a linearly varying potential that differs by $\Delta V$ between the outermost layers.}
\begin{equation}\label{eq:H_V}
    H_V = \text{diag}\left(-\frac{1}{2}I_2, - \frac{1}{6}I_2,  \frac{1}{6}I_2, \frac{1}{2}I_2\right)\Delta V,
\end{equation}
where $I_2$ is the $2\times 2 $ identity matrix in sublattice space.
In Eq.~\ref{eq:HK0_ABAB}, $\bm{k}_l = \bm{k} - \bm{K}_l$ is the momentum measured from the (rotated) Dirac point of layer $l = 1,2$. With $k_\theta = \frac{8\pi}{3\sqrt{3}a_{\text{CC}}}\sin(\theta/2)$ where $a_{\text{CC}} = 1.42 \times 10^{-10}\,$m is the length of C-C bond, we define the primitive vectors of the moir\'e reciprocal lattice as $\bm{G}_1 = k_\theta(\sqrt{3}, 0)$ and $\bm{G}_2 = k_\theta(-\frac{\sqrt{3}}{2}, \frac{3}{2})$. The Dirac points of the two layers are related by $\bm{K}_1 - \bm{K}_2 = \frac{1}{3}\bm{G}_1 + \frac{2}{3}\bm{G}_2$. In Eq.~\ref{eq:HK0_ABAB}, we also define
\begin{align}
    H_0 &= v\bm{k} \cdot \bm{\sigma} \\
H^A_0(\bm{k}) &= H_0(\bm{k}) + \frac{1}{2}(I_2 - \sigma^z)\Delta^\prime \\
    H^B_0(\bm{k}) &= H_0(\bm{k}) + \frac{1}{2}(I_2 + \sigma^z)\Delta^\prime \\
        g(\bm{k}) &= \begin{pmatrix}
        -\hbar v_4 k_- & -\hbar v_3 k_+ \\
       \gamma_1 & -\hbar v_4 k_-
    \end{pmatrix},
\end{align}
where $\bm{\sigma} = (\sigma^x, \sigma^y)$ are the Pauli matrices in sublattice space, $v$ is the graphene Fermi velocity, and $k_\pm = k_x \pm ik_y$.  $H^A_0$ and $H_0^B$ capture the physics within each layer of graphene, and $g$ represents hopping between the two layers within each Bernal bilayer stack. The parameters are given by $\gamma_0 = 2.610$~eV, $\gamma_1 = 0.361$~eV, $\gamma_3 = 0.283$~eV, $\gamma_4 = 0.138$~eV and $\Delta^\prime = 0.015\,$eV~\cite{jung_accurate_2014}. The $v_i$'s are related to the $\gamma_i$'s by $v_i = \frac{3a}{2\hbar}\gamma_i$ and $v \equiv v_0$. The interlayer moir\'e coupling is
\begin{equation}
    U = T_1 + T_2e^{i(\bm{G}_1 + \bm{G}_2) \cdot \bm{r}} + T_3e^{i\bm{G}_2 \cdot \bm{r}},
\end{equation}
where
\begin{equation}
    T_{n + 1} = w_{\text{AA}} + w_{\text{AB}}(\sigma^x \cos(n\phi) + \sigma^y \sin(n\phi)),
\end{equation}
with $\phi = 2\pi/3$. We take $w_{AB} = 0.11$~eV, $w_{AA} = 0.08$~eV, and consider a twist angle $\theta = 1.2^\circ$ unless otherwise specified.

The continuum Hamiltonian in valley $K^\prime$ is determined by $H_{K^\prime}(\bm{k}) = H^*_K(-\bm{k})$ due to TRS. Unless otherwise stated, we assume $SU(2)_\text{s}$ spin-rotation symmetry for  graphene-based materials due to the negligible spin-orbit coupling. An important exception is twisted homobilayer MoTe$_2$ in Sec.~\ref{subsec:MoTe2_model}.

\subsubsection{Twisted double-bilayer graphene (ABBA)}

TDBG can also be fabricated in an alternative ABBA stacking arrangement~\cite{du_ferroelectricity_2024, liu_isospin_2022, he_symmetry-broken_2023}. Using the conventions established in the preceding subsection,
the single-particle Hamiltonian in valley $K$ is $H_K = H_{K,0} + H_V$, where $H_{K,0}$ is given by~\cite{koshino_band_2019, liu_quantum_2019-1}
\begin{equation}
H_{K, 0} = \begin{pmatrix}
    H^A_0(\bm{k}_1) & g(\bm{k}_1) & & \\
    g^\dagger(\bm{k}_1) & H^B_0(\bm{k}_1) & U & \\
    & U^\dagger & H^B_0(\bm{k}_2) & g(\bm{k}_2) \\
    & & g^\dagger(\bm{k}_2) & H^A_0(\bm{k}_2)
\end{pmatrix},
\end{equation}
and $H_V$ is defined in Eq.~\ref{eq:H_V}. All single-particle parameters are identical to those used in the previous subsection.

\subsubsection{Twisted monolayer-bilayer graphene}

Twisted monolayer-bilayer graphene (TMBG)~\cite{chen_electrically_2021, he_competing_2021, li_imaging_2022, polshyn_electrical_2020, polshyn_topological_2022, xu_tunable_2021, zhang_local_2023} is $\hat{\mathcal{C}}_{2z}$-breaking heterostructure that consists of one graphene monolayer and one Bernal graphene bilayer stacked together with a twist. Using the conventions established in Sec.~\ref{subsec:ABAB_model}, the single-particle Hamiltonian for TMBG in valley $K$ is given by~\cite{liu_quantum_2019-1, ledwith_family_2022}
\begin{equation}
H_K = \begin{pmatrix}
    H_0(\bm{k}_1) - \frac{1}{2}\Delta V & U & \\
    U^\dagger & H^A_0(\bm{k}_2) & g(\bm{k}_2) \\
    & g^\dagger(\bm{k}_2) & H^B_0(\bm{k}_2) + \frac{1}{2}\Delta V
\end{pmatrix}.
\end{equation}
All single-particle parameters are identical to those in Sec.~\ref{subsec:ABAB_model}.

\subsubsection{Helical trilayer graphene}

Helical trilayer graphene (HTG) is a supermoir\'e material that consists of three layers of graphene that are rotated consecutively with identical twist angles $\theta$. While HTG preserves $\hat{\mathcal{C}}_{2z}$ on the global scale, relaxation calculations show that it relaxes into large moir\'e-periodic domains (referred to as h-HTG and $\bar{\text{h}}$-HTG) of linear dimension $\sim 100\,\text{nm}$ (near the magic angle $\theta~\sim 1.8^\circ$), that individually break $\hat{\mathcal{C}}_{2z}$ and are mapped into each other under $\hat{\mathcal{C}}_{2z}$~\cite{xia2023helical,devakul2023HTG,nakatsuji2023multiscale}. The interacting physics of HTG can then be understood by focusing on the properties of the individual domains.

We  closely follow the conventions of Ref.~\cite{kwan2024strong}. Considering valley $K$ and h-HTG without loss of generality, the continuum Hamiltonian is 
$H_K = H_{K,0} +H_V$, where
\begin{equation}
H_{K,0}=\begin{pmatrix}
    -iv_F\bm{\sigma}\cdot\nabla & T(\bm{r}-\bm{d}_t) & 0 \\
    T^\dagger(\bm{r}-\bm{d}_t) & -iv_F\bm{\sigma}\cdot\nabla & T(\bm{r}-\bm{d}_b)\\
    0 & T^\dagger(\bm{r}-\bm{d}_b) & -iv_F\bm{\sigma}\cdot\nabla 
\end{pmatrix}
\end{equation}
with $v_F=8.8\times 10^5\,\text{ms}^{-1}$, and $H_V = \text{diag}(-1/2,0,1/2)\Delta V $. Note that the wavefunction in each layer is written relative to its Dirac point (this is very slightly strained to accommodate a moir\'e periodicity), and as a consequence the relevant interlayer tunneling are shifted by $\bm{d}_{t,b}$ in the top and bottom layer. The unshifted tunneling matrix is given by\footnote{The interlayer tunneling here does not include effects of the momentum-dependence of the tunneling~\cite{xia2023helical}. See App.~\ref{secapp:HTG} for a discussion of this correction.}
\begin{equation}
\begin{gathered}
    T(\bm{r})=\begin{pmatrix}
        w_{AA}t_0(\bm{r}) & w_{AB}t_{-1}(\bm{r})\\
        w_{AB}t_1(\bm{r}) & w_{AA}t_0(\bm{r})
    \end{pmatrix}\\
    t_\alpha(\bm{r})=\sum_{n=0}^{2}e^{\frac{2\pi i}{3}n\alpha}e^{-i\bm{q}_n\cdot\bm{r}}\\
    q_{n,x}+iq_{n,y}=-ik_\theta e^{\frac{2\pi i}{3}n},
\end{gathered}
\end{equation}
where $\bm{K}_{1,3}=\mp \bm{q}_0+\bm{K}_2$. We use $w_{AA}=75\,\text{meV}$ and $w_{AB}=110\,\text{meV}$. The relative shifts of the layers in h-HTG are parameterized by $\bm{d}_t-\bm{d}_b=\bm{\delta}=\frac{1}{3}(\bm{a}_2-\bm{a}_1)$, where $\bm{a}_{1,2}=\frac{4\pi}{3k_\theta}(\pm\frac{\sqrt{3}}{2},\frac{1}{2})$.  

\subsubsection{Twisted homobilayer MoTe$_2$}
\label{subsec:MoTe2_model}
We closely follow the conventions of Refs.~\cite{Yu2024FCIversus,Jia2024MFCI1} in our description of the  continuum model~\cite{Wu2019tmds} for the valence bands of AA-stacked homobilayer MoTe$_2$ with small twist angle ($t$MoTe$_2$). This system also lacks $\hat{\mathcal{C}}_{2z}$, and due to strong spin-orbit coupling, the valence bands are spin-valley locked such that $\tau=+$ ($\tau=-$) is tied to spin-$\uparrow$ (spin-$\downarrow$). In valley $K$, the rotated $K$-point of the top layer is at $\bm{K}_t=\frac{4\pi}{3a_0}(\cos\theta/2,-\sin\theta/2)$, while that of the bottom layer is at $\bm{K}_t=\frac{4\pi}{3a_0}(\cos\theta/2,\sin\theta/2)$, where $a_0=0.352\,\text{nm}$ is the lattice constant. In terms of $\bm{q}_1=\bm{K}_b-\bm{K}_t$, we have the basis moir\'e RLVs $\bm{b}_1=(\hat{\mathcal{C}}_3^2-\hat{\mathcal{C}}_3)\bm{q}_1$ and $\bm{b}_2=-\hat{\mathcal{C}}_3^2\bm{b}_1$, where $\hat{\mathcal{C}}_3$ is a counterclockwise rotation by $2\pi/3$.

In layer space, the real-space continuum Hamiltonian takes the form
\begin{equation}
    H_K=\begin{pmatrix}
    h_{\tau,b}(\bm{r})&t_\tau(\bm{r})\\
    t_\tau^*(\bm{r})&h_{\tau,t}(\bm{r})
\end{pmatrix}.
\end{equation}
The intralayer term (in a layer-dependent frame centered at the corresponding Dirac momentum) is 
\begin{equation}
        h_{\tau,l}(\bm{r})=\frac{\hbar^2\nabla^2}{2m^*}+V_{\tau,l}(\bm{r})+(-1)^l\frac{\Delta V}{2},
\end{equation}
where $m^*$ is the effective mass. The intralayer moir\'e potential $V_{\tau,l}(\bm{r})$ and interlayer hopping $t_\tau(\bm{r})$ are, at the first harmonic level, given by
\begin{gather}
    V_{\tau,l}(\bm{r})=Ve^{-(-1)^li\psi}\sum_{i=1,2,3}e^{i\bm{g}_i\cdot\bm{r}}+
    \text{c.c.}\\ 
    t_\tau(\bm{r})=w\sum_{i=1,2,3}e^{-i\tau\bm{q}_i\cdot\bm{r}},
\end{gather}
where $(-1)^t=1$ and $(-1)^b=-1$, and $\bm{g}_i=\hat{\mathcal{C}}_3^{i-1}\bm{b}_1$.

In the main text, we focus on twist angles $\theta=3.4^\circ-4.0^\circ$, and use the parameters of Ref.~\cite{Wang2024FCIMoTe2}, i.e.~$m^*=0.6m_e,w=-23.8\,\text{meV},V=20.8\,\text{meV},\psi=-107.7^\circ$. We have also checked that our conclusions on the nature of the IVC state remain unchanged using the parameters of Ref.~\cite{Reddy2023FQAHMoTe2}, i.e.~$m^*=0.62m_e,w=-13.3\,\text{meV},V=11.2\,\text{meV},\psi=-91^\circ$. However, we have not exhaustively explored the extended parameter space of the relevant models in full, leaving this as a subject for future work.

\subsubsection{Twisted symmetric trilayer graphene}
Twisted symmetric trilayer graphene (TSTG)~\cite{Cao2021TTGPauli, hao_electric_2021, zhang2022promotion, kim_evidence_2022, kim2023imaging, li_observation_2022, lin_zero-field_2022, Liu2022TTGisospin, Park2021TTGSC, Shen2022TTGDirac, Twistons, yang_wafer-scale_2022, zhang_angle-resolved_2024} is a $\hat{\mathcal{C}}_{2z}$-symmetric heterostructure that consists of three sheets of monolayer graphene stacked together, with the middle layer twisted by $\theta$ relative to the top and bottom layers. Using the conventions established in Sec.~\ref{subsec:ABAB_model}, the single-particle Hamiltonian for  TSTG in valley $K$ is given by~\cite{li_electronic_2019, khalaf_magic_2019,calugaru2021TSTG1}
\begin{equation}
H_K = \begin{pmatrix}
    H_0(\bm{k}_1) - \frac{1}{2}\Delta V & U & \\
    U^\dagger & H_0(\bm{k}_2) & U^\dagger \\
    & U & H_0(\bm{k}_1) + \frac{1}{2}\Delta V
\end{pmatrix}.
\end{equation}
We use $w_{AB} = 0.11$~eV, $w_{AA} = 0.075$~eV, $\gamma_0 = 2.73$~eV (which defines the graphene Fermi velocity as $v = \frac{3a}{2\hbar}\gamma_0$) and twist angle $\theta = 1.56^\circ$. 

\subsection{Details of Hartree-Fock calculations}
\label{subsec:HFdetails}

We will be interested in electronic filling factors $\nu$ near charge neutrality, which corresponds to $\nu=0$. Since interaction-induced phenomena primarily involve just a few low energy moir\'e bands, which are often energetically separated from higher `remote' bands, we project our calculations into a subset of `active' bands. These are associated with band creation operators $c^\dagger_{\bm{k},f,a}$, where $\bm{k}$ is a momentum in the moir\'e BZ (mBZ), $f$ is a spin-valley flavor index, and $a$ is a band index. For both variants of TDBG, TMBG, and h-HTG, we take the active bands to be the central two moir\'e bands per spin and valley. For $t$MoTe$_2$, we project onto the highest three valence bands per valley. For TSTG, we project onto the central four bands per spin and valley.

For the interaction part of the Hamiltonian, we include density-density interactions with gate-screened Coulomb potential $V(q) = (e^2/2\epsilon_0\epsilon_rq) \tanh qd_\text{sc}$, where $d_\text{sc}$ is the screening length. For both variants of TDBG, TMBG, h-HTG, and TSTG, we use $d_\text{sc}=25\,\text{nm}$. For $t$MoTe$_2$, we use $d_\text{sc}=10\,\text{nm}$. To capture the uncertainty in the interaction strength, we take the relative permittivity $\epsilon_r$ to be a variable parameter. 

To avoid double-counting interactions, when writing down the quartic Coulomb coupling we normal order with respect to a background or reference density matrix $P^\text{ref}$, that describes the `vacuum' state around which we measure charge fluctuations. For both variants of TDBG, TMBG, and h-HTG, $P^\text{ref}$ consists of occupied remote valence bands, and occupation of $1/2$ within the active central bands (this is an example of the so-called `average' scheme). For $t$MoTe$_2$, the reference density matrix corresponds to filling the valence bands to charge neutrality. For TSTG, the two central bands have occupation $1/2$ and the remote valence bands are occupied in $P^\text{ref}$.

We address the resulting interacting problem using self-consistent Hartree-Fock (HF) mean field simulations. These have been shown to be a versatile and reliable tool for addressing the physics of moir\'e materials at or near integer fillings. The effectiveness of HF can be motivated in part from  strong coupling~\cite{KangVafekPRL,bultinck_ground_2020,TBG4}, where for example in the case of TBG it is possible to show that it is exact in certain specialized limits in which the ground states are exact Slater determinants. Indeed, in cases where has been possible to compare against alternative beyond-mean field approaches such as exact diagonalization or density-matrix renormalization group (DMRG), these have often shown good agreement with the  HF, which is numerically far less costly for a given system size. For example,  DMRG studies of strained TBG at integer filling $\nu=-3$~\cite{wang2022kekule} find IKS order whose properties are consistent with those of the mean-field state originally identified using HF in Ref.~\onlinecite{kwan_kekule_2021}.

For the graphene-based moir\'e materials, we restrict to spin-collinear calculations, i.e.~we constrain the density matrix $P_{f,a;f',b}(\bm{k},\bm{k}')=\langle c^\dagger_{\bm{k},f,a}c_{\bm{k}',f',b}\rangle$ to be diagonal in spin. Since we are interested in valley spirals, we enforce a generalized translation invariance parameterized by a boost wavevector $\bm{q}$. This means that IVC is only permitted at wavevector $\bm{q}$, such that $\bm{k}$ in valley $\tau=+$ hybridizes with $\bm{k}+\bm{q}$ in valley $\tau=-$, leading to an intervalley spiral. On the other hand, valley-diagonal observables yield moir\'e translation-invariant expectation values. We perform calculations over a range of $\bm{q}$, and select the HF solution with the lowest energy. For both variants of TDBG and TMBG, we sweep over all values of $\bm{q}$ on the momentum grid. For all other materials, we restrict to $\hat{\mathcal{C}}_{3z}$-symmetric boosts $\bm{q}=\Gamma_M,K_M,K'_M$ for simplicity.

\section{Results}\label{sec:results}
\subsection{Twisted double-bilayer graphene (ABAB)}\label{subsec:ABAB}

\begin{figure}
    \centering
    \includegraphics[width = \linewidth]{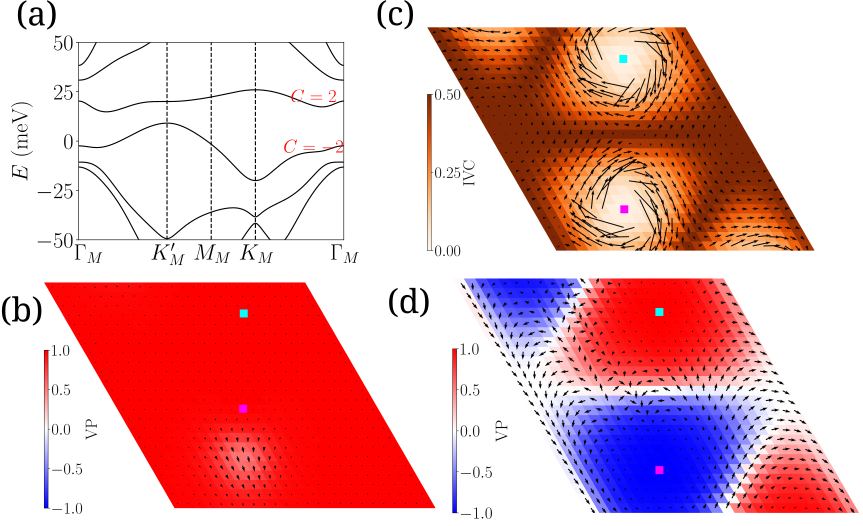}
    \caption{\textbf{ABAB-stacked twisted double-bilayer graphene (TDBG) at $\theta = 1.2^\circ$.} 
    a) Single-particle band structure at $\Delta V = 40$~meV in valley $K$ ($\tau=+$). 
    b) Momentum-dependent valley polarization (VP), shown as color, and intervalley coherence (IVC), shown with arrows, of the $\nu=+2$ tilted valley polarized (TVP) state at $\Delta V = 40$~meV and $\epsilon_r = 17$, from $24 \times 24$ HF. Cyan (magenta) square denotes $\Gamma_M$ point of valley $K$ ($K^\prime$). We also note that TVP has incommensurate $\bm{q}$, as the minimum of the non-interacting conduction band is not at a high-symmetry momentum. c) The IVC intensity, shown as color, and gauge-invariant velocity (Eq.~\ref{eq:invariant_current}), shown with arrows, of the $\nu=+2$ CTI$_2$ at $\Delta V = 40$~meV and $\epsilon_r = 20$, from $24 \times 24$ HF. Diverging velocities very close to vortex cores are not shown. d) Same as b) except for the CTI$_2$ shown in c). We observe that the intervalley order winds by $8\pi$ around the mBZ, which is consistent with the Chern number difference of 4 between the conduction band in the $K$ and $K^\prime$ valleys. For b,c,d), we only show the quantities in one spin sector.}
    \label{fig:ABAB}
\end{figure}

\begin{figure*}
    \centering
    \includegraphics[width = \linewidth]{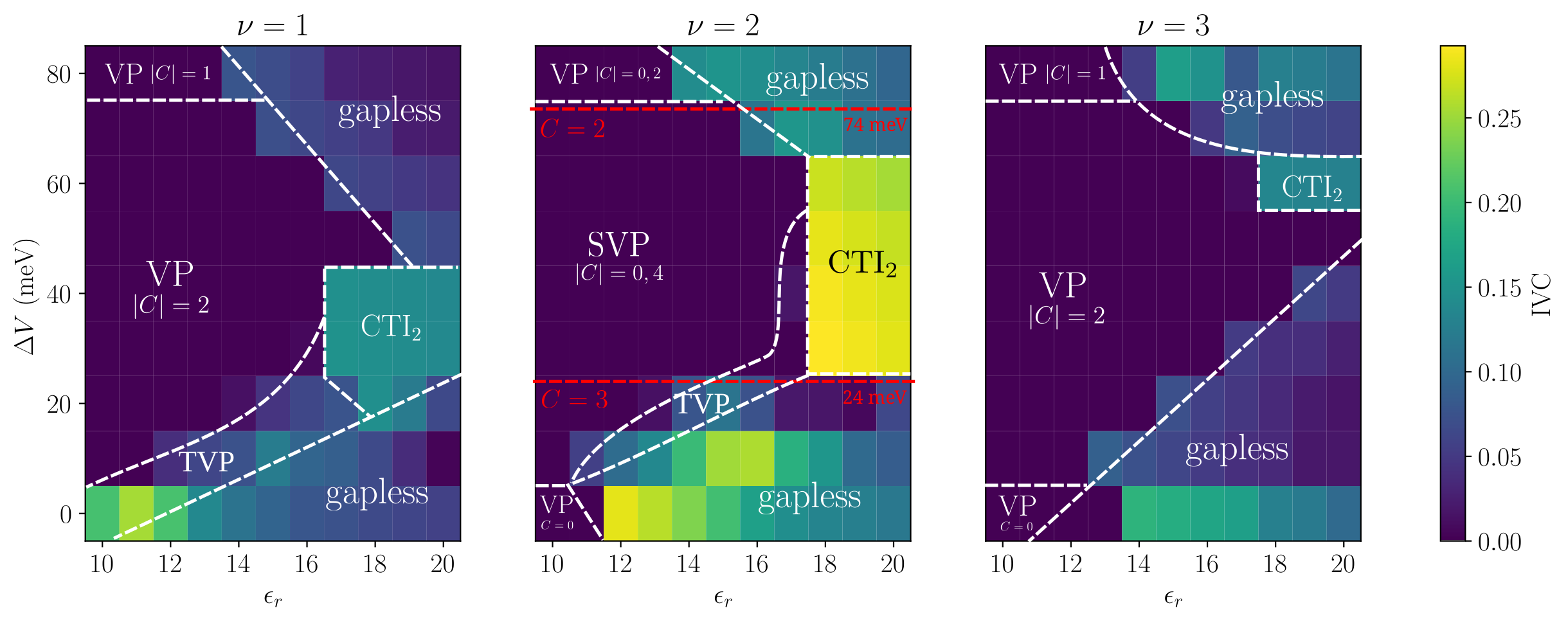}
    \caption{\textbf{Phase diagram of ABAB-stacked twisted double-bilayer graphene (TDBG) at $\theta = 1.2^\circ$.} The results are obtained using $12 \times 12$ HF at $\nu = 1, 2, 3$. VP denotes valley-polarized states, which also carry a non-zero spin polarization. SVP denotes spin- or valley-polarized states, which are degenerate in HF. TVP denotes tilted valley-polarized states. Topological transitions in the non-interacting band structure are marked in red, with the Chern number labels that of the central conduction band.}
    \label{fig:abab123}
\end{figure*}

As shown in Fig.~\ref{fig:ABAB}a, the single particle band structure of TDBG with ABAB stacking order at $\theta = 1.2^\circ$ has well-isolated Chern bands with $|C|=2$ under moderate interlayer potential $\Delta V$~\cite{koshino_band_2019, liu_quantum_2019-1, zhang_nearly_2019}. Within this window of $\Delta V$, self-consistent HF calculations at $\nu=1, 2, 3$ find flavor-polarized strong-coupling insulators at strong interaction strengths (i.e.~small relative permittivity $\epsilon_r$), as shown in Fig.~\ref{fig:abab123}. For weaker interactions, we find a greater tendency towards gapless metallic phases. However for all the fillings investigated here, we obtain insulating CTIs for a window of finite interlayer potentials.

We now focus on $\nu = 2$, where the CTI is observed over a large region of parameter space. The behaviour of the CTI at $\nu = 2$ is identical in both spin sectors, so we only show results for one spin sector. With electron doping and well-separated conduction and valence bands, we expect the non-trivial physics to occur within the non-interacting conduction bands. As such, we define the IVC order parameter $\Delta_{\bm{q}}(\bm{k})$ as
\begin{equation}\label{eq:IVC_OP_q}
    \Delta_{\bm{q}}(\bm{k})\equiv\langle c^\dagger_{\bm{k},+}c_{\bm{k}+\bm{q},-} \rangle
\end{equation}
where $c^\dagger_{\bm{k}\tau}$ is the spin-up electron creation operator in the conduction band of valley $\tau$. The order parameter, shown in Fig.~\ref{fig:ABAB}d, shows vortices with $8\pi$-winding around the moir\'e Brillouin zone (mBZ), which is topologically required due to the Chern numbers $C=\pm2$ of the constituent conduction bands. This leads us to identify the phase as a CTI$_2$. We note that the vorticity is carried by two vortices with winding number 2. While the order parameter is dependent on the gauge choice\footnote{App.~\ref{secapp:gauge_fixing} details how we implement a smooth but non-periodic gauge-fixing.}, we also define the gauge-independent velocity as
\begin{equation}\label{eq:invariant_current}
    \bm{j}_{\bm{k}} = \bm{\nabla}\theta_{\bm{k}} + \bm{A_{\bm{k},+}} -  \bm{A_{\bm{k}+\bm{q},-}}
\end{equation}
where $\theta_{\bm{k}} = \arg(\Delta_{\bm{q}}(\bm{k}))$ is the angle of the IVC order parameter and $\bm{A}_{\bm{k}\tau} = i\braket{u_{\bm{k}\tau}|\partial_{\bm{k}}u_{\bm{k}\tau}}$ is the Berry-connection. As shown in Fig.~\ref{fig:ABAB}c, the velocity diverges near the vortices, which is compensated by the vanishing IVC order parameter. In addition to the CTI$_2$, we have also found a sliver of the tilted valley-polarized (TVP) phase (Fig.~\ref{fig:ABAB}b), which was introduced in Ref.~\cite{companion}.

The physics of CTI in $\nu = 1, 3$ is essentially identical to that in $\nu = 2$, except that for these odd integer fillings, IVC is only present in one of the two spin sectors (namely, the spin sector with filling factor 1 relative to its charge neutrality). We note that the interaction-induced band normalization depends on the total filling, leading to the CTI being found at different regions of parameter space for different fillings.

\begin{figure}
    \centering
    \includegraphics[width = \linewidth]{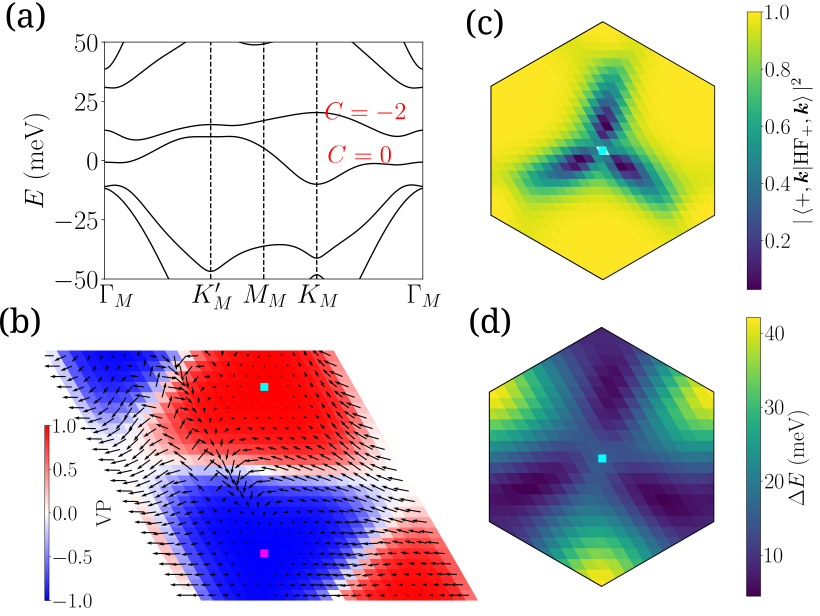}
    \caption{\textbf{ABBA-stacked twisted double-bilayer graphene (TDBG) at $\theta = 1.2^\circ$.} 
    a) Single-particle band structure at $\Delta V = 20$~meV in valley $K$ ($\tau=+$). The two central bands are separated by a direct gap, making the Chern number well-defined for each band, but they have a negative indirect gap. 
    b) Valley polarization (VP), shown as color, and intervalley coherence (IVC) in the valley-filtered basis, shown with arrows, of the $\nu=+2$ CTI$_{1}$ state at $\Delta V = 20$~meV and $\epsilon_r = 16$ from $24 \times 24$ HF. Cyan (magenta) square denotes the $\Gamma_M$ point of valley $K$ ($K^\prime$) in b,c,d). 
    c) The momentum-resolved overlap in valley $K$ of the valley-filtered basis in b) and the single-particle conduction basis. We observe that in some regions, the overlap vanishes due to mixing with the single-particle valence band, allowing the Chern number of the valley-filtered bands to differ from the single-particle conduction bands. 
    d) The direct band gap between the HF valence and conduction bands in valley $K$ for a symmetry-preserving $16 \times 16$ HF calculation at $\nu=+2$. Comparing with c), we observe that band hybridization occurs where the energy difference is small.
    For b,c,d), we only show the quantities in one spin sector.}
    \label{fig:ABBA}
\end{figure}

\begin{figure*}
    \centering
    \includegraphics[width = \linewidth]{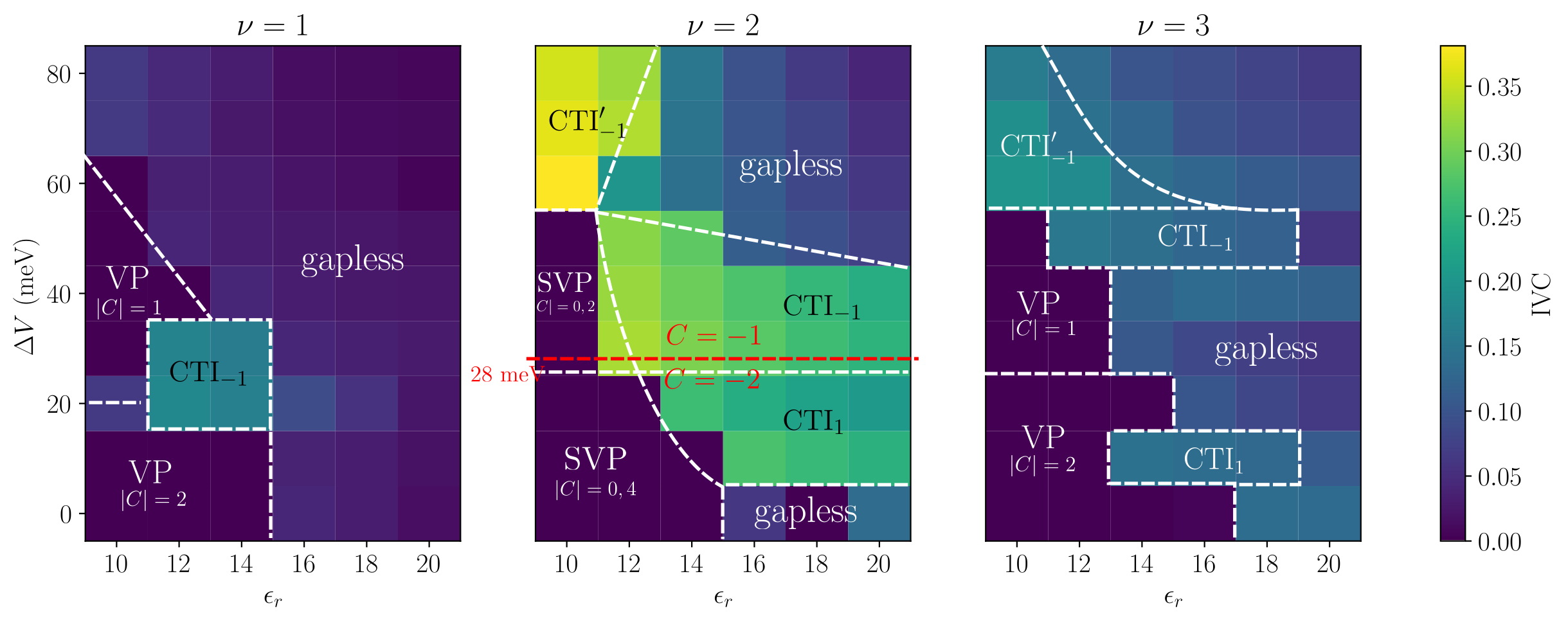}
    \caption{\textbf{Phase diagram of ABBA-stacked twisted double-bilayer graphene (TDBG) at $\theta = 1.2^\circ$.} The results are obtained using $12 \times 12$ HF at $\nu = 1, 2, 3$. VP denotes valley-polarized states, which also carry a non-zero spin polarization. SVP denotes spin- or valley-polarized states, which are degenerate in HF. TVP denotes tilted-valley-polarized states. Topological transitions in the non-interacting band structure are marked in red, with the Chern number refers to that of the central conduction band. CTI$'$ has a different intervalley spiral wavevector $\bm{q}$ from the unprimed CTI states.}
    \label{fig:abba123}
\end{figure*}

\subsection{Twisted double-bilayer graphene (ABBA)}\label{subsec:ABBA}

Similar to ABAB-stacked TDBG, we observe the presence of CTIs at all of $\nu = 1, 2, 3$ for ABBA-stacked TDBG (Fig.~\ref{fig:abba123}). We again focus on $\nu=+2$, which contains the largest region of gapped IVC insulators. However, we observe some complications not present in the simplest setting with just one band per valley. As discussed in the previous subsection, in the the regime where the CTI is observed in $\nu=+2$ ABAB-stacked TDBG, the conduction band is sufficiently separated from the valence band and other remote bands to prevent them from participating non-trivally in the IVC. In ABBA-stacked TDBG, at a small interlayer potential of $\Delta V=20$~meV, the central conduction bands are also topological with $|C|=2$~\cite{koshino_band_2019, liu_quantum_2019-1}, but they can significantly mix with the central valence bands owing to the lack of clear energetic separation (Fig.~\ref{fig:ABBA}a). 

To understand the nature of intervalley coherent phases obtained in HF calculations in such cases, we consider the following valley-filtered basis. If the HF spectrum contains an isolated\footnote{We briefly discuss the situation where multiple HF bands are proximate in energy in Sec.~\ref{subsec:TSTG}.} IVC band above or below $E_F$ (within a spin sector), we can unfold this HF band $\ket{\text{HF},\bm{k}}$ into two normalized valley-diagonal bands $\ket{\text{HF}_+,\bm{k}}$ and $\ket{\text{HF}_-,\bm{k}}$:
\begin{equation}\label{eq:IR_basis}
    \ket{\text{HF},\bm{k}}=\alpha(\bm{k})\ket{\text{HF}_+,\bm{k}}+\beta(\bm{k})\ket{\text{HF}_-,\bm{k}}.
\end{equation}
For the IVC states in ABBA-stacked TDBG at $\nu = 2$, we define the valley-filtered basis using the unfilled HF band immediately above $E_F$ in each spin sector, since it is isolated from the filled bands and higher remote conduction bands by charge gaps (the latter do not mix with the central bands in any significant way). We note that the valley-filtered basis for a CTI defined using Eq.~\ref{eq:IR_basis} is ambiguous at discrete points in each valley where $\ket{\text{HF},\bm{k}}$ is fully polarized into the other valley, but this can be resolved by assuming that the valley-filtered basis is continuous in $\bm{k}$-space. We also comment that even though we extract the valley-filtered basis from an unfilled HF band, we continue to define the order parameter as $\Delta_{\bm{q}}(\bm{k})\equiv\langle c^\dagger_{\bm{k},+}c_{\bm{k}+\bm{q},-} \rangle$, where $c^\dagger_{\bm{k}\tau}$ are \textit{electron} creation operators for the valley-filtered basis\footnote{The hole order parameter would differ by a minus sign.}.

At $\Delta V = 20$~meV, the single-particle conduction band has Chern number $C = -2$ $(+2)$ in valley $K$ ($K'$), while the valence bands are trivial. For strong interactions, HF calculations at $\nu=+2$ find flavor-polarized insulators (Fig.~\ref{fig:abba123}). For weaker interactions, we find a $\hat{\mathcal{T}}$ and $SU(2)_\text{s}$-symmetric IVC insulator\footnote{Within spin-collinear HF calculations, states with different IVC angles in the two spin sectors are degenerate with  $SU(2)_\text{s}$-symmetric IVC states.}. However, the corresponding valley-filtered basis is found to have $C = +1$ ($-1$) in valley $K$ $(K')$, which differs from that of the single-particle conduction bands, and the IVC order parameter winds by $4\pi$ instead of $-8\pi$ around the mBZ (Fig.~\ref{fig:ABBA}b). This means that we obtain a CTI$_1$, instead of a CTI$_{-2}$ that would be na\"ively expected by considering the non-interacting band topology. In Fig.~\ref{fig:ABBA}c, we plot the overlap of the valley-filtered band with the single-particle conduction band, which vanishes at three $\hat{\mathcal{C}}_{3z}$-related points in the mBZ.  This allows the valley Chern number of the valley-filtered basis to differ from that of the single-particle conduction basis. 

We note that the absolute value of the valley-filtered basis Chern number is lowered from 2 to 1. This is consistent with the notion that there is a tendency to reduce the winding (i.e.~`unfrustrate' the topological obstruction) of the order parameter, since this allows for more favorable exchange energetics. The system achieves this by substantially hybridizing with the single-particle valence band. Due to the kinetic energy penalty, such hybridization preferentially occurs at regions with small direct gap. We can roughly estimate the hybridization positions by performing symmetry-preserving HF at this filling, which partially accounts for interaction-induced renormalization effects (Fig.~\ref{fig:ABBA}d). In this specific example, the hybridization occurs away from high symmetry points. This restricts the valley-filtered basis Chern number to change by multiples of 3~\cite{fang_bulk_2012}, which does not permit a full lifting of the topological frustration. For larger $\Delta V$, we observe additional CTI$_n$ phases, including two CTI$_{-1}$ regions with distinct $\bm{q}$, highlighting the complex interplay between the kinetic energy and interaction effects. 

\begin{figure}
    \centering
    \includegraphics[width = \linewidth]{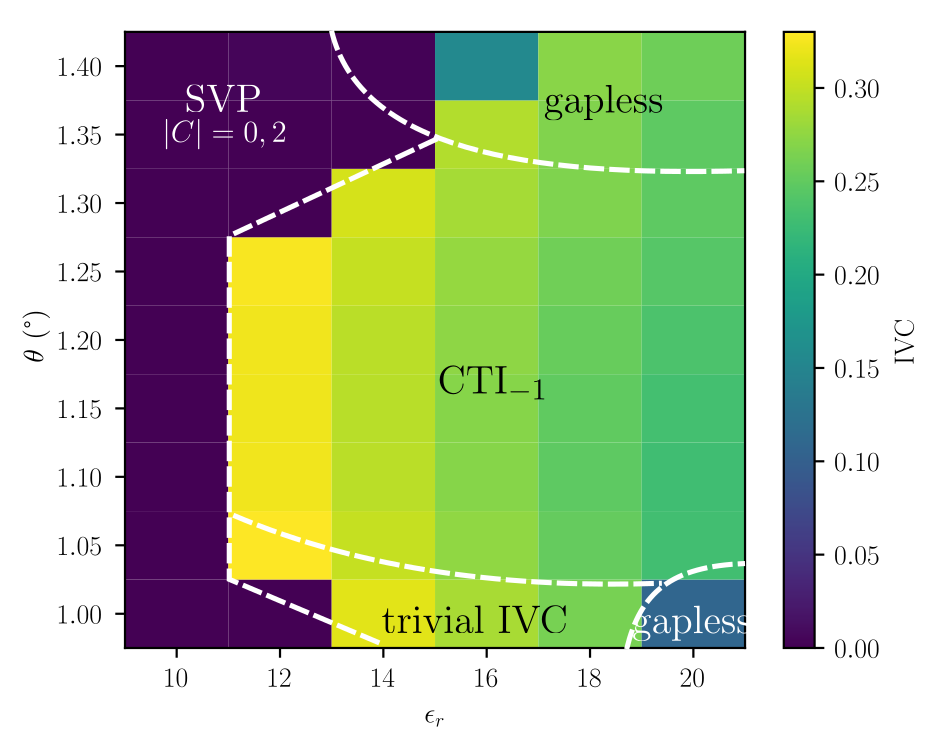}
    \caption{\textbf{Phase diagram of ABBA-stacked twisted double-bilayer graphene (TDBG) with variable twist angle.} The results are obtained using $12 \times 12$ HF at $\nu = 2$ and $\Delta V = 40$ meV. SVP denotes spin- or valley-polarized states, which are degenerate in HF. We observe that the CTI is stable under a range of twist angles.}
    \label{fig:abba_angle}
\end{figure}

We have also performed HF calculations at $\nu=2$ for different twist angles $\theta$ to confirm that the CTI in ABBA-stacked TDBG is stable under a range of twist angles (Fig.~\ref{fig:abba_angle}).

\subsection{Twisted monolayer-bilayer graphene}\label{subsec:TMBG}
\begin{figure}
    \centering
    \includegraphics[width = \linewidth]{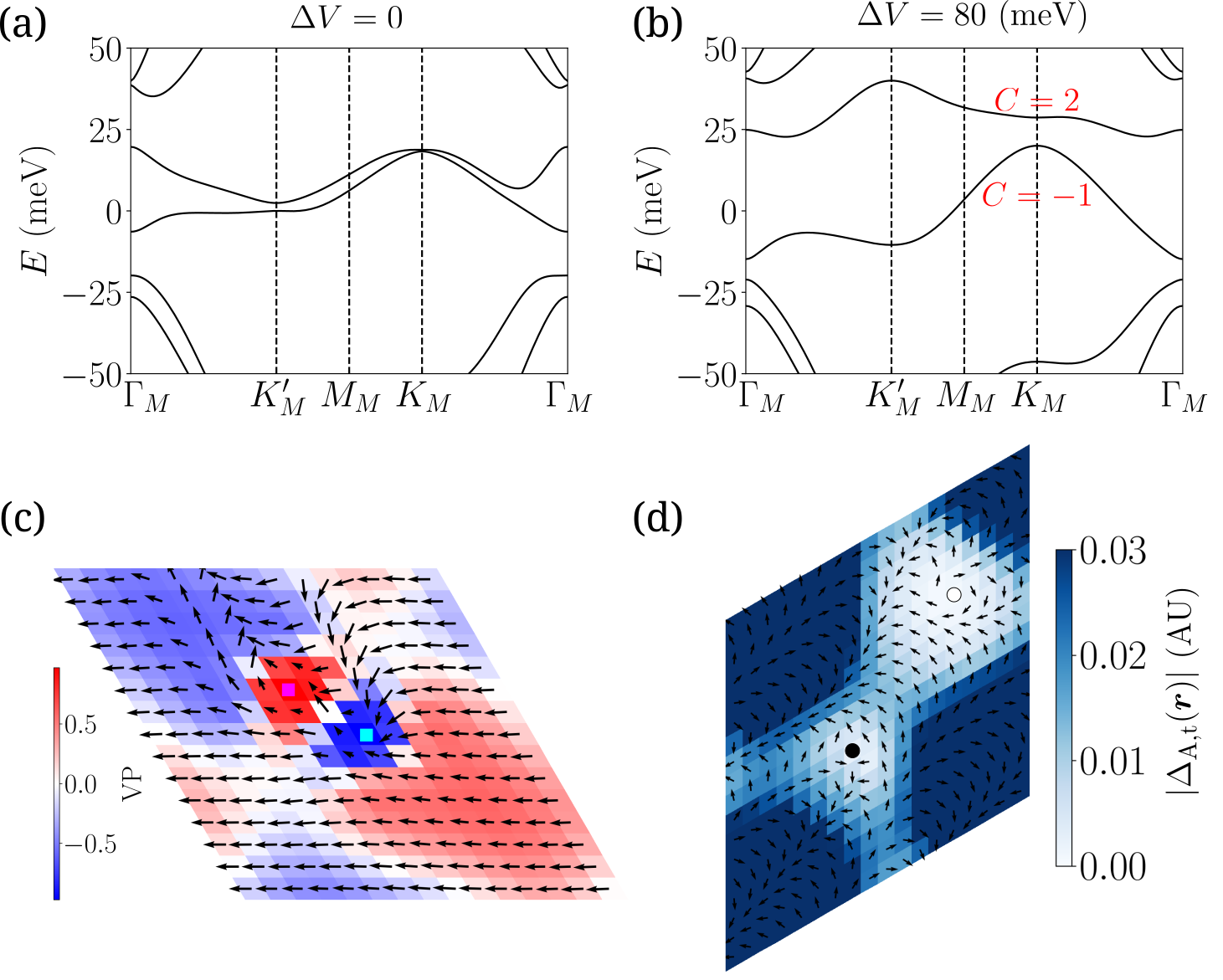}
    \caption{\textbf{Twisted monolayer-bilayer graphene (TMBG) at $\theta = 1.2^\circ$.} 
    a) Single-particle band structure at zero interlayer potential. The direct gap between the two central bands is nearly vanishing. 
    b) At finite interlayer potential of $\Delta V = 80$ (meV), the conduction and valence bands are well-isolated with Chern numbers $2$ and $-1$ respectively. 
    c) The \textit{momentum-space} valley polarization (VP), shown as color, and intervalley coherence (IVC) in the valley-filtered basis, shown with arrows, of the $\nu=+2$ trivial IVC state at $\Delta V = 0$ and $\epsilon_r = 12$ in $15 \times 15$ HF. Cyan (magenta) square denotes the $\Gamma_M$ point of valley $K$ ($K^\prime$). 
    d) The same state as in c), but showing the \textit{real-space} IVC order parameter \textit{within} a moir\'e unit cell for sublattice $A$ on the top layer. The color shows the IVC magnitude and the arrows show the IVC phase angle, where we have subtracted off the spiral modulation due to finite $\bm{q}$. There is a vortex-antivortex pair,  labelled with black and white circles. Note that the upper limit of the color scale has been clamped to show the regions near vortices more clearly.}
    \label{fig:tmbg_sp}
\end{figure}

\begin{figure*}
    \centering
    \includegraphics[width = \linewidth]{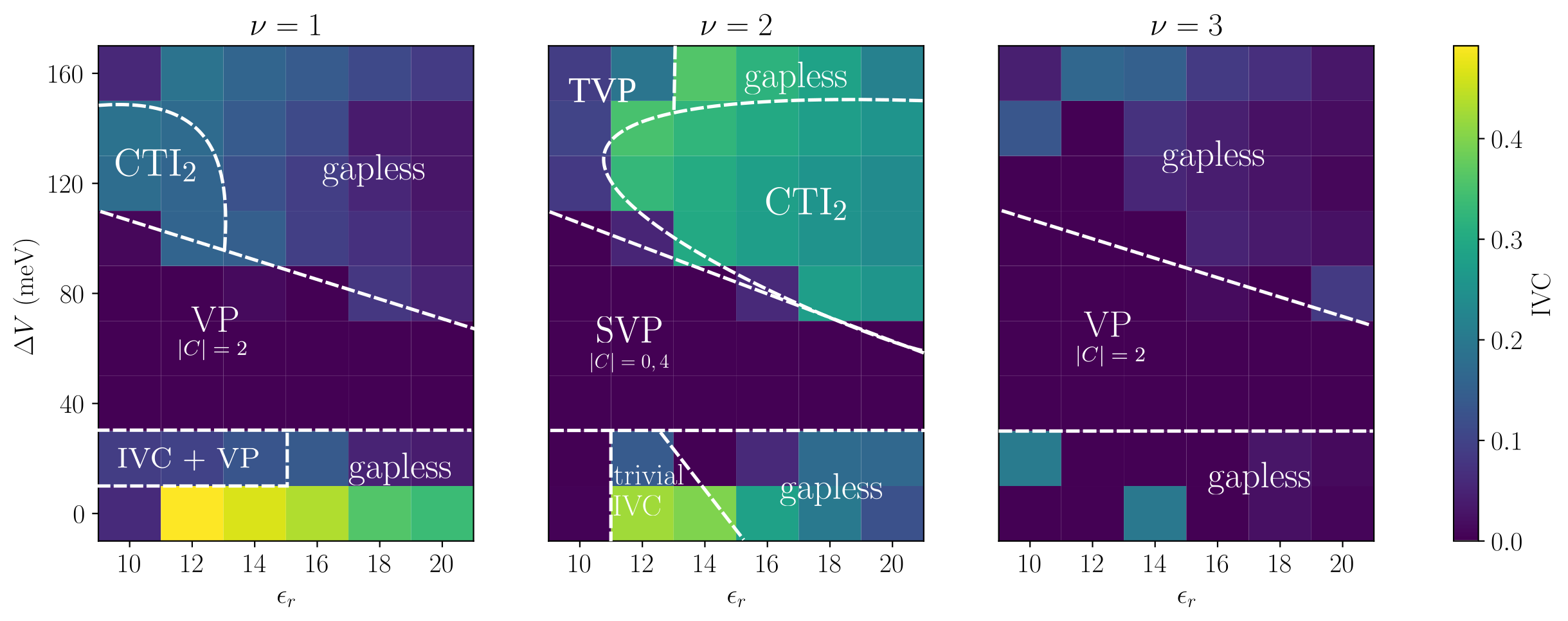}
    \caption{\textbf{Phase diagram of twisted monolayer-bilayer graphene (TMBG) at $\theta = 1.2^\circ$.} The results are obtained using $12 \times 12$ HF at $\nu = 1, 2, 3$. VP denotes valley-polarized states, which also carry a non-zero spin polarization. SVP denotes spin- or valley-polarized states, which are degenerate in HF. TVP denotes tilted-valley-polarized states. IVC + VP denotes a state that is IVC in one spin sector and VP in the other spin sector.}
    \label{fig:tmbg123}
\end{figure*}

Twisted monolayer-bilayer graphene (TMBG) is similar to ABAB-stacked TDBG in that it realizes an isolated conduction band with valley Chern number $2$ at finite interlayer potential~\cite{liu_quantum_2019-1,ledwith_family_2022} (Fig.~\ref{fig:tmbg_sp}). Its phase diagram is shown in Fig.~\ref{fig:tmbg123}. We find regions of CTI$_2$ at moderately high interlayer potential $\Delta V$ for $\nu = 1, 2$. At lower interlayer potentials, we also find a trivial IVC insulator with TRS at $\nu = 2$, where `trivial' means that the valley-filtered bands have vanishing Chern number (Fig.~\ref{fig:tmbg_sp}c). This allows the IVC to be non-vanishing across the mBZ. For such small $\Delta V$, the non-interacting conduction and valence bands are not isolated (Fig.~\ref{fig:tmbg_sp}a), leading to strong inter-band hybridization which alleviates the IVC obstructions. 

We stress that the IVC vortices discussed so far are momentum-space vortices, which are distinct from possible vortices in the real-space IVC order parameter. We define the latter as $\Delta_f(\bm{r}) = \braket{\psi^\dagger_{+, f}(\bm{r})\psi_{-, f}(\bm{r})}$, where $+,-$ are valley labels and $f$ indexes other degrees of freedom such as sublattice or layer. For the trivial IVC state in TMBG, we consider the real-space order parameter on sublattice $A$ of the top layer, and observe a vortex and anti-vortex within the moir\'e unit cell (Fig.~\ref{fig:tmbg_sp}d). Hence, we conclude that the presence or absence of real-space IVC vortices cannot be directly used to discriminate between a CTI and a trivial IVC state. However, detailed scanning of the intra-moir\'e cell IVC pattern is still invaluable in providing evidence for the existence of specific IVC states, as was done for the IKS in TBG in Ref.~\cite{nuckolls2023quantum}. 

In  Ref.~\cite{he_competing_2021}, insulating states with vanishing Chern number $C = 0$ were experimentally observed in TMBG at $\nu = 1$ with finite displacement field in multiple devices with twist angles $\theta\simeq 1.1^\circ - 1.2^\circ$. Since a fully flavor polarized state would have non-zero Chern number due to the single-particle band structure, while the CTI states have zero Chern number, this indirectly supports a CTI state. The theoretically calculated valley magnon instability with finite momentum transfer in Ref.~\cite{he_competing_2021} is also consistent with the finite-$\bm{q}$ nature our proposed CTI state\footnote{However, the authors of Ref.~\cite{he_competing_2021} conclude that the state breaks time-reversal symmetry, which differs from the CTI setting.}.

\subsection{Helical trilayer graphene}\label{subsec:hTTG}

\begin{figure*}
    \centering
    \includegraphics[width = \linewidth]{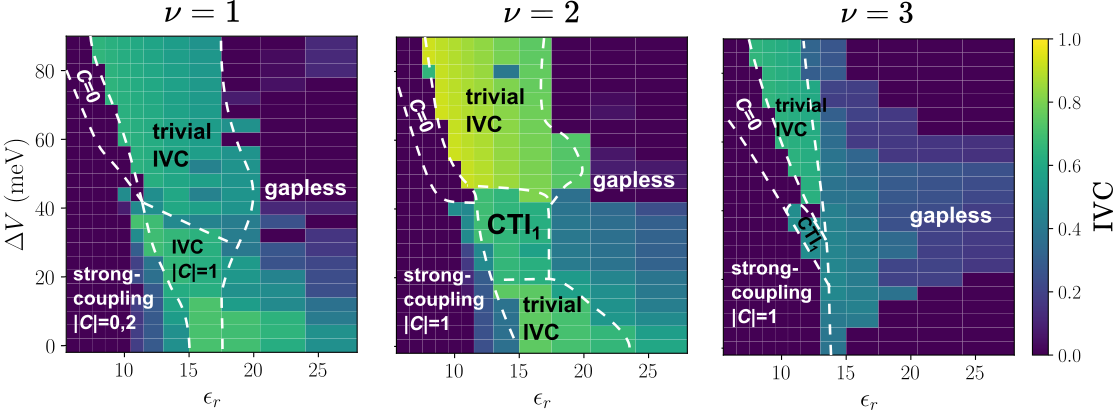}
    \caption{\textbf{Phase diagram of helical trilayer graphene (HTG) at $\theta=1.85^\circ$.} We focus on one h-HTG domain.  Phase diagram calculated using $12\times12$ HF at $\nu=1,2,3$. Momentum-dependent tunneling terms are not included.\label{fig:htg}}
\end{figure*}

Helical trilayer graphene (HTG) is a super-moir\'e platform that hosts large-scale moir\'e-periodic domains whose isolated central bands carry strong-coupling valley Chern numbers of up to 2~\cite{xia2023helical,devakul2023HTG,guerci2023chern,nakatsuji2023multiscale}. While HTG preserves $\hat{\mathcal{C}}_{2z}$ at the global scale, the domains  (labelled h-HTG and $\bar{\text{h}}$-TTG) individually break $\hat{\mathcal{C}}_{2z}$. The presence of Kekul\'e spirals with a commensurate IVC wavevector $\bm{q}$ was previously reported in HF studies at finite $\Delta V$~\cite{kwan2024strong}. 

In Fig.~\ref{fig:htg}, we show the phase diagram in the h-HTG domain of helical trilayer graphene at $\theta=1.85^\circ$. For strong interactions (small $\epsilon_r$), we have the `strong-coupling' phase which can be understood by fully occupying certain spin- and valley-diagonal Chern bands~\cite{kwan2024strong}. Near the phase boundary to the intermediate-coupling phases, there can be small amounts of IVC arising from momentum-local excitonic instabilities. For $\epsilon\sim 15$ and $\nu=+2$, we find three gapped IVC phases that preserve TRS. The two phases for small and large interlayer potential $\Delta V$ have trivial IVC, as diagnosed by the non-vanishing of IVC across the mBZ, and the fact that the valley-filtered basis has $C=0$. For intermediate $\Delta V$, we find a CTI$_1$ phase. For weaker interactions, HF yields gapless phases. We note that a CTI$_2$ can be obtained by applying an artifically large sublattice potential (projected to the central bands). This is because the Chern basis bands with predominant weight on sublattice $B$ ($A$) have $|C|=2$ ($|C|=1$). Hence by applying a large enough sublattice potential, the $B$ bands can be sufficiently separated from the $A$ bands so as to prevent interaction-induced mixing, and allow a CTI$_2$ constructed from the $B$ bands. 

For $\nu=3$, we find a much narrower CTI$_1$ region, while for $\nu=1$, we do not find any CTIs. In Sec.~\ref{secapp:HTG}, we compute the phase diagram for $\theta=1.75^\circ$, which yields sizable TR-symmetric insulating IVC regions, but none corresponding to CTIs. We also consider the inclusion of momentum-dependent tunneling in the single-particle continuum model, which captures the particle-hole asymmetry observed in the experimental phase diagram~\cite{xia2023helical}. Again, we do not find any CTIs, suggesting that such states may not be robustly present in HTG. 

\subsection{Twisted transition metal dichalcogenide homobilayers}\label{subsec:TMD}
\begin{figure}
    \centering
    \includegraphics[width = \linewidth]{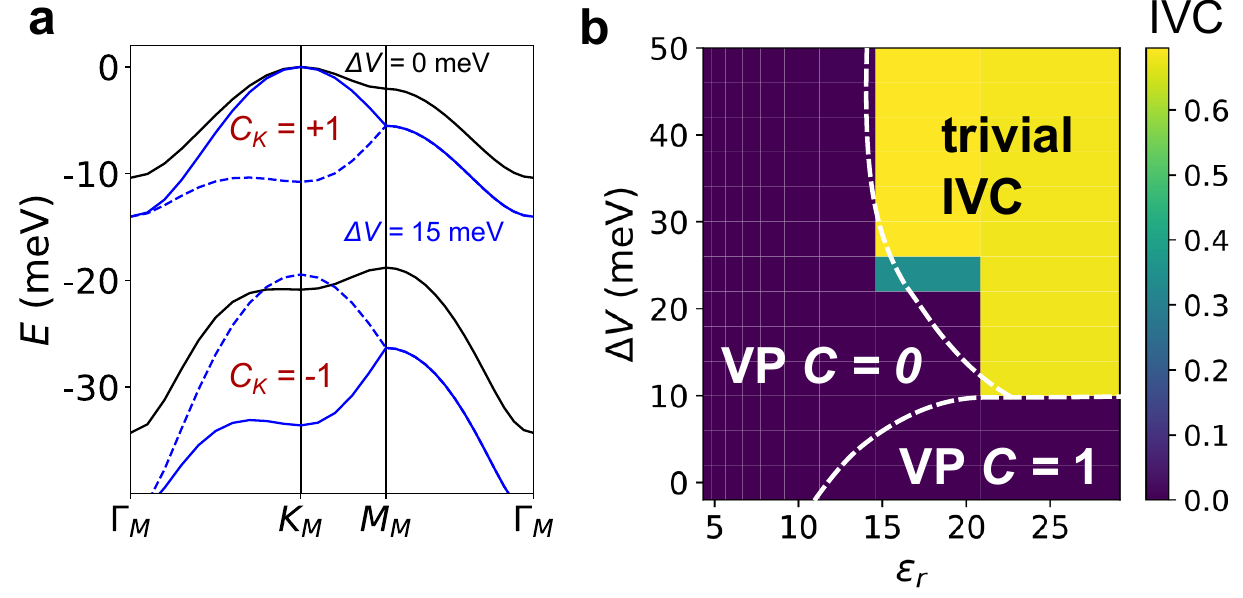}
    \caption{\textbf{Twisted bilayer MoTe$_2$ at $\theta=4^\circ$.} a) Single-particle valence band structure at $\Delta V=0\,$meV (black) and $15\,$meV (blue). Valley $K$ ($K'$) is shown with solid (dashed) lines. The band closest to charge neutrality has non-trivial valley Chern number for $\Delta V<26$\,meV. b) Phase diagram calculated using $12\times 12 $ HF at $\nu=-1$. VP indicates fully valley polarized (which also implies spin polarized due to spin-valley-locking). Trivial IVC denotes a $C=0$ layer-polarized phase with no intervalley frustration. Similar results are obtained for $\theta=3.4^\circ,3.7^\circ$.}
    \label{fig:tMoTe2}
\end{figure}

Twisted homobilayers of hexagonal TMDs such as MoTe$_2$ and WSe$_2$ are also ideal platforms for realizing narrow topological time-reversed bands. Unlike graphene-based systems, they have strong spin-orbit coupling such that the low-energy bands are spin-valley-locked to be either $K\uparrow$ or $K'\downarrow$~\cite{Wu2019}. Twisted bilayer MoTe$_2$ ($t$MoTe$_2$) has attracted considerable attention due to its displacement field-tuned~\cite{Anderson2023programming} Chern insulator at $\nu=-1$, and correlated topological phases, including fractional Chern insulators, at various other fillings~\cite{Cai2023y,Zeng2023,Xu2023,Park2023,kang2024observation}. 

We focus on $\theta=3.4^\circ-4.0^\circ$ $t$MoTe$_2$, where the lowest single-particle valence band has $C=+1$ ($-1$) in valley $K$ ($K'$) for small interlayer potentials $\Delta V$ as demonstrated with continuum model calculations (Fig.~\ref{fig:tMoTe2}a). This raises the question of whether a CTI$_1$ is feasible at $\nu=-1$. We note that the highest valence band is separated from the next valence band with opposite Chern number by $\lesssim 10$\,meV, and there is a topological phase transition that trivializes the highest valence band at around $\Delta V=26\,$meV. As shown in Fig.~\ref{fig:tMoTe2}b, HF yields a valley-polarized Chern insulator at $\Delta V=0$, while the ground state is a valley- and layer-polarized trivial insulator for strong interactions. For moderate interlayer potentials and weak/moderate interactions, we find a layer-polarized $\hat{\mathcal{T}}$-symmetric\footnote{Our definition of $\hat{\mathcal{T}}$ for this subsection interchanges $(K\uparrow)\leftrightarrow (K'\downarrow)$ and satisfies $\hat{\mathcal{T}}^2=1$.} IVC insulator~\cite{wang2023topological}. However, this is not a CTI, as evidenced by the non-vanishing of IVC across the mBZ and the trivial valley-filtered Chern number of the unfilled HF band. While this is not surprising for large $\Delta V$ above the single-particle topological transition, the fact that this IVC state remains unfrustrated for smaller $\Delta V<26$\,meV implies that interaction-induced hybridization with the next valence band removes the topological obstruction.

\subsection{Twisted symmetric trilayer graphene}\label{subsec:TSTG}

\begin{figure}
    \centering
    \includegraphics[width = \linewidth]{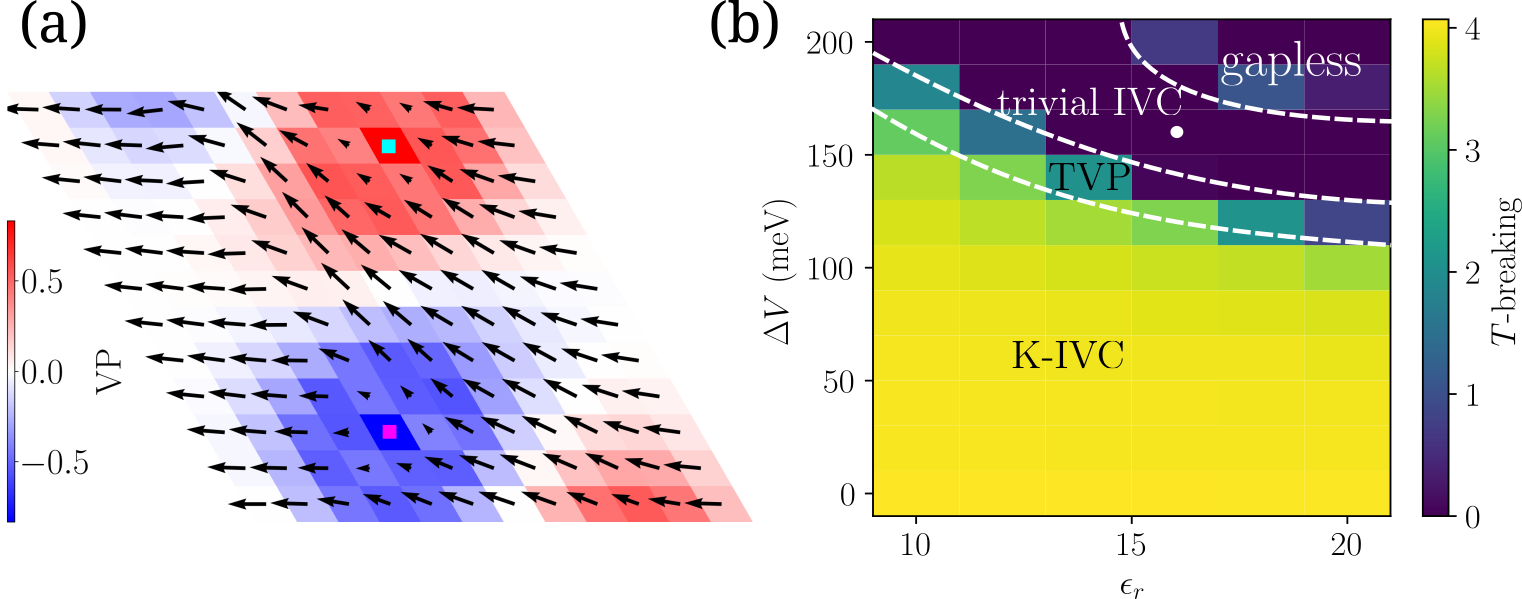}
    \caption{\textbf{Twisted symmetric trilayer graphene (TSTG) at $\theta = 1.56^\circ$.} (a) The order parameter of the IVC state at $\Delta V = 160$~meV and $\epsilon_r = 16$, from $12 \times 12$ HF. The extraction of the basis is explained in the text. No intervalley frustration is observed, as evidenced by lack of full valley polarization anywhere in the mBZ and zero winding of the order parameter around the mBZ. (b) Phase diagram from $12 \times 12$ HF at $\nu = 2$. TVP denoted tilted valley polarized state. Besides the region labelled `gapless', the K-IVC region is also not gapped. }
    \label{fig:TSTG}
\end{figure}

At the single-particle level, TSTG is $\hat{\mathcal{C}}_{2z}$-symmetric, and at zero or low interlayer potential, the HF ground states preserve this symmetry. However, at large interlayer potential, Ref.~\onlinecite{wang_kekule_2024} found a spontaneously $\hat{\mathcal{C}}_{2z}$-breaking intervalley coherent state with non-zero intervalley boost vector $\bm{q}$, which is commensurate in the absence of strain.  Fig.~\ref{fig:TSTG} shows the phase diagram of unstrained TSTG at $\nu=2$  in the displacement field-interaction plane, with the commensurate  Kekulé spiral region now characterized as `trivial IVC', for reasons we now describe.

Characterizing possible intervalley frustration in the IVC phase is more nuanced in TSTG than in the other materials discussed so far. In the previous cases, for $\nu>0$, there is only one empty band per spin sector that contains IVC (the higher remote bands do not qualitatively affect the physics), and the valley-filtered basis is naturally extracted from this band. In TSTG, since the single-particle band structure has no spectral gaps, there is no clear distinction between central and remote bands, and at the interacting level, there are multiple bands per spin sector both above and below the Fermi energy, many of which are not isolated. This makes the identification of a suitable valley-filtered basis difficult. Nevertheless, we find the \textit{first} band above Fermi energy to be isolated, at least for some parameters, from which we can extract a valley-filtered basis. The corresponding intervalley order parameter is shown in Fig.~\ref{fig:TSTG}a, which does not display winding of the IVC. As such, we conclude that the $\hat{\mathcal{C}}_{2z}$-breaking Kekul\'e spiral in TSTG is not a CTI.

\section{Discussion}\label{sec:discussion}

While we have investigated the HF phase diagrams for several moir\'e systems, as listed in Tab.~\ref{tab:moire_materials}, the moir\'e paradigm offers near limitless combinations of constituent materials and tuning knobs that can generate favorable conditions for realizing textured excitonic states. So far, we have only uncovered CTIs in graphene-based platforms. Our calculations within a narrow twist angle range $\theta=3.4^\circ-4.0^\circ$ for twisted homobilayer MoTe$_2$ in Sec.~\ref{subsec:TMD} have not yielded similar states. However, the single-particle models are known to vary significantly for different choices of TMDs and twist angles. For instance, for twisted bilayer WSe$_2$, as well as $t$MoTe$_2$ at smaller twist angles, density functional theory calculations have shown that the highest two valence moir\'e bands within a valley can have the same Chern numbers at zero displacement field~\cite{zhang2024polarizationdriven}. This may influence the robustness of CTIs against trivialization of the intervalley coherence. Topological phases have also been experimentally observed in TMD heterobilayers~\cite{li2021quantum}. Further exploration of the theoretical phase diagram is required to fully flesh out the nature of candidate IVC insulators in these TMD systems. Moir\'e platforms are natural potential hosts for CTIs because they often contain valleys which are mapped into each under time-reversal. More generally though, CTIs may arise as correlated insulators in other types of platforms if they are topological, are equipped with a time-reversal-odd $U(1)$ index, and exhibit competition between interactions and kinetic dispersion.

While we have found that several moir\'e materials favor the formation of CTIs in realistic parameter regimes, uncertainty in the modelling of the materials, in particular the effective interaction strength parameterized by $\epsilon_r$, precludes more detailed predictions in the absence of specific experimental details. Other single-particle effects not treated here, such as strain (which is essential for the IKS, an example of an ETI in twisted bilayer and trilayer graphene~\cite{companion}), would also impact the phase boundaries, including potentially widening the CTI regime. We also note that our calculations assume generalized translation symmetry on the moir\'e scale and hence ignores possible charge density wave (CDW) orders on this scale. Future work is required to establish the energetic competition of the CTI versus various moir\'e-scale density wave states.

CTIs are time-reversal invariant, meaning that, unlike Chern insulators, they do not produce a quantized charge Hall response, which makes their experimental detection more challenging. In Ref.~\onlinecite{he_competing_2021}, the authors detected a $C = 0$ insulating state at $\nu = 1$ in TMBG at finite interlayer potential. Since valley polarized states are expected to carry non-zero Chern numbers due to the non-interacting band topology, this provides an indirect piece of evidence for the existence of the CTI here. We note that the authors found non-vanishing anomalous Hall response at finite doping away from $\nu = 1$. An important direction therefore is to theoretically investigate whether time-reversal symmetry is broken when CTIs are doped away from integer fillings (as was studied in Ref.~\cite{wagner_global_2021} for the IKS in TBG, which is an ETI). In cases where time-reversal continues to be preserved, then the CTI may be a potential correlated insulating parent state that can be doped to yield superconductivity, although this seems to occur very rarely in  $\hat{C}_{2z}$-breaking moir\'e platforms.

Due to their finite intervalley coherence, CTIs in moir\'e systems are expected to produce translation symmetry-breaking patterns on the \emph{atomic} scale, such as $\sqrt{3} \times \sqrt{3}$ Kekul\'e charge patterns observed with scanning tunnelling microscopy in twisted graphene materials~\cite{nuckolls2023quantum, kim2023imaging,Princeton_STM_Theory,Berkeley_STM_Theory}. Detection of such patterns provides supporting evidence for the existence of IVC which is necessary for CTIs. Furthermore, the spiral wavevector $\bm{q}$ can be extracted from detailed analysis of the \emph{moir\'e}-scale modulation of the IVC~\cite{nuckolls2023quantum,kim2023imaging}, which can be compared with theoretical predictions.  We emphasize that, unlike the real-space IVC vortices detected in Ref.~\onlinecite{nuckolls2023quantum} \emph{within} a moir\'e unit cell, the valley pseudospin texturing we have discussed for CTIs occurs in momentum space. As we have illustrated with an explicit example from TMBG, it is possible for a state without momentum space vortices (a `trivial IVC' insulator by our criteria) to still carry vortices in the real-space IVC order parameter. Nevertheless, detailed real-space maps of the IVC texturing within a moir\'e cell can be fruitfully compared between experiment and theory to reinforce the interpretation of specific correlated states~\cite{nuckolls2023quantum}.

\begin{acknowledgements}
We thank Jonah Herzog-Arbeitmann, Andrei Bernevig, and Jiabin Yu for useful discussions. This work was supported by a Leverhulme Trust International Professorship (Grant Number LIP-202-014, ZW), by a University of Zurich postdoc grant (FK-23-134, GW),  by the European Research Council under the European Union Horizon 2020 Research and Innovation Programme via Grant Agreements No. 804213-TMCS (SAP) and No. 101076597-SIESS (NB), and by EPSRC Grants EP/S020527/1 and EP/X030881/1 (SHS). YHK is supported by a postdoctoral research fellowship
at the Princeton Center for Theoretical Science.

\end{acknowledgements}

\bibliography{refs.bib}

\begin{thebibliography}{95}%
\makeatletter
\providecommand \@ifxundefined [1]{%
 \@ifx{#1\undefined}
}%
\providecommand \@ifnum [1]{%
 \ifnum #1\expandafter \@firstoftwo
 \else \expandafter \@secondoftwo
 \fi
}%
\providecommand \@ifx [1]{%
 \ifx #1\expandafter \@firstoftwo
 \else \expandafter \@secondoftwo
 \fi
}%
\providecommand \natexlab [1]{#1}%
\providecommand \enquote  [1]{``#1''}%
\providecommand \bibnamefont  [1]{#1}%
\providecommand \bibfnamefont [1]{#1}%
\providecommand \citenamefont [1]{#1}%
\providecommand \href@noop [0]{\@secondoftwo}%
\providecommand \href [0]{\begingroup \@sanitize@url \@href}%
\providecommand \@href[1]{\@@startlink{#1}\@@href}%
\providecommand \@@href[1]{\endgroup#1\@@endlink}%
\providecommand \@sanitize@url [0]{\catcode `\\12\catcode `\$12\catcode `\&12\catcode `\#12\catcode `\^12\catcode `\_12\catcode `\%12\relax}%
\providecommand \@@startlink[1]{}%
\providecommand \@@endlink[0]{}%
\providecommand \url  [0]{\begingroup\@sanitize@url \@url }%
\providecommand \@url [1]{\endgroup\@href {#1}{\urlprefix }}%
\providecommand \urlprefix  [0]{URL }%
\providecommand \Eprint [0]{\href }%
\providecommand \doibase [0]{https://doi.org/}%
\providecommand \selectlanguage [0]{\@gobble}%
\providecommand \bibinfo  [0]{\@secondoftwo}%
\providecommand \bibfield  [0]{\@secondoftwo}%
\providecommand \translation [1]{[#1]}%
\providecommand \BibitemOpen [0]{}%
\providecommand \bibitemStop [0]{}%
\providecommand \bibitemNoStop [0]{.\EOS\space}%
\providecommand \EOS [0]{\spacefactor3000\relax}%
\providecommand \BibitemShut  [1]{\csname bibitem#1\endcsname}%
\let\auto@bib@innerbib\@empty
\bibitem [{\citenamefont {Kwan}\ \emph {et~al.}()\citenamefont {Kwan}, \citenamefont {Wang}, \citenamefont {Wagner}, \citenamefont {Simon}, \citenamefont {Parameswaran},\ and\ \citenamefont {Bultinck}}]{companion}%
  \BibitemOpen
  \bibfield  {author} {\bibinfo {author} {\bibfnamefont {Y.~H.}\ \bibnamefont {Kwan}}, \bibinfo {author} {\bibfnamefont {Z.}~\bibnamefont {Wang}}, \bibinfo {author} {\bibfnamefont {G.}~\bibnamefont {Wagner}}, \bibinfo {author} {\bibfnamefont {S.~H.}\ \bibnamefont {Simon}}, \bibinfo {author} {\bibfnamefont {S.~A.}\ \bibnamefont {Parameswaran}},\ and\ \bibinfo {author} {\bibfnamefont {N.}~\bibnamefont {Bultinck}},\ }\bibfield  {title} {\bibinfo {title} {Textured exciton insulators},\ }\bibinfo {note} {same arXiv posting}\BibitemShut {NoStop}%
\bibitem [{\citenamefont {Cao}\ \emph {et~al.}(2018{\natexlab{a}})\citenamefont {Cao}, \citenamefont {Fatemi}, \citenamefont {Fang}, \citenamefont {Watanabe}, \citenamefont {Taniguchi}, \citenamefont {Kaxiras},\ and\ \citenamefont {Jarillo-Herrero}}]{Cao2018b}%
  \BibitemOpen
  \bibfield  {author} {\bibinfo {author} {\bibfnamefont {Y.}~\bibnamefont {Cao}}, \bibinfo {author} {\bibfnamefont {V.}~\bibnamefont {Fatemi}}, \bibinfo {author} {\bibfnamefont {S.}~\bibnamefont {Fang}}, \bibinfo {author} {\bibfnamefont {K.}~\bibnamefont {Watanabe}}, \bibinfo {author} {\bibfnamefont {T.}~\bibnamefont {Taniguchi}}, \bibinfo {author} {\bibfnamefont {E.}~\bibnamefont {Kaxiras}},\ and\ \bibinfo {author} {\bibfnamefont {P.}~\bibnamefont {Jarillo-Herrero}},\ }\bibfield  {title} {\bibinfo {title} {Unconventional superconductivity in magic-angle graphene superlattices},\ }\href {https://doi.org/10.1038/nature26160} {\bibfield  {journal} {\bibinfo  {journal} {Nature}\ }\textbf {\bibinfo {volume} {556}},\ \bibinfo {pages} {43} (\bibinfo {year} {2018}{\natexlab{a}})}\BibitemShut {NoStop}%
\bibitem [{\citenamefont {Yankowitz}\ \emph {et~al.}(2019)\citenamefont {Yankowitz}, \citenamefont {Chen}, \citenamefont {Polshyn}, \citenamefont {Zhang}, \citenamefont {Watanabe}, \citenamefont {Taniguchi}, \citenamefont {Graf}, \citenamefont {Young},\ and\ \citenamefont {Dean}}]{Yankowitz_2019}%
  \BibitemOpen
  \bibfield  {author} {\bibinfo {author} {\bibfnamefont {M.}~\bibnamefont {Yankowitz}}, \bibinfo {author} {\bibfnamefont {S.}~\bibnamefont {Chen}}, \bibinfo {author} {\bibfnamefont {H.}~\bibnamefont {Polshyn}}, \bibinfo {author} {\bibfnamefont {Y.}~\bibnamefont {Zhang}}, \bibinfo {author} {\bibfnamefont {K.}~\bibnamefont {Watanabe}}, \bibinfo {author} {\bibfnamefont {T.}~\bibnamefont {Taniguchi}}, \bibinfo {author} {\bibfnamefont {D.}~\bibnamefont {Graf}}, \bibinfo {author} {\bibfnamefont {A.~F.}\ \bibnamefont {Young}},\ and\ \bibinfo {author} {\bibfnamefont {C.~R.}\ \bibnamefont {Dean}},\ }\bibfield  {title} {\bibinfo {title} {Tuning superconductivity in twisted bilayer graphene},\ }\href {https://doi.org/10.1126/science.aav1910} {\bibfield  {journal} {\bibinfo  {journal} {Science}\ }\textbf {\bibinfo {volume} {363}},\ \bibinfo {pages} {1059} (\bibinfo {year} {2019})}\BibitemShut {NoStop}%
\bibitem [{\citenamefont {Lu}\ \emph {et~al.}(2019)\citenamefont {Lu}, \citenamefont {Stepanov}, \citenamefont {Yang}, \citenamefont {Xie}, \citenamefont {Aamir}, \citenamefont {Das}, \citenamefont {Urgell}, \citenamefont {Watanabe}, \citenamefont {Taniguchi}, \citenamefont {Zhang}, \citenamefont {Bachtold}, \citenamefont {MacDonald},\ and\ \citenamefont {Efetov}}]{Lu2019}%
  \BibitemOpen
  \bibfield  {author} {\bibinfo {author} {\bibfnamefont {X.}~\bibnamefont {Lu}}, \bibinfo {author} {\bibfnamefont {P.}~\bibnamefont {Stepanov}}, \bibinfo {author} {\bibfnamefont {W.}~\bibnamefont {Yang}}, \bibinfo {author} {\bibfnamefont {M.}~\bibnamefont {Xie}}, \bibinfo {author} {\bibfnamefont {M.~A.}\ \bibnamefont {Aamir}}, \bibinfo {author} {\bibfnamefont {I.}~\bibnamefont {Das}}, \bibinfo {author} {\bibfnamefont {C.}~\bibnamefont {Urgell}}, \bibinfo {author} {\bibfnamefont {K.}~\bibnamefont {Watanabe}}, \bibinfo {author} {\bibfnamefont {T.}~\bibnamefont {Taniguchi}}, \bibinfo {author} {\bibfnamefont {G.}~\bibnamefont {Zhang}}, \bibinfo {author} {\bibfnamefont {A.}~\bibnamefont {Bachtold}}, \bibinfo {author} {\bibfnamefont {A.~H.}\ \bibnamefont {MacDonald}},\ and\ \bibinfo {author} {\bibfnamefont {D.~K.}\ \bibnamefont {Efetov}},\ }\bibfield  {title} {\bibinfo {title} {Superconductors, orbital magnets and correlated states in magic-angle bilayer graphene},\ }\href
  {https://doi.org/10.1038/s41586-019-1695-0} {\bibfield  {journal} {\bibinfo  {journal} {Nature}\ }\textbf {\bibinfo {volume} {574}},\ \bibinfo {pages} {653} (\bibinfo {year} {2019})}\BibitemShut {NoStop}%
\bibitem [{\citenamefont {Park}\ \emph {et~al.}(2021{\natexlab{a}})\citenamefont {Park}, \citenamefont {Cao}, \citenamefont {Watanabe}, \citenamefont {Taniguchi},\ and\ \citenamefont {Jarillo-Herrero}}]{Park2021TTGSC}%
  \BibitemOpen
  \bibfield  {author} {\bibinfo {author} {\bibfnamefont {J.~M.}\ \bibnamefont {Park}}, \bibinfo {author} {\bibfnamefont {Y.}~\bibnamefont {Cao}}, \bibinfo {author} {\bibfnamefont {K.}~\bibnamefont {Watanabe}}, \bibinfo {author} {\bibfnamefont {T.}~\bibnamefont {Taniguchi}},\ and\ \bibinfo {author} {\bibfnamefont {P.}~\bibnamefont {Jarillo-Herrero}},\ }\bibfield  {title} {\bibinfo {title} {Tunable strongly coupled superconductivity in magic-angle twisted trilayer graphene},\ }\href {https://doi.org/10.1038/s41586-021-03192-0} {\bibfield  {journal} {\bibinfo  {journal} {Nature}\ }\textbf {\bibinfo {volume} {590}},\ \bibinfo {pages} {249} (\bibinfo {year} {2021}{\natexlab{a}})}\BibitemShut {NoStop}%
\bibitem [{\citenamefont {Cao}\ \emph {et~al.}(2018{\natexlab{b}})\citenamefont {Cao}, \citenamefont {Fatemi}, \citenamefont {Demir}, \citenamefont {Fang}, \citenamefont {Tomarken}, \citenamefont {Luo}, \citenamefont {Sanchez-Yamagishi}, \citenamefont {Watanabe}, \citenamefont {Taniguchi}, \citenamefont {Kaxiras}, \citenamefont {Ashoori},\ and\ \citenamefont {Jarillo-Herrero}}]{Cao2018}%
  \BibitemOpen
  \bibfield  {author} {\bibinfo {author} {\bibfnamefont {Y.}~\bibnamefont {Cao}}, \bibinfo {author} {\bibfnamefont {V.}~\bibnamefont {Fatemi}}, \bibinfo {author} {\bibfnamefont {A.}~\bibnamefont {Demir}}, \bibinfo {author} {\bibfnamefont {S.}~\bibnamefont {Fang}}, \bibinfo {author} {\bibfnamefont {S.~L.}\ \bibnamefont {Tomarken}}, \bibinfo {author} {\bibfnamefont {J.~Y.}\ \bibnamefont {Luo}}, \bibinfo {author} {\bibfnamefont {J.~D.}\ \bibnamefont {Sanchez-Yamagishi}}, \bibinfo {author} {\bibfnamefont {K.}~\bibnamefont {Watanabe}}, \bibinfo {author} {\bibfnamefont {T.}~\bibnamefont {Taniguchi}}, \bibinfo {author} {\bibfnamefont {E.}~\bibnamefont {Kaxiras}}, \bibinfo {author} {\bibfnamefont {R.~C.}\ \bibnamefont {Ashoori}},\ and\ \bibinfo {author} {\bibfnamefont {P.}~\bibnamefont {Jarillo-Herrero}},\ }\bibfield  {title} {\bibinfo {title} {Correlated insulator behaviour at half-filling in magic-angle graphene superlattices},\ }\href {https://doi.org/10.1038/nature26154} {\bibfield  {journal} {\bibinfo  {journal}
  {Nature}\ }\textbf {\bibinfo {volume} {556}},\ \bibinfo {pages} {80} (\bibinfo {year} {2018}{\natexlab{b}})}\BibitemShut {NoStop}%
\bibitem [{\citenamefont {Sharpe}\ \emph {et~al.}(2019)\citenamefont {Sharpe}, \citenamefont {Fox}, \citenamefont {Barnard}, \citenamefont {Finney}, \citenamefont {Watanabe}, \citenamefont {Taniguchi}, \citenamefont {Kastner},\ and\ \citenamefont {Goldhaber-Gordon}}]{Sharpe_2019}%
  \BibitemOpen
  \bibfield  {author} {\bibinfo {author} {\bibfnamefont {A.~L.}\ \bibnamefont {Sharpe}}, \bibinfo {author} {\bibfnamefont {E.~J.}\ \bibnamefont {Fox}}, \bibinfo {author} {\bibfnamefont {A.~W.}\ \bibnamefont {Barnard}}, \bibinfo {author} {\bibfnamefont {J.}~\bibnamefont {Finney}}, \bibinfo {author} {\bibfnamefont {K.}~\bibnamefont {Watanabe}}, \bibinfo {author} {\bibfnamefont {T.}~\bibnamefont {Taniguchi}}, \bibinfo {author} {\bibfnamefont {M.~A.}\ \bibnamefont {Kastner}},\ and\ \bibinfo {author} {\bibfnamefont {D.}~\bibnamefont {Goldhaber-Gordon}},\ }\bibfield  {title} {\bibinfo {title} {Emergent ferromagnetism near three-quarters filling in twisted bilayer graphene},\ }\href {https://doi.org/10.1126/science.aaw3780} {\bibfield  {journal} {\bibinfo  {journal} {Science}\ }\textbf {\bibinfo {volume} {365}},\ \bibinfo {pages} {605} (\bibinfo {year} {2019})}\BibitemShut {NoStop}%
\bibitem [{\citenamefont {Serlin}\ \emph {et~al.}(2020)\citenamefont {Serlin}, \citenamefont {Tschirhart}, \citenamefont {Polshyn}, \citenamefont {Zhang}, \citenamefont {Zhu}, \citenamefont {Watanabe}, \citenamefont {Taniguchi}, \citenamefont {Balents},\ and\ \citenamefont {Young}}]{Serlin900}%
  \BibitemOpen
  \bibfield  {author} {\bibinfo {author} {\bibfnamefont {M.}~\bibnamefont {Serlin}}, \bibinfo {author} {\bibfnamefont {C.~L.}\ \bibnamefont {Tschirhart}}, \bibinfo {author} {\bibfnamefont {H.}~\bibnamefont {Polshyn}}, \bibinfo {author} {\bibfnamefont {Y.}~\bibnamefont {Zhang}}, \bibinfo {author} {\bibfnamefont {J.}~\bibnamefont {Zhu}}, \bibinfo {author} {\bibfnamefont {K.}~\bibnamefont {Watanabe}}, \bibinfo {author} {\bibfnamefont {T.}~\bibnamefont {Taniguchi}}, \bibinfo {author} {\bibfnamefont {L.}~\bibnamefont {Balents}},\ and\ \bibinfo {author} {\bibfnamefont {A.~F.}\ \bibnamefont {Young}},\ }\bibfield  {title} {\bibinfo {title} {Intrinsic quantized anomalous hall effect in a moir{\'e} heterostructure},\ }\href {https://doi.org/10.1126/science.aay5533} {\bibfield  {journal} {\bibinfo  {journal} {Science}\ }\textbf {\bibinfo {volume} {367}},\ \bibinfo {pages} {900} (\bibinfo {year} {2020})}\BibitemShut {NoStop}%
\bibitem [{\citenamefont {Park}\ \emph {et~al.}(2023)\citenamefont {Park}, \citenamefont {Cai}, \citenamefont {Anderson}, \citenamefont {Zhang}, \citenamefont {Zhu}, \citenamefont {Liu}, \citenamefont {Wang}, \citenamefont {Holtzmann}, \citenamefont {Hu}, \citenamefont {Liu}, \citenamefont {Taniguchi}, \citenamefont {Watanabe}, \citenamefont {Chu}, \citenamefont {Cao}, \citenamefont {Fu}, \citenamefont {Yao}, \citenamefont {Chang}, \citenamefont {Cobden}, \citenamefont {Xiao},\ and\ \citenamefont {Xu}}]{Park2023}%
  \BibitemOpen
  \bibfield  {author} {\bibinfo {author} {\bibfnamefont {H.}~\bibnamefont {Park}}, \bibinfo {author} {\bibfnamefont {J.}~\bibnamefont {Cai}}, \bibinfo {author} {\bibfnamefont {E.}~\bibnamefont {Anderson}}, \bibinfo {author} {\bibfnamefont {Y.}~\bibnamefont {Zhang}}, \bibinfo {author} {\bibfnamefont {J.}~\bibnamefont {Zhu}}, \bibinfo {author} {\bibfnamefont {X.}~\bibnamefont {Liu}}, \bibinfo {author} {\bibfnamefont {C.}~\bibnamefont {Wang}}, \bibinfo {author} {\bibfnamefont {W.}~\bibnamefont {Holtzmann}}, \bibinfo {author} {\bibfnamefont {C.}~\bibnamefont {Hu}}, \bibinfo {author} {\bibfnamefont {Z.}~\bibnamefont {Liu}}, \bibinfo {author} {\bibfnamefont {T.}~\bibnamefont {Taniguchi}}, \bibinfo {author} {\bibfnamefont {K.}~\bibnamefont {Watanabe}}, \bibinfo {author} {\bibfnamefont {J.-H.}\ \bibnamefont {Chu}}, \bibinfo {author} {\bibfnamefont {T.}~\bibnamefont {Cao}}, \bibinfo {author} {\bibfnamefont {L.}~\bibnamefont {Fu}}, \bibinfo {author} {\bibfnamefont {W.}~\bibnamefont {Yao}}, \bibinfo {author}
  {\bibfnamefont {C.-Z.}\ \bibnamefont {Chang}}, \bibinfo {author} {\bibfnamefont {D.}~\bibnamefont {Cobden}}, \bibinfo {author} {\bibfnamefont {D.}~\bibnamefont {Xiao}},\ and\ \bibinfo {author} {\bibfnamefont {X.}~\bibnamefont {Xu}},\ }\bibfield  {title} {\bibinfo {title} {Observation of fractionally quantized anomalous hall effect},\ }\href {https://doi.org/10.1038/s41586-023-06536-0} {\bibfield  {journal} {\bibinfo  {journal} {Nature}\ }\textbf {\bibinfo {volume} {622}},\ \bibinfo {pages} {74} (\bibinfo {year} {2023})}\BibitemShut {NoStop}%
\bibitem [{\citenamefont {Cai}\ \emph {et~al.}(2023)\citenamefont {Cai}, \citenamefont {Anderson}, \citenamefont {Wang}, \citenamefont {Zhang}, \citenamefont {Liu}, \citenamefont {Holtzmann}, \citenamefont {Zhang}, \citenamefont {Fan}, \citenamefont {Taniguchi}, \citenamefont {Watanabe}, \citenamefont {Ran}, \citenamefont {Cao}, \citenamefont {Fu}, \citenamefont {Xiao}, \citenamefont {Yao},\ and\ \citenamefont {Xu}}]{Cai2023y}%
  \BibitemOpen
  \bibfield  {author} {\bibinfo {author} {\bibfnamefont {J.}~\bibnamefont {Cai}}, \bibinfo {author} {\bibfnamefont {E.}~\bibnamefont {Anderson}}, \bibinfo {author} {\bibfnamefont {C.}~\bibnamefont {Wang}}, \bibinfo {author} {\bibfnamefont {X.}~\bibnamefont {Zhang}}, \bibinfo {author} {\bibfnamefont {X.}~\bibnamefont {Liu}}, \bibinfo {author} {\bibfnamefont {W.}~\bibnamefont {Holtzmann}}, \bibinfo {author} {\bibfnamefont {Y.}~\bibnamefont {Zhang}}, \bibinfo {author} {\bibfnamefont {F.}~\bibnamefont {Fan}}, \bibinfo {author} {\bibfnamefont {T.}~\bibnamefont {Taniguchi}}, \bibinfo {author} {\bibfnamefont {K.}~\bibnamefont {Watanabe}}, \bibinfo {author} {\bibfnamefont {Y.}~\bibnamefont {Ran}}, \bibinfo {author} {\bibfnamefont {T.}~\bibnamefont {Cao}}, \bibinfo {author} {\bibfnamefont {L.}~\bibnamefont {Fu}}, \bibinfo {author} {\bibfnamefont {D.}~\bibnamefont {Xiao}}, \bibinfo {author} {\bibfnamefont {W.}~\bibnamefont {Yao}},\ and\ \bibinfo {author} {\bibfnamefont {X.}~\bibnamefont {Xu}},\ }\bibfield  {title}
  {\bibinfo {title} {Signatures of fractional quantum anomalous hall states in twisted mote2},\ }\href {https://doi.org/10.1038/s41586-023-06289-w} {\bibfield  {journal} {\bibinfo  {journal} {Nature}\ }\textbf {\bibinfo {volume} {622}},\ \bibinfo {pages} {63} (\bibinfo {year} {2023})}\BibitemShut {NoStop}%
\bibitem [{\citenamefont {Zeng}\ \emph {et~al.}(2023)\citenamefont {Zeng}, \citenamefont {Xia}, \citenamefont {Kang}, \citenamefont {Zhu}, \citenamefont {Kn{\"u}ppel}, \citenamefont {Vaswani}, \citenamefont {Watanabe}, \citenamefont {Taniguchi}, \citenamefont {Mak},\ and\ \citenamefont {Shan}}]{Zeng2023}%
  \BibitemOpen
  \bibfield  {author} {\bibinfo {author} {\bibfnamefont {Y.}~\bibnamefont {Zeng}}, \bibinfo {author} {\bibfnamefont {Z.}~\bibnamefont {Xia}}, \bibinfo {author} {\bibfnamefont {K.}~\bibnamefont {Kang}}, \bibinfo {author} {\bibfnamefont {J.}~\bibnamefont {Zhu}}, \bibinfo {author} {\bibfnamefont {P.}~\bibnamefont {Kn{\"u}ppel}}, \bibinfo {author} {\bibfnamefont {C.}~\bibnamefont {Vaswani}}, \bibinfo {author} {\bibfnamefont {K.}~\bibnamefont {Watanabe}}, \bibinfo {author} {\bibfnamefont {T.}~\bibnamefont {Taniguchi}}, \bibinfo {author} {\bibfnamefont {K.~F.}\ \bibnamefont {Mak}},\ and\ \bibinfo {author} {\bibfnamefont {J.}~\bibnamefont {Shan}},\ }\bibfield  {title} {\bibinfo {title} {Thermodynamic evidence of fractional chern insulator in moir{\'e}mote2},\ }\href {https://doi.org/10.1038/s41586-023-06452-3} {\bibfield  {journal} {\bibinfo  {journal} {Nature}\ }\textbf {\bibinfo {volume} {622}},\ \bibinfo {pages} {69} (\bibinfo {year} {2023})}\BibitemShut {NoStop}%
\bibitem [{\citenamefont {Xu}\ \emph {et~al.}(2023)\citenamefont {Xu}, \citenamefont {Sun}, \citenamefont {Jia}, \citenamefont {Liu}, \citenamefont {Xu}, \citenamefont {Li}, \citenamefont {Gu}, \citenamefont {Watanabe}, \citenamefont {Taniguchi}, \citenamefont {Tong}, \citenamefont {Jia}, \citenamefont {Shi}, \citenamefont {Jiang}, \citenamefont {Zhang}, \citenamefont {Liu},\ and\ \citenamefont {Li}}]{Xu2023}%
  \BibitemOpen
  \bibfield  {author} {\bibinfo {author} {\bibfnamefont {F.}~\bibnamefont {Xu}}, \bibinfo {author} {\bibfnamefont {Z.}~\bibnamefont {Sun}}, \bibinfo {author} {\bibfnamefont {T.}~\bibnamefont {Jia}}, \bibinfo {author} {\bibfnamefont {C.}~\bibnamefont {Liu}}, \bibinfo {author} {\bibfnamefont {C.}~\bibnamefont {Xu}}, \bibinfo {author} {\bibfnamefont {C.}~\bibnamefont {Li}}, \bibinfo {author} {\bibfnamefont {Y.}~\bibnamefont {Gu}}, \bibinfo {author} {\bibfnamefont {K.}~\bibnamefont {Watanabe}}, \bibinfo {author} {\bibfnamefont {T.}~\bibnamefont {Taniguchi}}, \bibinfo {author} {\bibfnamefont {B.}~\bibnamefont {Tong}}, \bibinfo {author} {\bibfnamefont {J.}~\bibnamefont {Jia}}, \bibinfo {author} {\bibfnamefont {Z.}~\bibnamefont {Shi}}, \bibinfo {author} {\bibfnamefont {S.}~\bibnamefont {Jiang}}, \bibinfo {author} {\bibfnamefont {Y.}~\bibnamefont {Zhang}}, \bibinfo {author} {\bibfnamefont {X.}~\bibnamefont {Liu}},\ and\ \bibinfo {author} {\bibfnamefont {T.}~\bibnamefont {Li}},\ }\bibfield  {title} {\bibinfo {title}
  {Observation of integer and fractional quantum anomalous hall effects in twisted bilayer ${\mathrm{mote}}_{2}$},\ }\href {https://doi.org/10.1103/PhysRevX.13.031037} {\bibfield  {journal} {\bibinfo  {journal} {Phys. Rev. X}\ }\textbf {\bibinfo {volume} {13}},\ \bibinfo {pages} {031037} (\bibinfo {year} {2023})}\BibitemShut {NoStop}%
\bibitem [{\citenamefont {Lu}\ \emph {et~al.}(2023)\citenamefont {Lu}, \citenamefont {Han}, \citenamefont {Yao}, \citenamefont {Reddy}, \citenamefont {Yang}, \citenamefont {Seo}, \citenamefont {Watanabe}, \citenamefont {Taniguchi}, \citenamefont {Fu},\ and\ \citenamefont {Ju}}]{lu2023fractional}%
  \BibitemOpen
  \bibfield  {author} {\bibinfo {author} {\bibfnamefont {Z.}~\bibnamefont {Lu}}, \bibinfo {author} {\bibfnamefont {T.}~\bibnamefont {Han}}, \bibinfo {author} {\bibfnamefont {Y.}~\bibnamefont {Yao}}, \bibinfo {author} {\bibfnamefont {A.~P.}\ \bibnamefont {Reddy}}, \bibinfo {author} {\bibfnamefont {J.}~\bibnamefont {Yang}}, \bibinfo {author} {\bibfnamefont {J.}~\bibnamefont {Seo}}, \bibinfo {author} {\bibfnamefont {K.}~\bibnamefont {Watanabe}}, \bibinfo {author} {\bibfnamefont {T.}~\bibnamefont {Taniguchi}}, \bibinfo {author} {\bibfnamefont {L.}~\bibnamefont {Fu}},\ and\ \bibinfo {author} {\bibfnamefont {L.}~\bibnamefont {Ju}},\ }\href@noop {} {\bibinfo {title} {Fractional quantum anomalous hall effect in a graphene moire superlattice}} (\bibinfo {year} {2023}),\ \Eprint {https://arxiv.org/abs/2309.17436} {arXiv:2309.17436 [cond-mat.mes-hall]} \BibitemShut {NoStop}%
\bibitem [{\citenamefont {Xie}\ \emph {et~al.}(2021)\citenamefont {Xie}, \citenamefont {Pierce}, \citenamefont {Park}, \citenamefont {Parker}, \citenamefont {Khalaf}, \citenamefont {Ledwith}, \citenamefont {Cao}, \citenamefont {Lee}, \citenamefont {Chen}, \citenamefont {Forrester}, \citenamefont {Watanabe}, \citenamefont {Taniguchi}, \citenamefont {Vishwanath}, \citenamefont {Jarillo-Herrero},\ and\ \citenamefont {Yacoby}}]{Xie2021fractional}%
  \BibitemOpen
  \bibfield  {author} {\bibinfo {author} {\bibfnamefont {Y.}~\bibnamefont {Xie}}, \bibinfo {author} {\bibfnamefont {A.~T.}\ \bibnamefont {Pierce}}, \bibinfo {author} {\bibfnamefont {J.~M.}\ \bibnamefont {Park}}, \bibinfo {author} {\bibfnamefont {D.~E.}\ \bibnamefont {Parker}}, \bibinfo {author} {\bibfnamefont {E.}~\bibnamefont {Khalaf}}, \bibinfo {author} {\bibfnamefont {P.}~\bibnamefont {Ledwith}}, \bibinfo {author} {\bibfnamefont {Y.}~\bibnamefont {Cao}}, \bibinfo {author} {\bibfnamefont {S.~H.}\ \bibnamefont {Lee}}, \bibinfo {author} {\bibfnamefont {S.}~\bibnamefont {Chen}}, \bibinfo {author} {\bibfnamefont {P.~R.}\ \bibnamefont {Forrester}}, \bibinfo {author} {\bibfnamefont {K.}~\bibnamefont {Watanabe}}, \bibinfo {author} {\bibfnamefont {T.}~\bibnamefont {Taniguchi}}, \bibinfo {author} {\bibfnamefont {A.}~\bibnamefont {Vishwanath}}, \bibinfo {author} {\bibfnamefont {P.}~\bibnamefont {Jarillo-Herrero}},\ and\ \bibinfo {author} {\bibfnamefont {A.}~\bibnamefont {Yacoby}},\ }\bibfield  {title} {\bibinfo
  {title} {Fractional chern insulators in magic-angle twisted bilayer graphene},\ }\href {https://doi.org/10.1038/s41586-021-04002-3} {\bibfield  {journal} {\bibinfo  {journal} {Nature}\ }\textbf {\bibinfo {volume} {600}},\ \bibinfo {pages} {439} (\bibinfo {year} {2021})}\BibitemShut {NoStop}%
\bibitem [{\citenamefont {Jaoui}\ \emph {et~al.}(2022)\citenamefont {Jaoui}, \citenamefont {Das}, \citenamefont {Di~Battista}, \citenamefont {D{\'\i}ez-M{\'e}rida}, \citenamefont {Lu}, \citenamefont {Watanabe}, \citenamefont {Taniguchi}, \citenamefont {Ishizuka}, \citenamefont {Levitov},\ and\ \citenamefont {Efetov}}]{Jaoui2022}%
  \BibitemOpen
  \bibfield  {author} {\bibinfo {author} {\bibfnamefont {A.}~\bibnamefont {Jaoui}}, \bibinfo {author} {\bibfnamefont {I.}~\bibnamefont {Das}}, \bibinfo {author} {\bibfnamefont {G.}~\bibnamefont {Di~Battista}}, \bibinfo {author} {\bibfnamefont {J.}~\bibnamefont {D{\'\i}ez-M{\'e}rida}}, \bibinfo {author} {\bibfnamefont {X.}~\bibnamefont {Lu}}, \bibinfo {author} {\bibfnamefont {K.}~\bibnamefont {Watanabe}}, \bibinfo {author} {\bibfnamefont {T.}~\bibnamefont {Taniguchi}}, \bibinfo {author} {\bibfnamefont {H.}~\bibnamefont {Ishizuka}}, \bibinfo {author} {\bibfnamefont {L.}~\bibnamefont {Levitov}},\ and\ \bibinfo {author} {\bibfnamefont {D.~K.}\ \bibnamefont {Efetov}},\ }\bibfield  {title} {\bibinfo {title} {Quantum critical behaviour in magic-angle twisted bilayer graphene},\ }\href {https://doi.org/10.1038/s41567-022-01556-5} {\bibfield  {journal} {\bibinfo  {journal} {Nature Physics}\ }\textbf {\bibinfo {volume} {18}},\ \bibinfo {pages} {633} (\bibinfo {year} {2022})}\BibitemShut {NoStop}%
\bibitem [{\citenamefont {Polshyn}\ \emph {et~al.}(2019)\citenamefont {Polshyn}, \citenamefont {Yankowitz}, \citenamefont {Chen}, \citenamefont {Zhang}, \citenamefont {Watanabe}, \citenamefont {Taniguchi}, \citenamefont {Dean},\ and\ \citenamefont {Young}}]{polshyn2019large}%
  \BibitemOpen
  \bibfield  {author} {\bibinfo {author} {\bibfnamefont {H.}~\bibnamefont {Polshyn}}, \bibinfo {author} {\bibfnamefont {M.}~\bibnamefont {Yankowitz}}, \bibinfo {author} {\bibfnamefont {S.}~\bibnamefont {Chen}}, \bibinfo {author} {\bibfnamefont {Y.}~\bibnamefont {Zhang}}, \bibinfo {author} {\bibfnamefont {K.}~\bibnamefont {Watanabe}}, \bibinfo {author} {\bibfnamefont {T.}~\bibnamefont {Taniguchi}}, \bibinfo {author} {\bibfnamefont {C.~R.}\ \bibnamefont {Dean}},\ and\ \bibinfo {author} {\bibfnamefont {A.~F.}\ \bibnamefont {Young}},\ }\bibfield  {title} {\bibinfo {title} {Large linear-in-temperature resistivity in twisted bilayer graphene},\ }\href@noop {} {\bibfield  {journal} {\bibinfo  {journal} {Nature Physics}\ }\textbf {\bibinfo {volume} {15}},\ \bibinfo {pages} {1011} (\bibinfo {year} {2019})}\BibitemShut {NoStop}%
\bibitem [{\citenamefont {Regan}\ \emph {et~al.}(2020)\citenamefont {Regan}, \citenamefont {Wang}, \citenamefont {Jin}, \citenamefont {Bakti~Utama}, \citenamefont {Gao}, \citenamefont {Wei}, \citenamefont {Zhao}, \citenamefont {Zhao}, \citenamefont {Zhang}, \citenamefont {Yumigeta}, \citenamefont {Blei}, \citenamefont {Carlstr{\"o}m}, \citenamefont {Watanabe}, \citenamefont {Taniguchi}, \citenamefont {Tongay}, \citenamefont {Crommie}, \citenamefont {Zettl},\ and\ \citenamefont {Wang}}]{Regan2020}%
  \BibitemOpen
  \bibfield  {author} {\bibinfo {author} {\bibfnamefont {E.~C.}\ \bibnamefont {Regan}}, \bibinfo {author} {\bibfnamefont {D.}~\bibnamefont {Wang}}, \bibinfo {author} {\bibfnamefont {C.}~\bibnamefont {Jin}}, \bibinfo {author} {\bibfnamefont {M.~I.}\ \bibnamefont {Bakti~Utama}}, \bibinfo {author} {\bibfnamefont {B.}~\bibnamefont {Gao}}, \bibinfo {author} {\bibfnamefont {X.}~\bibnamefont {Wei}}, \bibinfo {author} {\bibfnamefont {S.}~\bibnamefont {Zhao}}, \bibinfo {author} {\bibfnamefont {W.}~\bibnamefont {Zhao}}, \bibinfo {author} {\bibfnamefont {Z.}~\bibnamefont {Zhang}}, \bibinfo {author} {\bibfnamefont {K.}~\bibnamefont {Yumigeta}}, \bibinfo {author} {\bibfnamefont {M.}~\bibnamefont {Blei}}, \bibinfo {author} {\bibfnamefont {J.~D.}\ \bibnamefont {Carlstr{\"o}m}}, \bibinfo {author} {\bibfnamefont {K.}~\bibnamefont {Watanabe}}, \bibinfo {author} {\bibfnamefont {T.}~\bibnamefont {Taniguchi}}, \bibinfo {author} {\bibfnamefont {S.}~\bibnamefont {Tongay}}, \bibinfo {author} {\bibfnamefont {M.}~\bibnamefont
  {Crommie}}, \bibinfo {author} {\bibfnamefont {A.}~\bibnamefont {Zettl}},\ and\ \bibinfo {author} {\bibfnamefont {F.}~\bibnamefont {Wang}},\ }\bibfield  {title} {\bibinfo {title} {Mott and generalized wigner crystal states in wse2/ws2 moir{\'e}superlattices},\ }\href {https://doi.org/10.1038/s41586-020-2092-4} {\bibfield  {journal} {\bibinfo  {journal} {Nature}\ }\textbf {\bibinfo {volume} {579}},\ \bibinfo {pages} {359} (\bibinfo {year} {2020})}\BibitemShut {NoStop}%
\bibitem [{\citenamefont {Park}\ \emph {et~al.}(2021{\natexlab{b}})\citenamefont {Park}, \citenamefont {Cao}, \citenamefont {Xia}, \citenamefont {Sun}, \citenamefont {Watanabe}, \citenamefont {Taniguchi},\ and\ \citenamefont {Jarillo-Herrero}}]{park2021magicangle}%
  \BibitemOpen
  \bibfield  {author} {\bibinfo {author} {\bibfnamefont {J.~M.}\ \bibnamefont {Park}}, \bibinfo {author} {\bibfnamefont {Y.}~\bibnamefont {Cao}}, \bibinfo {author} {\bibfnamefont {L.}~\bibnamefont {Xia}}, \bibinfo {author} {\bibfnamefont {S.}~\bibnamefont {Sun}}, \bibinfo {author} {\bibfnamefont {K.}~\bibnamefont {Watanabe}}, \bibinfo {author} {\bibfnamefont {T.}~\bibnamefont {Taniguchi}},\ and\ \bibinfo {author} {\bibfnamefont {P.}~\bibnamefont {Jarillo-Herrero}},\ }\href@noop {} {\bibinfo {title} {Magic-angle multilayer graphene: A robust family of moir\'e superconductors}} (\bibinfo {year} {2021}{\natexlab{b}}),\ \Eprint {https://arxiv.org/abs/2112.10760} {arXiv:2112.10760 [cond-mat.supr-con]} \BibitemShut {NoStop}%
\bibitem [{\citenamefont {Zhang}\ \emph {et~al.}(2022)\citenamefont {Zhang}, \citenamefont {Polski}, \citenamefont {Lewandowski}, \citenamefont {Thomson}, \citenamefont {Peng}, \citenamefont {Choi}, \citenamefont {Kim}, \citenamefont {Watanabe}, \citenamefont {Taniguchi}, \citenamefont {Alicea} \emph {et~al.}}]{zhang2022promotion}%
  \BibitemOpen
  \bibfield  {author} {\bibinfo {author} {\bibfnamefont {Y.}~\bibnamefont {Zhang}}, \bibinfo {author} {\bibfnamefont {R.}~\bibnamefont {Polski}}, \bibinfo {author} {\bibfnamefont {C.}~\bibnamefont {Lewandowski}}, \bibinfo {author} {\bibfnamefont {A.}~\bibnamefont {Thomson}}, \bibinfo {author} {\bibfnamefont {Y.}~\bibnamefont {Peng}}, \bibinfo {author} {\bibfnamefont {Y.}~\bibnamefont {Choi}}, \bibinfo {author} {\bibfnamefont {H.}~\bibnamefont {Kim}}, \bibinfo {author} {\bibfnamefont {K.}~\bibnamefont {Watanabe}}, \bibinfo {author} {\bibfnamefont {T.}~\bibnamefont {Taniguchi}}, \bibinfo {author} {\bibfnamefont {J.}~\bibnamefont {Alicea}}, \emph {et~al.},\ }\bibfield  {title} {\bibinfo {title} {Promotion of superconductivity in magic-angle graphene multilayers},\ }\href@noop {} {\bibfield  {journal} {\bibinfo  {journal} {Science}\ }\textbf {\bibinfo {volume} {377}},\ \bibinfo {pages} {1538} (\bibinfo {year} {2022})}\BibitemShut {NoStop}%
\bibitem [{\citenamefont {Mak}\ and\ \citenamefont {Shan}(2022)}]{Mak2022}%
  \BibitemOpen
  \bibfield  {author} {\bibinfo {author} {\bibfnamefont {K.~F.}\ \bibnamefont {Mak}}\ and\ \bibinfo {author} {\bibfnamefont {J.}~\bibnamefont {Shan}},\ }\bibfield  {title} {\bibinfo {title} {Semiconductor moir{\'e}materials},\ }\href {https://doi.org/10.1038/s41565-022-01165-6} {\bibfield  {journal} {\bibinfo  {journal} {Nature Nanotechnology}\ }\textbf {\bibinfo {volume} {17}},\ \bibinfo {pages} {686} (\bibinfo {year} {2022})}\BibitemShut {NoStop}%
\bibitem [{\citenamefont {Sondhi}\ \emph {et~al.}(1993)\citenamefont {Sondhi}, \citenamefont {Karlhede}, \citenamefont {Kivelson},\ and\ \citenamefont {Rezayi}}]{Sondhi1993}%
  \BibitemOpen
  \bibfield  {author} {\bibinfo {author} {\bibfnamefont {S.~L.}\ \bibnamefont {Sondhi}}, \bibinfo {author} {\bibfnamefont {A.}~\bibnamefont {Karlhede}}, \bibinfo {author} {\bibfnamefont {S.~A.}\ \bibnamefont {Kivelson}},\ and\ \bibinfo {author} {\bibfnamefont {E.~H.}\ \bibnamefont {Rezayi}},\ }\bibfield  {title} {\bibinfo {title} {Skyrmions and the crossover from the integer to fractional quantum hall effect at small zeeman energies},\ }\href {https://doi.org/10.1103/PhysRevB.47.16419} {\bibfield  {journal} {\bibinfo  {journal} {Phys. Rev. B}\ }\textbf {\bibinfo {volume} {47}},\ \bibinfo {pages} {16419} (\bibinfo {year} {1993})}\BibitemShut {NoStop}%
\bibitem [{\citenamefont {Bultinck}\ \emph {et~al.}(2020{\natexlab{a}})\citenamefont {Bultinck}, \citenamefont {Chatterjee},\ and\ \citenamefont {Zaletel}}]{Bultinck2019mechanism}%
  \BibitemOpen
  \bibfield  {author} {\bibinfo {author} {\bibfnamefont {N.}~\bibnamefont {Bultinck}}, \bibinfo {author} {\bibfnamefont {S.}~\bibnamefont {Chatterjee}},\ and\ \bibinfo {author} {\bibfnamefont {M.~P.}\ \bibnamefont {Zaletel}},\ }\bibfield  {title} {\bibinfo {title} {Mechanism for anomalous hall ferromagnetism in twisted bilayer graphene},\ }\href {https://doi.org/10.1103/PhysRevLett.124.166601} {\bibfield  {journal} {\bibinfo  {journal} {Phys. Rev. Lett.}\ }\textbf {\bibinfo {volume} {124}},\ \bibinfo {pages} {166601} (\bibinfo {year} {2020}{\natexlab{a}})}\BibitemShut {NoStop}%
\bibitem [{\citenamefont {Ahn}\ \emph {et~al.}(2019)\citenamefont {Ahn}, \citenamefont {Park},\ and\ \citenamefont {Yang}}]{Ahn2019}%
  \BibitemOpen
  \bibfield  {author} {\bibinfo {author} {\bibfnamefont {J.}~\bibnamefont {Ahn}}, \bibinfo {author} {\bibfnamefont {S.}~\bibnamefont {Park}},\ and\ \bibinfo {author} {\bibfnamefont {B.-J.}\ \bibnamefont {Yang}},\ }\bibfield  {title} {\bibinfo {title} {Failure of nielsen-ninomiya theorem and fragile topology in two-dimensional systems with space-time inversion symmetry: Application to twisted bilayer graphene at magic angle},\ }\href {https://doi.org/10.1103/PhysRevX.9.021013} {\bibfield  {journal} {\bibinfo  {journal} {Phys. Rev. X}\ }\textbf {\bibinfo {volume} {9}},\ \bibinfo {pages} {021013} (\bibinfo {year} {2019})}\BibitemShut {NoStop}%
\bibitem [{\citenamefont {Song}\ \emph {et~al.}(2019)\citenamefont {Song}, \citenamefont {Wang}, \citenamefont {Shi}, \citenamefont {Li}, \citenamefont {Fang},\ and\ \citenamefont {Bernevig}}]{Song2019PHS}%
  \BibitemOpen
  \bibfield  {author} {\bibinfo {author} {\bibfnamefont {Z.}~\bibnamefont {Song}}, \bibinfo {author} {\bibfnamefont {Z.}~\bibnamefont {Wang}}, \bibinfo {author} {\bibfnamefont {W.}~\bibnamefont {Shi}}, \bibinfo {author} {\bibfnamefont {G.}~\bibnamefont {Li}}, \bibinfo {author} {\bibfnamefont {C.}~\bibnamefont {Fang}},\ and\ \bibinfo {author} {\bibfnamefont {B.~A.}\ \bibnamefont {Bernevig}},\ }\bibfield  {title} {\bibinfo {title} {All magic angles in twisted bilayer graphene are topological},\ }\bibfield  {journal} {\bibinfo  {journal} {Physical Review Letters}\ }\textbf {\bibinfo {volume} {123}},\ \href {https://doi.org/10.1103/physrevlett.123.036401} {10.1103/physrevlett.123.036401} (\bibinfo {year} {2019})\BibitemShut {NoStop}%
\bibitem [{\citenamefont {Kwan}\ \emph {et~al.}(2021)\citenamefont {Kwan}, \citenamefont {Wagner}, \citenamefont {Soejima}, \citenamefont {Zaletel}, \citenamefont {Simon}, \citenamefont {Parameswaran},\ and\ \citenamefont {Bultinck}}]{kwan_kekule_2021}%
  \BibitemOpen
  \bibfield  {author} {\bibinfo {author} {\bibfnamefont {Y.~H.}\ \bibnamefont {Kwan}}, \bibinfo {author} {\bibfnamefont {G.}~\bibnamefont {Wagner}}, \bibinfo {author} {\bibfnamefont {T.}~\bibnamefont {Soejima}}, \bibinfo {author} {\bibfnamefont {M.~P.}\ \bibnamefont {Zaletel}}, \bibinfo {author} {\bibfnamefont {S.~H.}\ \bibnamefont {Simon}}, \bibinfo {author} {\bibfnamefont {S.~A.}\ \bibnamefont {Parameswaran}},\ and\ \bibinfo {author} {\bibfnamefont {N.}~\bibnamefont {Bultinck}},\ }\bibfield  {title} {\bibinfo {title} {Kekul\'e spiral order at all nonzero integer fillings in twisted bilayer graphene},\ }\href {https://doi.org/10.1103/PhysRevX.11.041063} {\bibfield  {journal} {\bibinfo  {journal} {Phys. Rev. X}\ }\textbf {\bibinfo {volume} {11}},\ \bibinfo {pages} {041063} (\bibinfo {year} {2021})}\BibitemShut {NoStop}%
\bibitem [{\citenamefont {Nuckolls}\ \emph {et~al.}(2023)\citenamefont {Nuckolls}, \citenamefont {Lee}, \citenamefont {Oh}, \citenamefont {Wong}, \citenamefont {Soejima}, \citenamefont {Hong}, \citenamefont {C{\u{a}}lug{\u{a}}ru}, \citenamefont {Herzog-Arbeitman}, \citenamefont {Bernevig}, \citenamefont {Watanabe}, \citenamefont {Taniguchi}, \citenamefont {Regnault}, \citenamefont {Zaletel},\ and\ \citenamefont {Yazdani}}]{nuckolls2023quantum}%
  \BibitemOpen
  \bibfield  {author} {\bibinfo {author} {\bibfnamefont {K.~P.}\ \bibnamefont {Nuckolls}}, \bibinfo {author} {\bibfnamefont {R.~L.}\ \bibnamefont {Lee}}, \bibinfo {author} {\bibfnamefont {M.}~\bibnamefont {Oh}}, \bibinfo {author} {\bibfnamefont {D.}~\bibnamefont {Wong}}, \bibinfo {author} {\bibfnamefont {T.}~\bibnamefont {Soejima}}, \bibinfo {author} {\bibfnamefont {J.~P.}\ \bibnamefont {Hong}}, \bibinfo {author} {\bibfnamefont {D.}~\bibnamefont {C{\u{a}}lug{\u{a}}ru}}, \bibinfo {author} {\bibfnamefont {J.}~\bibnamefont {Herzog-Arbeitman}}, \bibinfo {author} {\bibfnamefont {B.~A.}\ \bibnamefont {Bernevig}}, \bibinfo {author} {\bibfnamefont {K.}~\bibnamefont {Watanabe}}, \bibinfo {author} {\bibfnamefont {T.}~\bibnamefont {Taniguchi}}, \bibinfo {author} {\bibfnamefont {N.}~\bibnamefont {Regnault}}, \bibinfo {author} {\bibfnamefont {M.~P.}\ \bibnamefont {Zaletel}},\ and\ \bibinfo {author} {\bibfnamefont {A.}~\bibnamefont {Yazdani}},\ }\bibfield  {title} {\bibinfo {title} {Quantum textures of the many-body
  wavefunctions in magic-angle graphene},\ }\href {https://doi.org/10.1038/s41586-023-06226-x} {\bibfield  {journal} {\bibinfo  {journal} {Nature}\ }\textbf {\bibinfo {volume} {620}},\ \bibinfo {pages} {525} (\bibinfo {year} {2023})}\BibitemShut {NoStop}%
\bibitem [{\citenamefont {Kim}\ \emph {et~al.}(2023)\citenamefont {Kim}, \citenamefont {Choi}, \citenamefont {Lantagne-Hurtubise}, \citenamefont {Lewandowski}, \citenamefont {Thomson}, \citenamefont {Kong}, \citenamefont {Zhou}, \citenamefont {Baum}, \citenamefont {Zhang}, \citenamefont {Holleis} \emph {et~al.}}]{kim2023imaging}%
  \BibitemOpen
  \bibfield  {author} {\bibinfo {author} {\bibfnamefont {H.}~\bibnamefont {Kim}}, \bibinfo {author} {\bibfnamefont {Y.}~\bibnamefont {Choi}}, \bibinfo {author} {\bibfnamefont {{\'E}.}~\bibnamefont {Lantagne-Hurtubise}}, \bibinfo {author} {\bibfnamefont {C.}~\bibnamefont {Lewandowski}}, \bibinfo {author} {\bibfnamefont {A.}~\bibnamefont {Thomson}}, \bibinfo {author} {\bibfnamefont {L.}~\bibnamefont {Kong}}, \bibinfo {author} {\bibfnamefont {H.}~\bibnamefont {Zhou}}, \bibinfo {author} {\bibfnamefont {E.}~\bibnamefont {Baum}}, \bibinfo {author} {\bibfnamefont {Y.}~\bibnamefont {Zhang}}, \bibinfo {author} {\bibfnamefont {L.}~\bibnamefont {Holleis}}, \emph {et~al.},\ }\bibfield  {title} {\bibinfo {title} {Imaging inter-valley coherent order in magic-angle twisted trilayer graphene},\ }\href@noop {} {\bibfield  {journal} {\bibinfo  {journal} {Nature}\ }\textbf {\bibinfo {volume} {623}},\ \bibinfo {pages} {942} (\bibinfo {year} {2023})}\BibitemShut {NoStop}%
\bibitem [{\citenamefont {Wang}\ \emph {et~al.}(2024{\natexlab{a}})\citenamefont {Wang}, \citenamefont {Kwan}, \citenamefont {Wagner}, \citenamefont {Bultinck}, \citenamefont {Simon},\ and\ \citenamefont {Parameswaran}}]{wang_kekule_2024}%
  \BibitemOpen
  \bibfield  {author} {\bibinfo {author} {\bibfnamefont {Z.}~\bibnamefont {Wang}}, \bibinfo {author} {\bibfnamefont {Y.~H.}\ \bibnamefont {Kwan}}, \bibinfo {author} {\bibfnamefont {G.}~\bibnamefont {Wagner}}, \bibinfo {author} {\bibfnamefont {N.}~\bibnamefont {Bultinck}}, \bibinfo {author} {\bibfnamefont {S.~H.}\ \bibnamefont {Simon}},\ and\ \bibinfo {author} {\bibfnamefont {S.~A.}\ \bibnamefont {Parameswaran}},\ }\bibfield  {title} {\bibinfo {title} {Kekul{\'e} spirals and charge transfer cascades in twisted symmetric trilayer graphene},\ }\href {https://doi.org/10.1103/PhysRevB.109.L201119} {\bibfield  {journal} {\bibinfo  {journal} {Physical Review B}\ }\textbf {\bibinfo {volume} {109}},\ \bibinfo {pages} {L201119} (\bibinfo {year} {2024}{\natexlab{a}})}\BibitemShut {NoStop}%
\bibitem [{\citenamefont {Zhang}\ \emph {et~al.}(2019{\natexlab{a}})\citenamefont {Zhang}, \citenamefont {Mao},\ and\ \citenamefont {Senthil}}]{Zhang2019anomalous}%
  \BibitemOpen
  \bibfield  {author} {\bibinfo {author} {\bibfnamefont {Y.-H.}\ \bibnamefont {Zhang}}, \bibinfo {author} {\bibfnamefont {D.}~\bibnamefont {Mao}},\ and\ \bibinfo {author} {\bibfnamefont {T.}~\bibnamefont {Senthil}},\ }\bibfield  {title} {\bibinfo {title} {Twisted bilayer graphene aligned with hexagonal boron nitride: Anomalous hall effect and a lattice model},\ }\bibfield  {journal} {\bibinfo  {journal} {Physical Review Research}\ }\textbf {\bibinfo {volume} {1}},\ \href {https://doi.org/10.1103/physrevresearch.1.033126} {10.1103/physrevresearch.1.033126} (\bibinfo {year} {2019}{\natexlab{a}})\BibitemShut {NoStop}%
\bibitem [{\citenamefont {Zhang}\ \emph {et~al.}(2019{\natexlab{b}})\citenamefont {Zhang}, \citenamefont {Mao}, \citenamefont {Cao}, \citenamefont {Jarillo-Herrero},\ and\ \citenamefont {Senthil}}]{Zhang2019}%
  \BibitemOpen
  \bibfield  {author} {\bibinfo {author} {\bibfnamefont {Y.-H.}\ \bibnamefont {Zhang}}, \bibinfo {author} {\bibfnamefont {D.}~\bibnamefont {Mao}}, \bibinfo {author} {\bibfnamefont {Y.}~\bibnamefont {Cao}}, \bibinfo {author} {\bibfnamefont {P.}~\bibnamefont {Jarillo-Herrero}},\ and\ \bibinfo {author} {\bibfnamefont {T.}~\bibnamefont {Senthil}},\ }\bibfield  {title} {\bibinfo {title} {Nearly flat chern bands in moir\'e superlattices},\ }\href {https://doi.org/10.1103/PhysRevB.99.075127} {\bibfield  {journal} {\bibinfo  {journal} {Phys. Rev. B}\ }\textbf {\bibinfo {volume} {99}},\ \bibinfo {pages} {075127} (\bibinfo {year} {2019}{\natexlab{b}})}\BibitemShut {NoStop}%
\bibitem [{\citenamefont {Bistritzer}\ and\ \citenamefont {MacDonald}(2011)}]{Bistritzer2011}%
  \BibitemOpen
  \bibfield  {author} {\bibinfo {author} {\bibfnamefont {R.}~\bibnamefont {Bistritzer}}\ and\ \bibinfo {author} {\bibfnamefont {A.~H.}\ \bibnamefont {MacDonald}},\ }\bibfield  {title} {\bibinfo {title} {Moir{\'e} bands in twisted double-layer graphene},\ }\href {https://doi.org/10.1073/pnas.1108174108} {\bibfield  {journal} {\bibinfo  {journal} {Proceedings of the National Academy of Sciences}\ }\textbf {\bibinfo {volume} {108}},\ \bibinfo {pages} {12233} (\bibinfo {year} {2011})},\ \Eprint {https://arxiv.org/abs/https://www.pnas.org/content/108/30/12233.full.pdf} {https://www.pnas.org/content/108/30/12233.full.pdf} \BibitemShut {NoStop}%
\bibitem [{\citenamefont {Burg}\ \emph {et~al.}(2019)\citenamefont {Burg}, \citenamefont {Zhu}, \citenamefont {Taniguchi}, \citenamefont {Watanabe}, \citenamefont {MacDonald},\ and\ \citenamefont {Tutuc}}]{burg_correlated_2019}%
  \BibitemOpen
  \bibfield  {author} {\bibinfo {author} {\bibfnamefont {G.~W.}\ \bibnamefont {Burg}}, \bibinfo {author} {\bibfnamefont {J.}~\bibnamefont {Zhu}}, \bibinfo {author} {\bibfnamefont {T.}~\bibnamefont {Taniguchi}}, \bibinfo {author} {\bibfnamefont {K.}~\bibnamefont {Watanabe}}, \bibinfo {author} {\bibfnamefont {A.~H.}\ \bibnamefont {MacDonald}},\ and\ \bibinfo {author} {\bibfnamefont {E.}~\bibnamefont {Tutuc}},\ }\bibfield  {title} {\bibinfo {title} {Correlated {{Insulating States}} in {{Twisted Double Bilayer Graphene}}},\ }\href {https://doi.org/10.1103/PhysRevLett.123.197702} {\bibfield  {journal} {\bibinfo  {journal} {Physical Review Letters}\ }\textbf {\bibinfo {volume} {123}},\ \bibinfo {pages} {197702} (\bibinfo {year} {2019})}\BibitemShut {NoStop}%
\bibitem [{\citenamefont {Cao}\ \emph {et~al.}(2020)\citenamefont {Cao}, \citenamefont {{Rodan-Legrain}}, \citenamefont {{Rubies-Bigorda}}, \citenamefont {Park}, \citenamefont {Watanabe}, \citenamefont {Taniguchi},\ and\ \citenamefont {{Jarillo-Herrero}}}]{cao_tunable_2020}%
  \BibitemOpen
  \bibfield  {author} {\bibinfo {author} {\bibfnamefont {Y.}~\bibnamefont {Cao}}, \bibinfo {author} {\bibfnamefont {D.}~\bibnamefont {{Rodan-Legrain}}}, \bibinfo {author} {\bibfnamefont {O.}~\bibnamefont {{Rubies-Bigorda}}}, \bibinfo {author} {\bibfnamefont {J.~M.}\ \bibnamefont {Park}}, \bibinfo {author} {\bibfnamefont {K.}~\bibnamefont {Watanabe}}, \bibinfo {author} {\bibfnamefont {T.}~\bibnamefont {Taniguchi}},\ and\ \bibinfo {author} {\bibfnamefont {P.}~\bibnamefont {{Jarillo-Herrero}}},\ }\bibfield  {title} {\bibinfo {title} {Tunable correlated states and spin-polarized phases in twisted bilayer--bilayer graphene},\ }\href {https://doi.org/10.1038/s41586-020-2260-6} {\bibfield  {journal} {\bibinfo  {journal} {Nature}\ }\textbf {\bibinfo {volume} {583}},\ \bibinfo {pages} {215} (\bibinfo {year} {2020})}\BibitemShut {NoStop}%
\bibitem [{\citenamefont {Du}\ \emph {et~al.}(2024)\citenamefont {Du}, \citenamefont {Xiao}, \citenamefont {Zhang}, \citenamefont {Cai}, \citenamefont {Jiang}, \citenamefont {Lian}, \citenamefont {Watanabe}, \citenamefont {Taniguchi}, \citenamefont {Wang},\ and\ \citenamefont {Yu}}]{du_ferroelectricity_2024}%
  \BibitemOpen
  \bibfield  {author} {\bibinfo {author} {\bibfnamefont {R.}~\bibnamefont {Du}}, \bibinfo {author} {\bibfnamefont {J.}~\bibnamefont {Xiao}}, \bibinfo {author} {\bibfnamefont {D.}~\bibnamefont {Zhang}}, \bibinfo {author} {\bibfnamefont {X.}~\bibnamefont {Cai}}, \bibinfo {author} {\bibfnamefont {S.}~\bibnamefont {Jiang}}, \bibinfo {author} {\bibfnamefont {F.}~\bibnamefont {Lian}}, \bibinfo {author} {\bibfnamefont {K.}~\bibnamefont {Watanabe}}, \bibinfo {author} {\bibfnamefont {T.}~\bibnamefont {Taniguchi}}, \bibinfo {author} {\bibfnamefont {L.}~\bibnamefont {Wang}},\ and\ \bibinfo {author} {\bibfnamefont {G.}~\bibnamefont {Yu}},\ }\bibfield  {title} {\bibinfo {title} {Ferroelectricity in twisted double bilayer graphene},\ }\href {https://doi.org/10.1088/2053-1583/ad2107} {\bibfield  {journal} {\bibinfo  {journal} {2D Materials}\ }\textbf {\bibinfo {volume} {11}},\ \bibinfo {pages} {025015} (\bibinfo {year} {2024})},\ \Eprint {https://arxiv.org/abs/2404.18059} {arxiv:2404.18059 [cond-mat]} \BibitemShut {NoStop}%
\bibitem [{\citenamefont {He}\ \emph {et~al.}(2021{\natexlab{a}})\citenamefont {He}, \citenamefont {Li}, \citenamefont {Cai}, \citenamefont {Liu}, \citenamefont {Watanabe}, \citenamefont {Taniguchi}, \citenamefont {Xu},\ and\ \citenamefont {Yankowitz}}]{he_symmetry_2021}%
  \BibitemOpen
  \bibfield  {author} {\bibinfo {author} {\bibfnamefont {M.}~\bibnamefont {He}}, \bibinfo {author} {\bibfnamefont {Y.}~\bibnamefont {Li}}, \bibinfo {author} {\bibfnamefont {J.}~\bibnamefont {Cai}}, \bibinfo {author} {\bibfnamefont {Y.}~\bibnamefont {Liu}}, \bibinfo {author} {\bibfnamefont {K.}~\bibnamefont {Watanabe}}, \bibinfo {author} {\bibfnamefont {T.}~\bibnamefont {Taniguchi}}, \bibinfo {author} {\bibfnamefont {X.}~\bibnamefont {Xu}},\ and\ \bibinfo {author} {\bibfnamefont {M.}~\bibnamefont {Yankowitz}},\ }\bibfield  {title} {\bibinfo {title} {Symmetry breaking in twisted double bilayer graphene},\ }\href {https://doi.org/10.1038/s41567-020-1030-6} {\bibfield  {journal} {\bibinfo  {journal} {Nature Physics}\ }\textbf {\bibinfo {volume} {17}},\ \bibinfo {pages} {26} (\bibinfo {year} {2021}{\natexlab{a}})}\BibitemShut {NoStop}%
\bibitem [{\citenamefont {Liu}\ \emph {et~al.}(2022{\natexlab{a}})\citenamefont {Liu}, \citenamefont {Zhang}, \citenamefont {Chu}, \citenamefont {Shen}, \citenamefont {Huang}, \citenamefont {Yuan}, \citenamefont {Tian}, \citenamefont {Tang}, \citenamefont {Ji}, \citenamefont {Yang}, \citenamefont {Watanabe}, \citenamefont {Taniguchi}, \citenamefont {Shi}, \citenamefont {Liu}, \citenamefont {Yang},\ and\ \citenamefont {Zhang}}]{liu_isospin_2022}%
  \BibitemOpen
  \bibfield  {author} {\bibinfo {author} {\bibfnamefont {L.}~\bibnamefont {Liu}}, \bibinfo {author} {\bibfnamefont {S.}~\bibnamefont {Zhang}}, \bibinfo {author} {\bibfnamefont {Y.}~\bibnamefont {Chu}}, \bibinfo {author} {\bibfnamefont {C.}~\bibnamefont {Shen}}, \bibinfo {author} {\bibfnamefont {Y.}~\bibnamefont {Huang}}, \bibinfo {author} {\bibfnamefont {Y.}~\bibnamefont {Yuan}}, \bibinfo {author} {\bibfnamefont {J.}~\bibnamefont {Tian}}, \bibinfo {author} {\bibfnamefont {J.}~\bibnamefont {Tang}}, \bibinfo {author} {\bibfnamefont {Y.}~\bibnamefont {Ji}}, \bibinfo {author} {\bibfnamefont {R.}~\bibnamefont {Yang}}, \bibinfo {author} {\bibfnamefont {K.}~\bibnamefont {Watanabe}}, \bibinfo {author} {\bibfnamefont {T.}~\bibnamefont {Taniguchi}}, \bibinfo {author} {\bibfnamefont {D.}~\bibnamefont {Shi}}, \bibinfo {author} {\bibfnamefont {J.}~\bibnamefont {Liu}}, \bibinfo {author} {\bibfnamefont {W.}~\bibnamefont {Yang}},\ and\ \bibinfo {author} {\bibfnamefont {G.}~\bibnamefont {Zhang}},\ }\bibfield  {title}
  {\bibinfo {title} {Isospin competitions and valley polarized correlated insulators in twisted double bilayer graphene},\ }\href {https://doi.org/10.1038/s41467-022-30998-x} {\bibfield  {journal} {\bibinfo  {journal} {Nature Communications}\ }\textbf {\bibinfo {volume} {13}},\ \bibinfo {pages} {3292} (\bibinfo {year} {2022}{\natexlab{a}})}\BibitemShut {NoStop}%
\bibitem [{\citenamefont {Liu}\ \emph {et~al.}(2020)\citenamefont {Liu}, \citenamefont {Hao}, \citenamefont {Khalaf}, \citenamefont {Lee}, \citenamefont {Ronen}, \citenamefont {Yoo}, \citenamefont {Haei~Najafabadi}, \citenamefont {Watanabe}, \citenamefont {Taniguchi}, \citenamefont {Vishwanath},\ and\ \citenamefont {Kim}}]{liu_tunable_2020}%
  \BibitemOpen
  \bibfield  {author} {\bibinfo {author} {\bibfnamefont {X.}~\bibnamefont {Liu}}, \bibinfo {author} {\bibfnamefont {Z.}~\bibnamefont {Hao}}, \bibinfo {author} {\bibfnamefont {E.}~\bibnamefont {Khalaf}}, \bibinfo {author} {\bibfnamefont {J.~Y.}\ \bibnamefont {Lee}}, \bibinfo {author} {\bibfnamefont {Y.}~\bibnamefont {Ronen}}, \bibinfo {author} {\bibfnamefont {H.}~\bibnamefont {Yoo}}, \bibinfo {author} {\bibfnamefont {D.}~\bibnamefont {Haei~Najafabadi}}, \bibinfo {author} {\bibfnamefont {K.}~\bibnamefont {Watanabe}}, \bibinfo {author} {\bibfnamefont {T.}~\bibnamefont {Taniguchi}}, \bibinfo {author} {\bibfnamefont {A.}~\bibnamefont {Vishwanath}},\ and\ \bibinfo {author} {\bibfnamefont {P.}~\bibnamefont {Kim}},\ }\bibfield  {title} {\bibinfo {title} {Tunable spin-polarized correlated states in twisted double bilayer graphene},\ }\href {https://doi.org/10.1038/s41586-020-2458-7} {\bibfield  {journal} {\bibinfo  {journal} {Nature}\ }\textbf {\bibinfo {volume} {583}},\ \bibinfo {pages} {221} (\bibinfo {year}
  {2020})}\BibitemShut {NoStop}%
\bibitem [{\citenamefont {Rickhaus}\ \emph {et~al.}(2021)\citenamefont {Rickhaus}, \citenamefont {{de Vries}}, \citenamefont {Zhu}, \citenamefont {Portoles}, \citenamefont {Zheng}, \citenamefont {Masseroni}, \citenamefont {Kurzmann}, \citenamefont {Taniguchi}, \citenamefont {Watanabe}, \citenamefont {MacDonald}, \citenamefont {Ihn},\ and\ \citenamefont {Ensslin}}]{rickhaus_correlated_2021}%
  \BibitemOpen
  \bibfield  {author} {\bibinfo {author} {\bibfnamefont {P.}~\bibnamefont {Rickhaus}}, \bibinfo {author} {\bibfnamefont {F.~K.}\ \bibnamefont {{de Vries}}}, \bibinfo {author} {\bibfnamefont {J.}~\bibnamefont {Zhu}}, \bibinfo {author} {\bibfnamefont {E.}~\bibnamefont {Portoles}}, \bibinfo {author} {\bibfnamefont {G.}~\bibnamefont {Zheng}}, \bibinfo {author} {\bibfnamefont {M.}~\bibnamefont {Masseroni}}, \bibinfo {author} {\bibfnamefont {A.}~\bibnamefont {Kurzmann}}, \bibinfo {author} {\bibfnamefont {T.}~\bibnamefont {Taniguchi}}, \bibinfo {author} {\bibfnamefont {K.}~\bibnamefont {Watanabe}}, \bibinfo {author} {\bibfnamefont {A.~H.}\ \bibnamefont {MacDonald}}, \bibinfo {author} {\bibfnamefont {T.}~\bibnamefont {Ihn}},\ and\ \bibinfo {author} {\bibfnamefont {K.}~\bibnamefont {Ensslin}},\ }\bibfield  {title} {\bibinfo {title} {Correlated electron-hole state in twisted double-bilayer graphene},\ }\href {https://doi.org/10.1126/science.abc3534} {\bibfield  {journal} {\bibinfo  {journal} {Science}\ }\textbf
  {\bibinfo {volume} {373}},\ \bibinfo {pages} {1257} (\bibinfo {year} {2021})}\BibitemShut {NoStop}%
\bibitem [{\citenamefont {{Rubio-Verd{\'u}}}\ \emph {et~al.}(2022)\citenamefont {{Rubio-Verd{\'u}}}, \citenamefont {Turkel}, \citenamefont {Song}, \citenamefont {Klebl}, \citenamefont {Samajdar}, \citenamefont {Scheurer}, \citenamefont {Venderbos}, \citenamefont {Watanabe}, \citenamefont {Taniguchi}, \citenamefont {Ochoa}, \citenamefont {Xian}, \citenamefont {Kennes}, \citenamefont {Fernandes}, \citenamefont {Rubio},\ and\ \citenamefont {Pasupathy}}]{rubio-verdu_moire_2022}%
  \BibitemOpen
  \bibfield  {author} {\bibinfo {author} {\bibfnamefont {C.}~\bibnamefont {{Rubio-Verd{\'u}}}}, \bibinfo {author} {\bibfnamefont {S.}~\bibnamefont {Turkel}}, \bibinfo {author} {\bibfnamefont {Y.}~\bibnamefont {Song}}, \bibinfo {author} {\bibfnamefont {L.}~\bibnamefont {Klebl}}, \bibinfo {author} {\bibfnamefont {R.}~\bibnamefont {Samajdar}}, \bibinfo {author} {\bibfnamefont {M.~S.}\ \bibnamefont {Scheurer}}, \bibinfo {author} {\bibfnamefont {J.~W.~F.}\ \bibnamefont {Venderbos}}, \bibinfo {author} {\bibfnamefont {K.}~\bibnamefont {Watanabe}}, \bibinfo {author} {\bibfnamefont {T.}~\bibnamefont {Taniguchi}}, \bibinfo {author} {\bibfnamefont {H.}~\bibnamefont {Ochoa}}, \bibinfo {author} {\bibfnamefont {L.}~\bibnamefont {Xian}}, \bibinfo {author} {\bibfnamefont {D.~M.}\ \bibnamefont {Kennes}}, \bibinfo {author} {\bibfnamefont {R.~M.}\ \bibnamefont {Fernandes}}, \bibinfo {author} {\bibfnamefont {{\'A}.}~\bibnamefont {Rubio}},\ and\ \bibinfo {author} {\bibfnamefont {A.~N.}\ \bibnamefont {Pasupathy}},\ }\bibfield
  {title} {\bibinfo {title} {Moir{\'e} nematic phase in twisted double bilayer graphene},\ }\href {https://doi.org/10.1038/s41567-021-01438-2} {\bibfield  {journal} {\bibinfo  {journal} {Nature Physics}\ }\textbf {\bibinfo {volume} {18}},\ \bibinfo {pages} {196} (\bibinfo {year} {2022})}\BibitemShut {NoStop}%
\bibitem [{\citenamefont {Shen}\ \emph {et~al.}(2020)\citenamefont {Shen}, \citenamefont {Chu}, \citenamefont {Wu}, \citenamefont {Li}, \citenamefont {Wang}, \citenamefont {Zhao}, \citenamefont {Tang}, \citenamefont {Liu}, \citenamefont {Tian}, \citenamefont {Watanabe}, \citenamefont {Taniguchi}, \citenamefont {Yang}, \citenamefont {Meng}, \citenamefont {Shi}, \citenamefont {Yazyev},\ and\ \citenamefont {Zhang}}]{shen_correlated_2020}%
  \BibitemOpen
  \bibfield  {author} {\bibinfo {author} {\bibfnamefont {C.}~\bibnamefont {Shen}}, \bibinfo {author} {\bibfnamefont {Y.}~\bibnamefont {Chu}}, \bibinfo {author} {\bibfnamefont {Q.}~\bibnamefont {Wu}}, \bibinfo {author} {\bibfnamefont {N.}~\bibnamefont {Li}}, \bibinfo {author} {\bibfnamefont {S.}~\bibnamefont {Wang}}, \bibinfo {author} {\bibfnamefont {Y.}~\bibnamefont {Zhao}}, \bibinfo {author} {\bibfnamefont {J.}~\bibnamefont {Tang}}, \bibinfo {author} {\bibfnamefont {J.}~\bibnamefont {Liu}}, \bibinfo {author} {\bibfnamefont {J.}~\bibnamefont {Tian}}, \bibinfo {author} {\bibfnamefont {K.}~\bibnamefont {Watanabe}}, \bibinfo {author} {\bibfnamefont {T.}~\bibnamefont {Taniguchi}}, \bibinfo {author} {\bibfnamefont {R.}~\bibnamefont {Yang}}, \bibinfo {author} {\bibfnamefont {Z.~Y.}\ \bibnamefont {Meng}}, \bibinfo {author} {\bibfnamefont {D.}~\bibnamefont {Shi}}, \bibinfo {author} {\bibfnamefont {O.~V.}\ \bibnamefont {Yazyev}},\ and\ \bibinfo {author} {\bibfnamefont {G.}~\bibnamefont {Zhang}},\ }\bibfield  {title}
  {\bibinfo {title} {Correlated states in twisted double bilayer graphene},\ }\href {https://doi.org/10.1038/s41567-020-0825-9} {\bibfield  {journal} {\bibinfo  {journal} {Nature Physics}\ }\textbf {\bibinfo {volume} {16}},\ \bibinfo {pages} {520} (\bibinfo {year} {2020})}\BibitemShut {NoStop}%
\bibitem [{\citenamefont {Zhang}\ \emph {et~al.}(2021)\citenamefont {Zhang}, \citenamefont {Zhu}, \citenamefont {Kahn}, \citenamefont {Li}, \citenamefont {Yang}, \citenamefont {Herbig}, \citenamefont {Wu}, \citenamefont {Li}, \citenamefont {Watanabe}, \citenamefont {Taniguchi}, \citenamefont {Cabrini}, \citenamefont {Zettl}, \citenamefont {Zaletel}, \citenamefont {Wang},\ and\ \citenamefont {Crommie}}]{zhang_visualizing_2021}%
  \BibitemOpen
  \bibfield  {author} {\bibinfo {author} {\bibfnamefont {C.}~\bibnamefont {Zhang}}, \bibinfo {author} {\bibfnamefont {T.}~\bibnamefont {Zhu}}, \bibinfo {author} {\bibfnamefont {S.}~\bibnamefont {Kahn}}, \bibinfo {author} {\bibfnamefont {S.}~\bibnamefont {Li}}, \bibinfo {author} {\bibfnamefont {B.}~\bibnamefont {Yang}}, \bibinfo {author} {\bibfnamefont {C.}~\bibnamefont {Herbig}}, \bibinfo {author} {\bibfnamefont {X.}~\bibnamefont {Wu}}, \bibinfo {author} {\bibfnamefont {H.}~\bibnamefont {Li}}, \bibinfo {author} {\bibfnamefont {K.}~\bibnamefont {Watanabe}}, \bibinfo {author} {\bibfnamefont {T.}~\bibnamefont {Taniguchi}}, \bibinfo {author} {\bibfnamefont {S.}~\bibnamefont {Cabrini}}, \bibinfo {author} {\bibfnamefont {A.}~\bibnamefont {Zettl}}, \bibinfo {author} {\bibfnamefont {M.~P.}\ \bibnamefont {Zaletel}}, \bibinfo {author} {\bibfnamefont {F.}~\bibnamefont {Wang}},\ and\ \bibinfo {author} {\bibfnamefont {M.~F.}\ \bibnamefont {Crommie}},\ }\bibfield  {title} {\bibinfo {title} {Visualizing delocalized
  correlated electronic states in twisted double bilayer graphene},\ }\href {https://doi.org/10.1038/s41467-021-22711-1} {\bibfield  {journal} {\bibinfo  {journal} {Nature Communications}\ }\textbf {\bibinfo {volume} {12}},\ \bibinfo {pages} {2516} (\bibinfo {year} {2021})}\BibitemShut {NoStop}%
\bibitem [{\citenamefont {Zhu}\ \emph {et~al.}(2022)\citenamefont {Zhu}, \citenamefont {Chen}, \citenamefont {Li}, \citenamefont {Chen}, \citenamefont {Huang}, \citenamefont {Zhu}, \citenamefont {An}, \citenamefont {Song}, \citenamefont {Gan}, \citenamefont {Wang}, \citenamefont {Wei}, \citenamefont {Zong}, \citenamefont {Watanabe}, \citenamefont {Taniguchi}, \citenamefont {Wang}, \citenamefont {Huang}, \citenamefont {Xian}, \citenamefont {Sun}, \citenamefont {Yu},\ and\ \citenamefont {Wang}}]{zhu_tunable_2022}%
  \BibitemOpen
  \bibfield  {author} {\bibinfo {author} {\bibfnamefont {Y.}~\bibnamefont {Zhu}}, \bibinfo {author} {\bibfnamefont {Y.}~\bibnamefont {Chen}}, \bibinfo {author} {\bibfnamefont {Q.}~\bibnamefont {Li}}, \bibinfo {author} {\bibfnamefont {Y.}~\bibnamefont {Chen}}, \bibinfo {author} {\bibfnamefont {Y.}~\bibnamefont {Huang}}, \bibinfo {author} {\bibfnamefont {W.}~\bibnamefont {Zhu}}, \bibinfo {author} {\bibfnamefont {D.}~\bibnamefont {An}}, \bibinfo {author} {\bibfnamefont {J.}~\bibnamefont {Song}}, \bibinfo {author} {\bibfnamefont {Q.}~\bibnamefont {Gan}}, \bibinfo {author} {\bibfnamefont {K.}~\bibnamefont {Wang}}, \bibinfo {author} {\bibfnamefont {L.}~\bibnamefont {Wei}}, \bibinfo {author} {\bibfnamefont {Q.}~\bibnamefont {Zong}}, \bibinfo {author} {\bibfnamefont {K.}~\bibnamefont {Watanabe}}, \bibinfo {author} {\bibfnamefont {T.}~\bibnamefont {Taniguchi}}, \bibinfo {author} {\bibfnamefont {H.}~\bibnamefont {Wang}}, \bibinfo {author} {\bibfnamefont {L.}~\bibnamefont {Huang}}, \bibinfo {author} {\bibfnamefont
  {L.}~\bibnamefont {Xian}}, \bibinfo {author} {\bibfnamefont {L.}~\bibnamefont {Sun}}, \bibinfo {author} {\bibfnamefont {G.}~\bibnamefont {Yu}},\ and\ \bibinfo {author} {\bibfnamefont {L.}~\bibnamefont {Wang}},\ }\bibfield  {title} {\bibinfo {title} {Tunable multi-bands in twisted double bilayer graphene},\ }\href {https://doi.org/10.1088/2053-1583/ac69bb} {\bibfield  {journal} {\bibinfo  {journal} {2D Materials}\ }\textbf {\bibinfo {volume} {9}},\ \bibinfo {pages} {034001} (\bibinfo {year} {2022})}\BibitemShut {NoStop}%
\bibitem [{\citenamefont {He}\ \emph {et~al.}(2023)\citenamefont {He}, \citenamefont {Cai}, \citenamefont {Zhang}, \citenamefont {Liu}, \citenamefont {Li}, \citenamefont {Taniguchi}, \citenamefont {Watanabe}, \citenamefont {Cobden}, \citenamefont {Yankowitz},\ and\ \citenamefont {Xu}}]{he_symmetry-broken_2023}%
  \BibitemOpen
  \bibfield  {author} {\bibinfo {author} {\bibfnamefont {M.}~\bibnamefont {He}}, \bibinfo {author} {\bibfnamefont {J.}~\bibnamefont {Cai}}, \bibinfo {author} {\bibfnamefont {Y.-H.}\ \bibnamefont {Zhang}}, \bibinfo {author} {\bibfnamefont {Y.}~\bibnamefont {Liu}}, \bibinfo {author} {\bibfnamefont {Y.}~\bibnamefont {Li}}, \bibinfo {author} {\bibfnamefont {T.}~\bibnamefont {Taniguchi}}, \bibinfo {author} {\bibfnamefont {K.}~\bibnamefont {Watanabe}}, \bibinfo {author} {\bibfnamefont {D.~H.}\ \bibnamefont {Cobden}}, \bibinfo {author} {\bibfnamefont {M.}~\bibnamefont {Yankowitz}},\ and\ \bibinfo {author} {\bibfnamefont {X.}~\bibnamefont {Xu}},\ }\bibfield  {title} {\bibinfo {title} {Symmetry-{{Broken Chern Insulators}} in {{Twisted Double Bilayer Graphene}}},\ }\href {https://doi.org/10.1021/acs.nanolett.3c03414} {\bibfield  {journal} {\bibinfo  {journal} {Nano Letters}\ }\textbf {\bibinfo {volume} {23}},\ \bibinfo {pages} {11066} (\bibinfo {year} {2023})}\BibitemShut {NoStop}%
\bibitem [{\citenamefont {Koshino}(2019)}]{koshino_band_2019}%
  \BibitemOpen
  \bibfield  {author} {\bibinfo {author} {\bibfnamefont {M.}~\bibnamefont {Koshino}},\ }\bibfield  {title} {\bibinfo {title} {Band structure and topological properties of twisted double bilayer graphene},\ }\href {https://doi.org/10.1103/PhysRevB.99.235406} {\bibfield  {journal} {\bibinfo  {journal} {Physical Review B}\ }\textbf {\bibinfo {volume} {99}},\ \bibinfo {pages} {235406} (\bibinfo {year} {2019})}\BibitemShut {NoStop}%
\bibitem [{\citenamefont {Liu}\ \emph {et~al.}(2019)\citenamefont {Liu}, \citenamefont {Ma}, \citenamefont {Gao},\ and\ \citenamefont {Dai}}]{liu_quantum_2019-1}%
  \BibitemOpen
  \bibfield  {author} {\bibinfo {author} {\bibfnamefont {J.}~\bibnamefont {Liu}}, \bibinfo {author} {\bibfnamefont {Z.}~\bibnamefont {Ma}}, \bibinfo {author} {\bibfnamefont {J.}~\bibnamefont {Gao}},\ and\ \bibinfo {author} {\bibfnamefont {X.}~\bibnamefont {Dai}},\ }\bibfield  {title} {\bibinfo {title} {Quantum {{Valley Hall Effect}}, {{Orbital Magnetism}}, and {{Anomalous Hall Effect}} in {{Twisted Multilayer Graphene Systems}}},\ }\href {https://doi.org/10.1103/PhysRevX.9.031021} {\bibfield  {journal} {\bibinfo  {journal} {Physical Review X}\ }\textbf {\bibinfo {volume} {9}},\ \bibinfo {pages} {031021} (\bibinfo {year} {2019})}\BibitemShut {NoStop}%
\bibitem [{\citenamefont {Chebrolu}\ \emph {et~al.}(2019)\citenamefont {Chebrolu}, \citenamefont {Chittari},\ and\ \citenamefont {Jung}}]{chebrolu_flat_2019}%
  \BibitemOpen
  \bibfield  {author} {\bibinfo {author} {\bibfnamefont {N.~R.}\ \bibnamefont {Chebrolu}}, \bibinfo {author} {\bibfnamefont {B.~L.}\ \bibnamefont {Chittari}},\ and\ \bibinfo {author} {\bibfnamefont {J.}~\bibnamefont {Jung}},\ }\bibfield  {title} {\bibinfo {title} {Flat bands in twisted double bilayer graphene},\ }\href {https://doi.org/10.1103/PhysRevB.99.235417} {\bibfield  {journal} {\bibinfo  {journal} {Physical Review B}\ }\textbf {\bibinfo {volume} {99}},\ \bibinfo {pages} {235417} (\bibinfo {year} {2019})}\BibitemShut {NoStop}%
\bibitem [{\citenamefont {Lee}\ \emph {et~al.}(2019)\citenamefont {Lee}, \citenamefont {Khalaf}, \citenamefont {Liu}, \citenamefont {Liu}, \citenamefont {Hao}, \citenamefont {Kim},\ and\ \citenamefont {Vishwanath}}]{Lee2019}%
  \BibitemOpen
  \bibfield  {author} {\bibinfo {author} {\bibfnamefont {J.~Y.}\ \bibnamefont {Lee}}, \bibinfo {author} {\bibfnamefont {E.}~\bibnamefont {Khalaf}}, \bibinfo {author} {\bibfnamefont {S.}~\bibnamefont {Liu}}, \bibinfo {author} {\bibfnamefont {X.}~\bibnamefont {Liu}}, \bibinfo {author} {\bibfnamefont {Z.}~\bibnamefont {Hao}}, \bibinfo {author} {\bibfnamefont {P.}~\bibnamefont {Kim}},\ and\ \bibinfo {author} {\bibfnamefont {A.}~\bibnamefont {Vishwanath}},\ }\bibfield  {title} {\bibinfo {title} {Theory of correlated insulating behaviour and spin-triplet superconductivity in twisted double bilayer graphene},\ }\href {https://doi.org/10.1038/s41467-019-12981-1} {\bibfield  {journal} {\bibinfo  {journal} {Nature Communications}\ }\textbf {\bibinfo {volume} {10}},\ \bibinfo {pages} {5333} (\bibinfo {year} {2019})}\BibitemShut {NoStop}%
\bibitem [{\citenamefont {Haddadi}\ \emph {et~al.}(2020)\citenamefont {Haddadi}, \citenamefont {Wu}, \citenamefont {Kruchkov},\ and\ \citenamefont {Yazyev}}]{haddadi_moire_2020}%
  \BibitemOpen
  \bibfield  {author} {\bibinfo {author} {\bibfnamefont {F.}~\bibnamefont {Haddadi}}, \bibinfo {author} {\bibfnamefont {Q.}~\bibnamefont {Wu}}, \bibinfo {author} {\bibfnamefont {A.~J.}\ \bibnamefont {Kruchkov}},\ and\ \bibinfo {author} {\bibfnamefont {O.~V.}\ \bibnamefont {Yazyev}},\ }\bibfield  {title} {\bibinfo {title} {Moir{\'e} {{Flat Bands}} in {{Twisted Double Bilayer Graphene}}},\ }\href {https://doi.org/10.1021/acs.nanolett.9b05117} {\bibfield  {journal} {\bibinfo  {journal} {Nano Letters}\ }\textbf {\bibinfo {volume} {20}},\ \bibinfo {pages} {2410} (\bibinfo {year} {2020})}\BibitemShut {NoStop}%
\bibitem [{\citenamefont {Jung}\ and\ \citenamefont {MacDonald}(2014)}]{jung_accurate_2014}%
  \BibitemOpen
  \bibfield  {author} {\bibinfo {author} {\bibfnamefont {J.}~\bibnamefont {Jung}}\ and\ \bibinfo {author} {\bibfnamefont {A.~H.}\ \bibnamefont {MacDonald}},\ }\bibfield  {title} {\bibinfo {title} {Accurate tight-binding models for the \${\textbackslash}ensuremath\{{\textbackslash}pi\}\$ bands of bilayer graphene},\ }\href {https://doi.org/10.1103/PhysRevB.89.035405} {\bibfield  {journal} {\bibinfo  {journal} {Physical Review B}\ }\textbf {\bibinfo {volume} {89}},\ \bibinfo {pages} {035405} (\bibinfo {year} {2014})}\BibitemShut {NoStop}%
\bibitem [{\citenamefont {Chen}\ \emph {et~al.}(2021)\citenamefont {Chen}, \citenamefont {He}, \citenamefont {Zhang}, \citenamefont {Hsieh}, \citenamefont {Fei}, \citenamefont {Watanabe}, \citenamefont {Taniguchi}, \citenamefont {Cobden}, \citenamefont {Xu}, \citenamefont {Dean},\ and\ \citenamefont {Yankowitz}}]{chen_electrically_2021}%
  \BibitemOpen
  \bibfield  {author} {\bibinfo {author} {\bibfnamefont {S.}~\bibnamefont {Chen}}, \bibinfo {author} {\bibfnamefont {M.}~\bibnamefont {He}}, \bibinfo {author} {\bibfnamefont {Y.-H.}\ \bibnamefont {Zhang}}, \bibinfo {author} {\bibfnamefont {V.}~\bibnamefont {Hsieh}}, \bibinfo {author} {\bibfnamefont {Z.}~\bibnamefont {Fei}}, \bibinfo {author} {\bibfnamefont {K.}~\bibnamefont {Watanabe}}, \bibinfo {author} {\bibfnamefont {T.}~\bibnamefont {Taniguchi}}, \bibinfo {author} {\bibfnamefont {D.~H.}\ \bibnamefont {Cobden}}, \bibinfo {author} {\bibfnamefont {X.}~\bibnamefont {Xu}}, \bibinfo {author} {\bibfnamefont {C.~R.}\ \bibnamefont {Dean}},\ and\ \bibinfo {author} {\bibfnamefont {M.}~\bibnamefont {Yankowitz}},\ }\bibfield  {title} {\bibinfo {title} {Electrically tunable correlated and topological states in twisted monolayer--bilayer graphene},\ }\href {https://doi.org/10.1038/s41567-020-01062-6} {\bibfield  {journal} {\bibinfo  {journal} {Nature Physics}\ }\textbf {\bibinfo {volume} {17}},\ \bibinfo {pages} {374}
  (\bibinfo {year} {2021})}\BibitemShut {NoStop}%
\bibitem [{\citenamefont {He}\ \emph {et~al.}(2021{\natexlab{b}})\citenamefont {He}, \citenamefont {Zhang}, \citenamefont {Li}, \citenamefont {Fei}, \citenamefont {Watanabe}, \citenamefont {Taniguchi}, \citenamefont {Xu},\ and\ \citenamefont {Yankowitz}}]{he_competing_2021}%
  \BibitemOpen
  \bibfield  {author} {\bibinfo {author} {\bibfnamefont {M.}~\bibnamefont {He}}, \bibinfo {author} {\bibfnamefont {Y.-H.}\ \bibnamefont {Zhang}}, \bibinfo {author} {\bibfnamefont {Y.}~\bibnamefont {Li}}, \bibinfo {author} {\bibfnamefont {Z.}~\bibnamefont {Fei}}, \bibinfo {author} {\bibfnamefont {K.}~\bibnamefont {Watanabe}}, \bibinfo {author} {\bibfnamefont {T.}~\bibnamefont {Taniguchi}}, \bibinfo {author} {\bibfnamefont {X.}~\bibnamefont {Xu}},\ and\ \bibinfo {author} {\bibfnamefont {M.}~\bibnamefont {Yankowitz}},\ }\bibfield  {title} {\bibinfo {title} {Competing correlated states and abundant orbital magnetism in twisted monolayer-bilayer graphene},\ }\href {https://doi.org/10.1038/s41467-021-25044-1} {\bibfield  {journal} {\bibinfo  {journal} {Nature Communications}\ }\textbf {\bibinfo {volume} {12}},\ \bibinfo {pages} {4727} (\bibinfo {year} {2021}{\natexlab{b}})}\BibitemShut {NoStop}%
\bibitem [{\citenamefont {Li}\ \emph {et~al.}(2022{\natexlab{a}})\citenamefont {Li}, \citenamefont {Wang}, \citenamefont {Xue}, \citenamefont {Wang}, \citenamefont {Zhang}, \citenamefont {Liu}, \citenamefont {Zhu}, \citenamefont {Watanabe}, \citenamefont {Taniguchi}, \citenamefont {Gao}, \citenamefont {Jiang},\ and\ \citenamefont {Mao}}]{li_imaging_2022}%
  \BibitemOpen
  \bibfield  {author} {\bibinfo {author} {\bibfnamefont {S.-y.}\ \bibnamefont {Li}}, \bibinfo {author} {\bibfnamefont {Z.}~\bibnamefont {Wang}}, \bibinfo {author} {\bibfnamefont {Y.}~\bibnamefont {Xue}}, \bibinfo {author} {\bibfnamefont {Y.}~\bibnamefont {Wang}}, \bibinfo {author} {\bibfnamefont {S.}~\bibnamefont {Zhang}}, \bibinfo {author} {\bibfnamefont {J.}~\bibnamefont {Liu}}, \bibinfo {author} {\bibfnamefont {Z.}~\bibnamefont {Zhu}}, \bibinfo {author} {\bibfnamefont {K.}~\bibnamefont {Watanabe}}, \bibinfo {author} {\bibfnamefont {T.}~\bibnamefont {Taniguchi}}, \bibinfo {author} {\bibfnamefont {H.-j.}\ \bibnamefont {Gao}}, \bibinfo {author} {\bibfnamefont {Y.}~\bibnamefont {Jiang}},\ and\ \bibinfo {author} {\bibfnamefont {J.}~\bibnamefont {Mao}},\ }\bibfield  {title} {\bibinfo {title} {Imaging topological and correlated insulating states in twisted monolayer-bilayer graphene},\ }\href {https://doi.org/10.1038/s41467-022-31851-x} {\bibfield  {journal} {\bibinfo  {journal} {Nature Communications}\ }\textbf
  {\bibinfo {volume} {13}},\ \bibinfo {pages} {4225} (\bibinfo {year} {2022}{\natexlab{a}})}\BibitemShut {NoStop}%
\bibitem [{\citenamefont {Polshyn}\ \emph {et~al.}(2020)\citenamefont {Polshyn}, \citenamefont {Zhu}, \citenamefont {Kumar}, \citenamefont {Zhang}, \citenamefont {Yang}, \citenamefont {Tschirhart}, \citenamefont {Serlin}, \citenamefont {Watanabe}, \citenamefont {Taniguchi}, \citenamefont {MacDonald},\ and\ \citenamefont {Young}}]{polshyn_electrical_2020}%
  \BibitemOpen
  \bibfield  {author} {\bibinfo {author} {\bibfnamefont {H.}~\bibnamefont {Polshyn}}, \bibinfo {author} {\bibfnamefont {J.}~\bibnamefont {Zhu}}, \bibinfo {author} {\bibfnamefont {M.~A.}\ \bibnamefont {Kumar}}, \bibinfo {author} {\bibfnamefont {Y.}~\bibnamefont {Zhang}}, \bibinfo {author} {\bibfnamefont {F.}~\bibnamefont {Yang}}, \bibinfo {author} {\bibfnamefont {C.~L.}\ \bibnamefont {Tschirhart}}, \bibinfo {author} {\bibfnamefont {M.}~\bibnamefont {Serlin}}, \bibinfo {author} {\bibfnamefont {K.}~\bibnamefont {Watanabe}}, \bibinfo {author} {\bibfnamefont {T.}~\bibnamefont {Taniguchi}}, \bibinfo {author} {\bibfnamefont {A.~H.}\ \bibnamefont {MacDonald}},\ and\ \bibinfo {author} {\bibfnamefont {A.~F.}\ \bibnamefont {Young}},\ }\bibfield  {title} {\bibinfo {title} {Electrical switching of magnetic order in an orbital {{Chern}} insulator},\ }\href {https://doi.org/10.1038/s41586-020-2963-8} {\bibfield  {journal} {\bibinfo  {journal} {Nature}\ }\textbf {\bibinfo {volume} {588}},\ \bibinfo {pages} {66} (\bibinfo
  {year} {2020})}\BibitemShut {NoStop}%
\bibitem [{\citenamefont {Polshyn}\ \emph {et~al.}(2022)\citenamefont {Polshyn}, \citenamefont {Zhang}, \citenamefont {Kumar}, \citenamefont {Soejima}, \citenamefont {Ledwith}, \citenamefont {Watanabe}, \citenamefont {Taniguchi}, \citenamefont {Vishwanath}, \citenamefont {Zaletel},\ and\ \citenamefont {Young}}]{polshyn_topological_2022}%
  \BibitemOpen
  \bibfield  {author} {\bibinfo {author} {\bibfnamefont {H.}~\bibnamefont {Polshyn}}, \bibinfo {author} {\bibfnamefont {Y.}~\bibnamefont {Zhang}}, \bibinfo {author} {\bibfnamefont {M.~A.}\ \bibnamefont {Kumar}}, \bibinfo {author} {\bibfnamefont {T.}~\bibnamefont {Soejima}}, \bibinfo {author} {\bibfnamefont {P.}~\bibnamefont {Ledwith}}, \bibinfo {author} {\bibfnamefont {K.}~\bibnamefont {Watanabe}}, \bibinfo {author} {\bibfnamefont {T.}~\bibnamefont {Taniguchi}}, \bibinfo {author} {\bibfnamefont {A.}~\bibnamefont {Vishwanath}}, \bibinfo {author} {\bibfnamefont {M.~P.}\ \bibnamefont {Zaletel}},\ and\ \bibinfo {author} {\bibfnamefont {A.~F.}\ \bibnamefont {Young}},\ }\bibfield  {title} {\bibinfo {title} {Topological charge density waves at half-integer filling of a moir{\'e} superlattice},\ }\href {https://doi.org/10.1038/s41567-021-01418-6} {\bibfield  {journal} {\bibinfo  {journal} {Nature Physics}\ }\textbf {\bibinfo {volume} {18}},\ \bibinfo {pages} {42} (\bibinfo {year} {2022})}\BibitemShut {NoStop}%
\bibitem [{\citenamefont {Xu}\ \emph {et~al.}(2021)\citenamefont {Xu}, \citenamefont {Al~Ezzi}, \citenamefont {Balakrishnan}, \citenamefont {{Garcia-Ruiz}}, \citenamefont {Tsim}, \citenamefont {Mullan}, \citenamefont {Barrier}, \citenamefont {Xin}, \citenamefont {Piot}, \citenamefont {Taniguchi}, \citenamefont {Watanabe}, \citenamefont {Carvalho}, \citenamefont {Mishchenko}, \citenamefont {Geim}, \citenamefont {Fal'ko}, \citenamefont {Adam}, \citenamefont {Neto}, \citenamefont {Novoselov},\ and\ \citenamefont {Shi}}]{xu_tunable_2021}%
  \BibitemOpen
  \bibfield  {author} {\bibinfo {author} {\bibfnamefont {S.}~\bibnamefont {Xu}}, \bibinfo {author} {\bibfnamefont {M.~M.}\ \bibnamefont {Al~Ezzi}}, \bibinfo {author} {\bibfnamefont {N.}~\bibnamefont {Balakrishnan}}, \bibinfo {author} {\bibfnamefont {A.}~\bibnamefont {{Garcia-Ruiz}}}, \bibinfo {author} {\bibfnamefont {B.}~\bibnamefont {Tsim}}, \bibinfo {author} {\bibfnamefont {C.}~\bibnamefont {Mullan}}, \bibinfo {author} {\bibfnamefont {J.}~\bibnamefont {Barrier}}, \bibinfo {author} {\bibfnamefont {N.}~\bibnamefont {Xin}}, \bibinfo {author} {\bibfnamefont {B.~A.}\ \bibnamefont {Piot}}, \bibinfo {author} {\bibfnamefont {T.}~\bibnamefont {Taniguchi}}, \bibinfo {author} {\bibfnamefont {K.}~\bibnamefont {Watanabe}}, \bibinfo {author} {\bibfnamefont {A.}~\bibnamefont {Carvalho}}, \bibinfo {author} {\bibfnamefont {A.}~\bibnamefont {Mishchenko}}, \bibinfo {author} {\bibfnamefont {A.~K.}\ \bibnamefont {Geim}}, \bibinfo {author} {\bibfnamefont {V.~I.}\ \bibnamefont {Fal'ko}}, \bibinfo {author} {\bibfnamefont
  {S.}~\bibnamefont {Adam}}, \bibinfo {author} {\bibfnamefont {A.~H.~C.}\ \bibnamefont {Neto}}, \bibinfo {author} {\bibfnamefont {K.~S.}\ \bibnamefont {Novoselov}},\ and\ \bibinfo {author} {\bibfnamefont {Y.}~\bibnamefont {Shi}},\ }\bibfield  {title} {\bibinfo {title} {Tunable van {{Hove}} singularities and correlated states in twisted monolayer--bilayer graphene},\ }\href {https://doi.org/10.1038/s41567-021-01172-9} {\bibfield  {journal} {\bibinfo  {journal} {Nature Physics}\ }\textbf {\bibinfo {volume} {17}},\ \bibinfo {pages} {619} (\bibinfo {year} {2021})}\BibitemShut {NoStop}%
\bibitem [{\citenamefont {Zhang}\ \emph {et~al.}(2023)\citenamefont {Zhang}, \citenamefont {Zhu}, \citenamefont {Soejima}, \citenamefont {Kahn}, \citenamefont {Watanabe}, \citenamefont {Taniguchi}, \citenamefont {Zettl}, \citenamefont {Wang}, \citenamefont {Zaletel},\ and\ \citenamefont {Crommie}}]{zhang_local_2023}%
  \BibitemOpen
  \bibfield  {author} {\bibinfo {author} {\bibfnamefont {C.}~\bibnamefont {Zhang}}, \bibinfo {author} {\bibfnamefont {T.}~\bibnamefont {Zhu}}, \bibinfo {author} {\bibfnamefont {T.}~\bibnamefont {Soejima}}, \bibinfo {author} {\bibfnamefont {S.}~\bibnamefont {Kahn}}, \bibinfo {author} {\bibfnamefont {K.}~\bibnamefont {Watanabe}}, \bibinfo {author} {\bibfnamefont {T.}~\bibnamefont {Taniguchi}}, \bibinfo {author} {\bibfnamefont {A.}~\bibnamefont {Zettl}}, \bibinfo {author} {\bibfnamefont {F.}~\bibnamefont {Wang}}, \bibinfo {author} {\bibfnamefont {M.~P.}\ \bibnamefont {Zaletel}},\ and\ \bibinfo {author} {\bibfnamefont {M.~F.}\ \bibnamefont {Crommie}},\ }\bibfield  {title} {\bibinfo {title} {Local spectroscopy of a gate-switchable moir{\'e} quantum anomalous {{Hall}} insulator},\ }\href {https://doi.org/10.1038/s41467-023-39110-3} {\bibfield  {journal} {\bibinfo  {journal} {Nature Communications}\ }\textbf {\bibinfo {volume} {14}},\ \bibinfo {pages} {3595} (\bibinfo {year} {2023})}\BibitemShut {NoStop}%
\bibitem [{\citenamefont {Ledwith}\ \emph {et~al.}(2022)\citenamefont {Ledwith}, \citenamefont {Vishwanath},\ and\ \citenamefont {Khalaf}}]{ledwith_family_2022}%
  \BibitemOpen
  \bibfield  {author} {\bibinfo {author} {\bibfnamefont {P.~J.}\ \bibnamefont {Ledwith}}, \bibinfo {author} {\bibfnamefont {A.}~\bibnamefont {Vishwanath}},\ and\ \bibinfo {author} {\bibfnamefont {E.}~\bibnamefont {Khalaf}},\ }\bibfield  {title} {\bibinfo {title} {Family of {{Ideal Chern Flatbands}} with {{Arbitrary Chern Number}} in {{Chiral Twisted Graphene Multilayers}}},\ }\href {https://doi.org/10.1103/PhysRevLett.128.176404} {\bibfield  {journal} {\bibinfo  {journal} {Physical Review Letters}\ }\textbf {\bibinfo {volume} {128}},\ \bibinfo {pages} {176404} (\bibinfo {year} {2022})}\BibitemShut {NoStop}%
\bibitem [{\citenamefont {Xia}\ \emph {et~al.}(2023)\citenamefont {Xia}, \citenamefont {de~la Barrera}, \citenamefont {Uri}, \citenamefont {Sharpe}, \citenamefont {Kwan}, \citenamefont {Zhu}, \citenamefont {Watanabe}, \citenamefont {Taniguchi}, \citenamefont {Goldhaber-Gordon}, \citenamefont {Fu}, \citenamefont {Devakul},\ and\ \citenamefont {Jarillo-Herrero}}]{xia2023helical}%
  \BibitemOpen
  \bibfield  {author} {\bibinfo {author} {\bibfnamefont {L.-Q.}\ \bibnamefont {Xia}}, \bibinfo {author} {\bibfnamefont {S.~C.}\ \bibnamefont {de~la Barrera}}, \bibinfo {author} {\bibfnamefont {A.}~\bibnamefont {Uri}}, \bibinfo {author} {\bibfnamefont {A.}~\bibnamefont {Sharpe}}, \bibinfo {author} {\bibfnamefont {Y.~H.}\ \bibnamefont {Kwan}}, \bibinfo {author} {\bibfnamefont {Z.}~\bibnamefont {Zhu}}, \bibinfo {author} {\bibfnamefont {K.}~\bibnamefont {Watanabe}}, \bibinfo {author} {\bibfnamefont {T.}~\bibnamefont {Taniguchi}}, \bibinfo {author} {\bibfnamefont {D.}~\bibnamefont {Goldhaber-Gordon}}, \bibinfo {author} {\bibfnamefont {L.}~\bibnamefont {Fu}}, \bibinfo {author} {\bibfnamefont {T.}~\bibnamefont {Devakul}},\ and\ \bibinfo {author} {\bibfnamefont {P.}~\bibnamefont {Jarillo-Herrero}},\ }\href@noop {} {\bibinfo {title} {Helical trilayer graphene: a moir\'e platform for strongly-interacting topological bands}} (\bibinfo {year} {2023}),\ \Eprint {https://arxiv.org/abs/2310.12204} {arXiv:2310.12204
  [cond-mat.mes-hall]} \BibitemShut {NoStop}%
\bibitem [{\citenamefont {Devakul}\ \emph {et~al.}(2023)\citenamefont {Devakul}, \citenamefont {Ledwith}, \citenamefont {Xia}, \citenamefont {Uri}, \citenamefont {de~la Barrera}, \citenamefont {Jarillo-Herrero},\ and\ \citenamefont {Fu}}]{devakul2023HTG}%
  \BibitemOpen
  \bibfield  {author} {\bibinfo {author} {\bibfnamefont {T.}~\bibnamefont {Devakul}}, \bibinfo {author} {\bibfnamefont {P.~J.}\ \bibnamefont {Ledwith}}, \bibinfo {author} {\bibfnamefont {L.-Q.}\ \bibnamefont {Xia}}, \bibinfo {author} {\bibfnamefont {A.}~\bibnamefont {Uri}}, \bibinfo {author} {\bibfnamefont {S.~C.}\ \bibnamefont {de~la Barrera}}, \bibinfo {author} {\bibfnamefont {P.}~\bibnamefont {Jarillo-Herrero}},\ and\ \bibinfo {author} {\bibfnamefont {L.}~\bibnamefont {Fu}},\ }\bibfield  {title} {\bibinfo {title} {Magic-angle helical trilayer graphene},\ }\href {https://doi.org/10.1126/sciadv.adi6063} {\bibfield  {journal} {\bibinfo  {journal} {Science Advances}\ }\textbf {\bibinfo {volume} {9}},\ \bibinfo {pages} {eadi6063} (\bibinfo {year} {2023})},\ \Eprint {https://arxiv.org/abs/https://www.science.org/doi/pdf/10.1126/sciadv.adi6063} {https://www.science.org/doi/pdf/10.1126/sciadv.adi6063} \BibitemShut {NoStop}%
\bibitem [{\citenamefont {Nakatsuji}\ \emph {et~al.}(2023)\citenamefont {Nakatsuji}, \citenamefont {Kawakami},\ and\ \citenamefont {Koshino}}]{nakatsuji2023multiscale}%
  \BibitemOpen
  \bibfield  {author} {\bibinfo {author} {\bibfnamefont {N.}~\bibnamefont {Nakatsuji}}, \bibinfo {author} {\bibfnamefont {T.}~\bibnamefont {Kawakami}},\ and\ \bibinfo {author} {\bibfnamefont {M.}~\bibnamefont {Koshino}},\ }\bibfield  {title} {\bibinfo {title} {Multiscale lattice relaxation in general twisted trilayer graphenes},\ }\href {https://doi.org/10.1103/PhysRevX.13.041007} {\bibfield  {journal} {\bibinfo  {journal} {Phys. Rev. X}\ }\textbf {\bibinfo {volume} {13}},\ \bibinfo {pages} {041007} (\bibinfo {year} {2023})}\BibitemShut {NoStop}%
\bibitem [{\citenamefont {Kwan}\ \emph {et~al.}(2024)\citenamefont {Kwan}, \citenamefont {Ledwith}, \citenamefont {Lo},\ and\ \citenamefont {Devakul}}]{kwan2024strong}%
  \BibitemOpen
  \bibfield  {author} {\bibinfo {author} {\bibfnamefont {Y.~H.}\ \bibnamefont {Kwan}}, \bibinfo {author} {\bibfnamefont {P.~J.}\ \bibnamefont {Ledwith}}, \bibinfo {author} {\bibfnamefont {C.~F.~B.}\ \bibnamefont {Lo}},\ and\ \bibinfo {author} {\bibfnamefont {T.}~\bibnamefont {Devakul}},\ }\bibfield  {title} {\bibinfo {title} {Strong-coupling topological states and phase transitions in helical trilayer graphene},\ }\href {https://doi.org/10.1103/PhysRevB.109.125141} {\bibfield  {journal} {\bibinfo  {journal} {Phys. Rev. B}\ }\textbf {\bibinfo {volume} {109}},\ \bibinfo {pages} {125141} (\bibinfo {year} {2024})}\BibitemShut {NoStop}%
\bibitem [{\citenamefont {Yu}\ \emph {et~al.}(2024)\citenamefont {Yu}, \citenamefont {Herzog-Arbeitman}, \citenamefont {Wang}, \citenamefont {Vafek}, \citenamefont {Bernevig},\ and\ \citenamefont {Regnault}}]{Yu2024FCIversus}%
  \BibitemOpen
  \bibfield  {author} {\bibinfo {author} {\bibfnamefont {J.}~\bibnamefont {Yu}}, \bibinfo {author} {\bibfnamefont {J.}~\bibnamefont {Herzog-Arbeitman}}, \bibinfo {author} {\bibfnamefont {M.}~\bibnamefont {Wang}}, \bibinfo {author} {\bibfnamefont {O.}~\bibnamefont {Vafek}}, \bibinfo {author} {\bibfnamefont {B.~A.}\ \bibnamefont {Bernevig}},\ and\ \bibinfo {author} {\bibfnamefont {N.}~\bibnamefont {Regnault}},\ }\bibfield  {title} {\bibinfo {title} {Fractional chern insulators versus nonmagnetic states in twisted bilayer ${\mathrm{mote}}_{2}$},\ }\href {https://doi.org/10.1103/PhysRevB.109.045147} {\bibfield  {journal} {\bibinfo  {journal} {Phys. Rev. B}\ }\textbf {\bibinfo {volume} {109}},\ \bibinfo {pages} {045147} (\bibinfo {year} {2024})}\BibitemShut {NoStop}%
\bibitem [{\citenamefont {Jia}\ \emph {et~al.}(2024)\citenamefont {Jia}, \citenamefont {Yu}, \citenamefont {Liu}, \citenamefont {Herzog-Arbeitman}, \citenamefont {Qi}, \citenamefont {Pi}, \citenamefont {Regnault}, \citenamefont {Weng}, \citenamefont {Bernevig},\ and\ \citenamefont {Wu}}]{Jia2024MFCI1}%
  \BibitemOpen
  \bibfield  {author} {\bibinfo {author} {\bibfnamefont {Y.}~\bibnamefont {Jia}}, \bibinfo {author} {\bibfnamefont {J.}~\bibnamefont {Yu}}, \bibinfo {author} {\bibfnamefont {J.}~\bibnamefont {Liu}}, \bibinfo {author} {\bibfnamefont {J.}~\bibnamefont {Herzog-Arbeitman}}, \bibinfo {author} {\bibfnamefont {Z.}~\bibnamefont {Qi}}, \bibinfo {author} {\bibfnamefont {H.}~\bibnamefont {Pi}}, \bibinfo {author} {\bibfnamefont {N.}~\bibnamefont {Regnault}}, \bibinfo {author} {\bibfnamefont {H.}~\bibnamefont {Weng}}, \bibinfo {author} {\bibfnamefont {B.~A.}\ \bibnamefont {Bernevig}},\ and\ \bibinfo {author} {\bibfnamefont {Q.}~\bibnamefont {Wu}},\ }\bibfield  {title} {\bibinfo {title} {Moir\'e fractional chern insulators. i. first-principles calculations and continuum models of twisted bilayer ${\mathrm{mote}}_{2}$},\ }\href {https://doi.org/10.1103/PhysRevB.109.205121} {\bibfield  {journal} {\bibinfo  {journal} {Phys. Rev. B}\ }\textbf {\bibinfo {volume} {109}},\ \bibinfo {pages} {205121} (\bibinfo {year}
  {2024})}\BibitemShut {NoStop}%
\bibitem [{\citenamefont {Wu}\ \emph {et~al.}(2019{\natexlab{a}})\citenamefont {Wu}, \citenamefont {Lovorn}, \citenamefont {Tutuc}, \citenamefont {Martin},\ and\ \citenamefont {MacDonald}}]{Wu2019tmds}%
  \BibitemOpen
  \bibfield  {author} {\bibinfo {author} {\bibfnamefont {F.}~\bibnamefont {Wu}}, \bibinfo {author} {\bibfnamefont {T.}~\bibnamefont {Lovorn}}, \bibinfo {author} {\bibfnamefont {E.}~\bibnamefont {Tutuc}}, \bibinfo {author} {\bibfnamefont {I.}~\bibnamefont {Martin}},\ and\ \bibinfo {author} {\bibfnamefont {A.~H.}\ \bibnamefont {MacDonald}},\ }\bibfield  {title} {\bibinfo {title} {Topological insulators in twisted transition metal dichalcogenide homobilayers},\ }\href {https://doi.org/10.1103/PhysRevLett.122.086402} {\bibfield  {journal} {\bibinfo  {journal} {Phys. Rev. Lett.}\ }\textbf {\bibinfo {volume} {122}},\ \bibinfo {pages} {086402} (\bibinfo {year} {2019}{\natexlab{a}})}\BibitemShut {NoStop}%
\bibitem [{\citenamefont {Wang}\ \emph {et~al.}(2024{\natexlab{b}})\citenamefont {Wang}, \citenamefont {Zhang}, \citenamefont {Liu}, \citenamefont {He}, \citenamefont {Xu}, \citenamefont {Ran}, \citenamefont {Cao},\ and\ \citenamefont {Xiao}}]{Wang2024FCIMoTe2}%
  \BibitemOpen
  \bibfield  {author} {\bibinfo {author} {\bibfnamefont {C.}~\bibnamefont {Wang}}, \bibinfo {author} {\bibfnamefont {X.-W.}\ \bibnamefont {Zhang}}, \bibinfo {author} {\bibfnamefont {X.}~\bibnamefont {Liu}}, \bibinfo {author} {\bibfnamefont {Y.}~\bibnamefont {He}}, \bibinfo {author} {\bibfnamefont {X.}~\bibnamefont {Xu}}, \bibinfo {author} {\bibfnamefont {Y.}~\bibnamefont {Ran}}, \bibinfo {author} {\bibfnamefont {T.}~\bibnamefont {Cao}},\ and\ \bibinfo {author} {\bibfnamefont {D.}~\bibnamefont {Xiao}},\ }\bibfield  {title} {\bibinfo {title} {Fractional chern insulator in twisted bilayer ${\mathrm{mote}}_{2}$},\ }\href {https://doi.org/10.1103/PhysRevLett.132.036501} {\bibfield  {journal} {\bibinfo  {journal} {Phys. Rev. Lett.}\ }\textbf {\bibinfo {volume} {132}},\ \bibinfo {pages} {036501} (\bibinfo {year} {2024}{\natexlab{b}})}\BibitemShut {NoStop}%
\bibitem [{\citenamefont {Reddy}\ \emph {et~al.}(2023)\citenamefont {Reddy}, \citenamefont {Alsallom}, \citenamefont {Zhang}, \citenamefont {Devakul},\ and\ \citenamefont {Fu}}]{Reddy2023FQAHMoTe2}%
  \BibitemOpen
  \bibfield  {author} {\bibinfo {author} {\bibfnamefont {A.~P.}\ \bibnamefont {Reddy}}, \bibinfo {author} {\bibfnamefont {F.}~\bibnamefont {Alsallom}}, \bibinfo {author} {\bibfnamefont {Y.}~\bibnamefont {Zhang}}, \bibinfo {author} {\bibfnamefont {T.}~\bibnamefont {Devakul}},\ and\ \bibinfo {author} {\bibfnamefont {L.}~\bibnamefont {Fu}},\ }\bibfield  {title} {\bibinfo {title} {Fractional quantum anomalous hall states in twisted bilayer ${\mathrm{mote}}_{2}$ and ${\mathrm{wse}}_{2}$},\ }\href {https://doi.org/10.1103/PhysRevB.108.085117} {\bibfield  {journal} {\bibinfo  {journal} {Phys. Rev. B}\ }\textbf {\bibinfo {volume} {108}},\ \bibinfo {pages} {085117} (\bibinfo {year} {2023})}\BibitemShut {NoStop}%
\bibitem [{\citenamefont {Cao}\ \emph {et~al.}(2021)\citenamefont {Cao}, \citenamefont {Park}, \citenamefont {Watanabe}, \citenamefont {Taniguchi},\ and\ \citenamefont {Jarillo-Herrero}}]{Cao2021TTGPauli}%
  \BibitemOpen
  \bibfield  {author} {\bibinfo {author} {\bibfnamefont {Y.}~\bibnamefont {Cao}}, \bibinfo {author} {\bibfnamefont {J.~M.}\ \bibnamefont {Park}}, \bibinfo {author} {\bibfnamefont {K.}~\bibnamefont {Watanabe}}, \bibinfo {author} {\bibfnamefont {T.}~\bibnamefont {Taniguchi}},\ and\ \bibinfo {author} {\bibfnamefont {P.}~\bibnamefont {Jarillo-Herrero}},\ }\bibfield  {title} {\bibinfo {title} {Pauli-limit violation and re-entrant superconductivity in moir{\'{e}} graphene},\ }\href {https://doi.org/10.1038/s41586-021-03685-y} {\bibfield  {journal} {\bibinfo  {journal} {Nature}\ }\textbf {\bibinfo {volume} {595}},\ \bibinfo {pages} {526} (\bibinfo {year} {2021})}\BibitemShut {NoStop}%
\bibitem [{\citenamefont {Hao}\ \emph {et~al.}(2021)\citenamefont {Hao}, \citenamefont {Zimmerman}, \citenamefont {Ledwith}, \citenamefont {Khalaf}, \citenamefont {Najafabadi}, \citenamefont {Watanabe}, \citenamefont {Taniguchi}, \citenamefont {Vishwanath},\ and\ \citenamefont {Kim}}]{hao_electric_2021}%
  \BibitemOpen
  \bibfield  {author} {\bibinfo {author} {\bibfnamefont {Z.}~\bibnamefont {Hao}}, \bibinfo {author} {\bibfnamefont {A.~M.}\ \bibnamefont {Zimmerman}}, \bibinfo {author} {\bibfnamefont {P.}~\bibnamefont {Ledwith}}, \bibinfo {author} {\bibfnamefont {E.}~\bibnamefont {Khalaf}}, \bibinfo {author} {\bibfnamefont {D.~H.}\ \bibnamefont {Najafabadi}}, \bibinfo {author} {\bibfnamefont {K.}~\bibnamefont {Watanabe}}, \bibinfo {author} {\bibfnamefont {T.}~\bibnamefont {Taniguchi}}, \bibinfo {author} {\bibfnamefont {A.}~\bibnamefont {Vishwanath}},\ and\ \bibinfo {author} {\bibfnamefont {P.}~\bibnamefont {Kim}},\ }\bibfield  {title} {\bibinfo {title} {Electric field--tunable superconductivity in alternating-twist magic-angle trilayer graphene},\ }\href {https://doi.org/10.1126/science.abg0399} {\bibfield  {journal} {\bibinfo  {journal} {Science}\ }\textbf {\bibinfo {volume} {371}},\ \bibinfo {pages} {1133} (\bibinfo {year} {2021})}\BibitemShut {NoStop}%
\bibitem [{\citenamefont {Kim}\ \emph {et~al.}(2022)\citenamefont {Kim}, \citenamefont {Choi}, \citenamefont {Lewandowski}, \citenamefont {Thomson}, \citenamefont {Zhang}, \citenamefont {Polski}, \citenamefont {Watanabe}, \citenamefont {Taniguchi}, \citenamefont {Alicea},\ and\ \citenamefont {{Nadj-Perge}}}]{kim_evidence_2022}%
  \BibitemOpen
  \bibfield  {author} {\bibinfo {author} {\bibfnamefont {H.}~\bibnamefont {Kim}}, \bibinfo {author} {\bibfnamefont {Y.}~\bibnamefont {Choi}}, \bibinfo {author} {\bibfnamefont {C.}~\bibnamefont {Lewandowski}}, \bibinfo {author} {\bibfnamefont {A.}~\bibnamefont {Thomson}}, \bibinfo {author} {\bibfnamefont {Y.}~\bibnamefont {Zhang}}, \bibinfo {author} {\bibfnamefont {R.}~\bibnamefont {Polski}}, \bibinfo {author} {\bibfnamefont {K.}~\bibnamefont {Watanabe}}, \bibinfo {author} {\bibfnamefont {T.}~\bibnamefont {Taniguchi}}, \bibinfo {author} {\bibfnamefont {J.}~\bibnamefont {Alicea}},\ and\ \bibinfo {author} {\bibfnamefont {S.}~\bibnamefont {{Nadj-Perge}}},\ }\bibfield  {title} {\bibinfo {title} {Evidence for unconventional superconductivity in twisted trilayer graphene},\ }\href {https://doi.org/10.1038/s41586-022-04715-z} {\bibfield  {journal} {\bibinfo  {journal} {Nature}\ }\textbf {\bibinfo {volume} {606}},\ \bibinfo {pages} {494} (\bibinfo {year} {2022})}\BibitemShut {NoStop}%
\bibitem [{\citenamefont {Li}\ \emph {et~al.}(2022{\natexlab{b}})\citenamefont {Li}, \citenamefont {Zhang}, \citenamefont {Chen}, \citenamefont {Wei}, \citenamefont {Zhang}, \citenamefont {Xiao}, \citenamefont {Gao}, \citenamefont {Chen}, \citenamefont {Liang}, \citenamefont {Pei}, \citenamefont {Xu}, \citenamefont {Watanabe}, \citenamefont {Taniguchi}, \citenamefont {Yang}, \citenamefont {Miao}, \citenamefont {Liu}, \citenamefont {Cheng}, \citenamefont {Wang}, \citenamefont {Chen},\ and\ \citenamefont {Liu}}]{li_observation_2022}%
  \BibitemOpen
  \bibfield  {author} {\bibinfo {author} {\bibfnamefont {Y.}~\bibnamefont {Li}}, \bibinfo {author} {\bibfnamefont {S.}~\bibnamefont {Zhang}}, \bibinfo {author} {\bibfnamefont {F.}~\bibnamefont {Chen}}, \bibinfo {author} {\bibfnamefont {L.}~\bibnamefont {Wei}}, \bibinfo {author} {\bibfnamefont {Z.}~\bibnamefont {Zhang}}, \bibinfo {author} {\bibfnamefont {H.}~\bibnamefont {Xiao}}, \bibinfo {author} {\bibfnamefont {H.}~\bibnamefont {Gao}}, \bibinfo {author} {\bibfnamefont {M.}~\bibnamefont {Chen}}, \bibinfo {author} {\bibfnamefont {S.}~\bibnamefont {Liang}}, \bibinfo {author} {\bibfnamefont {D.}~\bibnamefont {Pei}}, \bibinfo {author} {\bibfnamefont {L.}~\bibnamefont {Xu}}, \bibinfo {author} {\bibfnamefont {K.}~\bibnamefont {Watanabe}}, \bibinfo {author} {\bibfnamefont {T.}~\bibnamefont {Taniguchi}}, \bibinfo {author} {\bibfnamefont {L.}~\bibnamefont {Yang}}, \bibinfo {author} {\bibfnamefont {F.}~\bibnamefont {Miao}}, \bibinfo {author} {\bibfnamefont {J.}~\bibnamefont {Liu}}, \bibinfo {author} {\bibfnamefont
  {B.}~\bibnamefont {Cheng}}, \bibinfo {author} {\bibfnamefont {M.}~\bibnamefont {Wang}}, \bibinfo {author} {\bibfnamefont {Y.}~\bibnamefont {Chen}},\ and\ \bibinfo {author} {\bibfnamefont {Z.}~\bibnamefont {Liu}},\ }\bibfield  {title} {\bibinfo {title} {Observation of {{Coexisting Dirac Bands}} and {{Moir{\'e} Flat Bands}} in {{Magic-Angle Twisted Trilayer Graphene}}},\ }\href {https://doi.org/10.1002/adma.202205996} {\bibfield  {journal} {\bibinfo  {journal} {Advanced Materials}\ }\textbf {\bibinfo {volume} {34}},\ \bibinfo {pages} {2205996} (\bibinfo {year} {2022}{\natexlab{b}})}\BibitemShut {NoStop}%
\bibitem [{\citenamefont {Lin}\ \emph {et~al.}(2022)\citenamefont {Lin}, \citenamefont {Siriviboon}, \citenamefont {Scammell}, \citenamefont {Liu}, \citenamefont {Rhodes}, \citenamefont {Watanabe}, \citenamefont {Taniguchi}, \citenamefont {Hone}, \citenamefont {Scheurer},\ and\ \citenamefont {Li}}]{lin_zero-field_2022}%
  \BibitemOpen
  \bibfield  {author} {\bibinfo {author} {\bibfnamefont {J.-X.}\ \bibnamefont {Lin}}, \bibinfo {author} {\bibfnamefont {P.}~\bibnamefont {Siriviboon}}, \bibinfo {author} {\bibfnamefont {H.~D.}\ \bibnamefont {Scammell}}, \bibinfo {author} {\bibfnamefont {S.}~\bibnamefont {Liu}}, \bibinfo {author} {\bibfnamefont {D.}~\bibnamefont {Rhodes}}, \bibinfo {author} {\bibfnamefont {K.}~\bibnamefont {Watanabe}}, \bibinfo {author} {\bibfnamefont {T.}~\bibnamefont {Taniguchi}}, \bibinfo {author} {\bibfnamefont {J.}~\bibnamefont {Hone}}, \bibinfo {author} {\bibfnamefont {M.~S.}\ \bibnamefont {Scheurer}},\ and\ \bibinfo {author} {\bibfnamefont {J.~I.~A.}\ \bibnamefont {Li}},\ }\bibfield  {title} {\bibinfo {title} {Zero-field superconducting diode effect in small-twist-angle trilayer graphene},\ }\href {https://doi.org/10.1038/s41567-022-01700-1} {\bibfield  {journal} {\bibinfo  {journal} {Nature Physics}\ }\textbf {\bibinfo {volume} {18}},\ \bibinfo {pages} {1221} (\bibinfo {year} {2022})}\BibitemShut {NoStop}%
\bibitem [{\citenamefont {Liu}\ \emph {et~al.}(2022{\natexlab{b}})\citenamefont {Liu}, \citenamefont {Zhang}, \citenamefont {Watanabe}, \citenamefont {Taniguchi},\ and\ \citenamefont {Li}}]{Liu2022TTGisospin}%
  \BibitemOpen
  \bibfield  {author} {\bibinfo {author} {\bibfnamefont {X.}~\bibnamefont {Liu}}, \bibinfo {author} {\bibfnamefont {N.~J.}\ \bibnamefont {Zhang}}, \bibinfo {author} {\bibfnamefont {K.}~\bibnamefont {Watanabe}}, \bibinfo {author} {\bibfnamefont {T.}~\bibnamefont {Taniguchi}},\ and\ \bibinfo {author} {\bibfnamefont {J.~I.~A.}\ \bibnamefont {Li}},\ }\bibfield  {title} {\bibinfo {title} {Isospin order in superconducting magic-angle twisted trilayer graphene},\ }\href {https://doi.org/10.1038/s41567-022-01515-0} {\bibfield  {journal} {\bibinfo  {journal} {Nature Physics}\ }\textbf {\bibinfo {volume} {18}},\ \bibinfo {pages} {522} (\bibinfo {year} {2022}{\natexlab{b}})}\BibitemShut {NoStop}%
\bibitem [{\citenamefont {Shen}\ \emph {et~al.}(2022)\citenamefont {Shen}, \citenamefont {Ledwith}, \citenamefont {Watanabe}, \citenamefont {Taniguchi}, \citenamefont {Khalaf}, \citenamefont {Vishwanath},\ and\ \citenamefont {Efetov}}]{Shen2022TTGDirac}%
  \BibitemOpen
  \bibfield  {author} {\bibinfo {author} {\bibfnamefont {C.}~\bibnamefont {Shen}}, \bibinfo {author} {\bibfnamefont {P.~J.}\ \bibnamefont {Ledwith}}, \bibinfo {author} {\bibfnamefont {K.}~\bibnamefont {Watanabe}}, \bibinfo {author} {\bibfnamefont {T.}~\bibnamefont {Taniguchi}}, \bibinfo {author} {\bibfnamefont {E.}~\bibnamefont {Khalaf}}, \bibinfo {author} {\bibfnamefont {A.}~\bibnamefont {Vishwanath}},\ and\ \bibinfo {author} {\bibfnamefont {D.~K.}\ \bibnamefont {Efetov}},\ }\bibfield  {title} {\bibinfo {title} {Dirac spectroscopy of strongly correlated phases in twisted trilayer graphene},\ }\href {https://doi.org/10.1038/s41563-022-01428-6} {\bibfield  {journal} {\bibinfo  {journal} {Nature Materials}\ }\textbf {\bibinfo {volume} {22}},\ \bibinfo {pages} {316} (\bibinfo {year} {2022})}\BibitemShut {NoStop}%
\bibitem [{\citenamefont {Turkel}\ \emph {et~al.}(2022)\citenamefont {Turkel}, \citenamefont {Swann}, \citenamefont {Zhu}, \citenamefont {Christos}, \citenamefont {Watanabe}, \citenamefont {Taniguchi}, \citenamefont {Sachdev}, \citenamefont {Scheurer}, \citenamefont {Kaxiras}, \citenamefont {Dean},\ and\ \citenamefont {Pasupathy}}]{Twistons}%
  \BibitemOpen
  \bibfield  {author} {\bibinfo {author} {\bibfnamefont {S.}~\bibnamefont {Turkel}}, \bibinfo {author} {\bibfnamefont {J.}~\bibnamefont {Swann}}, \bibinfo {author} {\bibfnamefont {Z.}~\bibnamefont {Zhu}}, \bibinfo {author} {\bibfnamefont {M.}~\bibnamefont {Christos}}, \bibinfo {author} {\bibfnamefont {K.}~\bibnamefont {Watanabe}}, \bibinfo {author} {\bibfnamefont {T.}~\bibnamefont {Taniguchi}}, \bibinfo {author} {\bibfnamefont {S.}~\bibnamefont {Sachdev}}, \bibinfo {author} {\bibfnamefont {M.~S.}\ \bibnamefont {Scheurer}}, \bibinfo {author} {\bibfnamefont {E.}~\bibnamefont {Kaxiras}}, \bibinfo {author} {\bibfnamefont {C.~R.}\ \bibnamefont {Dean}},\ and\ \bibinfo {author} {\bibfnamefont {A.~N.}\ \bibnamefont {Pasupathy}},\ }\bibfield  {title} {\bibinfo {title} {Orderly disorder in magic-angle twisted trilayer graphene},\ }\href {https://doi.org/10.1126/science.abk1895} {\bibfield  {journal} {\bibinfo  {journal} {Science}\ }\textbf {\bibinfo {volume} {376}},\ \bibinfo {pages} {193} (\bibinfo {year} {2022})},\
  \Eprint {https://arxiv.org/abs/https://www.science.org/doi/pdf/10.1126/science.abk1895} {https://www.science.org/doi/pdf/10.1126/science.abk1895} \BibitemShut {NoStop}%
\bibitem [{\citenamefont {Yang}\ \emph {et~al.}(2022)\citenamefont {Yang}, \citenamefont {Jung}, \citenamefont {Lee}, \citenamefont {Han}, \citenamefont {Choi}, \citenamefont {Jung}, \citenamefont {Choi}, \citenamefont {Park}, \citenamefont {Oh}, \citenamefont {Noh}, \citenamefont {Kim}, \citenamefont {Huang}, \citenamefont {Hwang},\ and\ \citenamefont {Kim}}]{yang_wafer-scale_2022}%
  \BibitemOpen
  \bibfield  {author} {\bibinfo {author} {\bibfnamefont {S.-J.}\ \bibnamefont {Yang}}, \bibinfo {author} {\bibfnamefont {J.-H.}\ \bibnamefont {Jung}}, \bibinfo {author} {\bibfnamefont {E.}~\bibnamefont {Lee}}, \bibinfo {author} {\bibfnamefont {E.}~\bibnamefont {Han}}, \bibinfo {author} {\bibfnamefont {M.-Y.}\ \bibnamefont {Choi}}, \bibinfo {author} {\bibfnamefont {D.}~\bibnamefont {Jung}}, \bibinfo {author} {\bibfnamefont {S.}~\bibnamefont {Choi}}, \bibinfo {author} {\bibfnamefont {J.-H.}\ \bibnamefont {Park}}, \bibinfo {author} {\bibfnamefont {D.}~\bibnamefont {Oh}}, \bibinfo {author} {\bibfnamefont {S.}~\bibnamefont {Noh}}, \bibinfo {author} {\bibfnamefont {K.-J.}\ \bibnamefont {Kim}}, \bibinfo {author} {\bibfnamefont {P.~Y.}\ \bibnamefont {Huang}}, \bibinfo {author} {\bibfnamefont {C.-C.}\ \bibnamefont {Hwang}},\ and\ \bibinfo {author} {\bibfnamefont {C.-J.}\ \bibnamefont {Kim}},\ }\bibfield  {title} {\bibinfo {title} {Wafer-{{Scale Programmed Assembly}} of {{One-Atom-Thick Crystals}}},\ }\href
  {https://doi.org/10.1021/acs.nanolett.1c04139} {\bibfield  {journal} {\bibinfo  {journal} {Nano Letters}\ }\textbf {\bibinfo {volume} {22}},\ \bibinfo {pages} {1518} (\bibinfo {year} {2022})}\BibitemShut {NoStop}%
\bibitem [{\citenamefont {Zhang}\ \emph {et~al.}(2024{\natexlab{a}})\citenamefont {Zhang}, \citenamefont {Lin}, \citenamefont {Chichinadze}, \citenamefont {Wang}, \citenamefont {Watanabe}, \citenamefont {Taniguchi}, \citenamefont {Fu},\ and\ \citenamefont {Li}}]{zhang_angle-resolved_2024}%
  \BibitemOpen
  \bibfield  {author} {\bibinfo {author} {\bibfnamefont {N.~J.}\ \bibnamefont {Zhang}}, \bibinfo {author} {\bibfnamefont {J.-X.}\ \bibnamefont {Lin}}, \bibinfo {author} {\bibfnamefont {D.~V.}\ \bibnamefont {Chichinadze}}, \bibinfo {author} {\bibfnamefont {Y.}~\bibnamefont {Wang}}, \bibinfo {author} {\bibfnamefont {K.}~\bibnamefont {Watanabe}}, \bibinfo {author} {\bibfnamefont {T.}~\bibnamefont {Taniguchi}}, \bibinfo {author} {\bibfnamefont {L.}~\bibnamefont {Fu}},\ and\ \bibinfo {author} {\bibfnamefont {J.~I.~A.}\ \bibnamefont {Li}},\ }\bibfield  {title} {\bibinfo {title} {Angle-resolved transport non-reciprocity and spontaneous symmetry breaking in twisted trilayer graphene},\ }\href {https://doi.org/10.1038/s41563-024-01809-z} {\bibfield  {journal} {\bibinfo  {journal} {Nature Materials}\ }\textbf {\bibinfo {volume} {23}},\ \bibinfo {pages} {356} (\bibinfo {year} {2024}{\natexlab{a}})}\BibitemShut {NoStop}%
\bibitem [{\citenamefont {Li}\ \emph {et~al.}(2019)\citenamefont {Li}, \citenamefont {Wu},\ and\ \citenamefont {MacDonald}}]{li_electronic_2019}%
  \BibitemOpen
  \bibfield  {author} {\bibinfo {author} {\bibfnamefont {X.}~\bibnamefont {Li}}, \bibinfo {author} {\bibfnamefont {F.}~\bibnamefont {Wu}},\ and\ \bibinfo {author} {\bibfnamefont {A.~H.}\ \bibnamefont {MacDonald}},\ }\href {https://doi.org/10.48550/arXiv.1907.12338} {\bibinfo {title} {Electronic {{Structure}} of {{Single-Twist Trilayer Graphene}}}} (\bibinfo {year} {2019}),\ \Eprint {https://arxiv.org/abs/1907.12338} {arXiv:1907.12338 [cond-mat]} \BibitemShut {NoStop}%
\bibitem [{\citenamefont {Khalaf}\ \emph {et~al.}(2019)\citenamefont {Khalaf}, \citenamefont {Kruchkov}, \citenamefont {Tarnopolsky},\ and\ \citenamefont {Vishwanath}}]{khalaf_magic_2019}%
  \BibitemOpen
  \bibfield  {author} {\bibinfo {author} {\bibfnamefont {E.}~\bibnamefont {Khalaf}}, \bibinfo {author} {\bibfnamefont {A.~J.}\ \bibnamefont {Kruchkov}}, \bibinfo {author} {\bibfnamefont {G.}~\bibnamefont {Tarnopolsky}},\ and\ \bibinfo {author} {\bibfnamefont {A.}~\bibnamefont {Vishwanath}},\ }\bibfield  {title} {\bibinfo {title} {Magic angle hierarchy in twisted graphene multilayers},\ }\href {https://doi.org/10.1103/PhysRevB.100.085109} {\bibfield  {journal} {\bibinfo  {journal} {Physical Review B}\ }\textbf {\bibinfo {volume} {100}},\ \bibinfo {pages} {085109} (\bibinfo {year} {2019})}\BibitemShut {NoStop}%
\bibitem [{\citenamefont {Calugaru}\ \emph {et~al.}(2021)\citenamefont {Calugaru}, \citenamefont {Xie}, \citenamefont {Song}, \citenamefont {Lian}, \citenamefont {Regnault},\ and\ \citenamefont {Bernevig}}]{calugaru2021TSTG1}%
  \BibitemOpen
  \bibfield  {author} {\bibinfo {author} {\bibfnamefont {D.}~\bibnamefont {Calugaru}}, \bibinfo {author} {\bibfnamefont {F.}~\bibnamefont {Xie}}, \bibinfo {author} {\bibfnamefont {Z.-D.}\ \bibnamefont {Song}}, \bibinfo {author} {\bibfnamefont {B.}~\bibnamefont {Lian}}, \bibinfo {author} {\bibfnamefont {N.}~\bibnamefont {Regnault}},\ and\ \bibinfo {author} {\bibfnamefont {B.~A.}\ \bibnamefont {Bernevig}},\ }\bibfield  {title} {\bibinfo {title} {Twisted symmetric trilayer graphene: Single-particle and many-body hamiltonians and hidden nonlocal symmetries of trilayer moir\'e systems with and without displacement field},\ }\href {https://doi.org/10.1103/PhysRevB.103.195411} {\bibfield  {journal} {\bibinfo  {journal} {Phys. Rev. B}\ }\textbf {\bibinfo {volume} {103}},\ \bibinfo {pages} {195411} (\bibinfo {year} {2021})}\BibitemShut {NoStop}%
\bibitem [{\citenamefont {Kang}\ and\ \citenamefont {Vafek}(2019)}]{KangVafekPRL}%
  \BibitemOpen
  \bibfield  {author} {\bibinfo {author} {\bibfnamefont {J.}~\bibnamefont {Kang}}\ and\ \bibinfo {author} {\bibfnamefont {O.}~\bibnamefont {Vafek}},\ }\bibfield  {title} {\bibinfo {title} {Strong coupling phases of partially filled twisted bilayer graphene narrow bands},\ }\href {https://doi.org/10.1103/PhysRevLett.122.246401} {\bibfield  {journal} {\bibinfo  {journal} {Phys. Rev. Lett.}\ }\textbf {\bibinfo {volume} {122}},\ \bibinfo {pages} {246401} (\bibinfo {year} {2019})}\BibitemShut {NoStop}%
\bibitem [{\citenamefont {Bultinck}\ \emph {et~al.}(2020{\natexlab{b}})\citenamefont {Bultinck}, \citenamefont {Khalaf}, \citenamefont {Liu}, \citenamefont {Chatterjee}, \citenamefont {Vishwanath},\ and\ \citenamefont {Zaletel}}]{bultinck_ground_2020}%
  \BibitemOpen
  \bibfield  {author} {\bibinfo {author} {\bibfnamefont {N.}~\bibnamefont {Bultinck}}, \bibinfo {author} {\bibfnamefont {E.}~\bibnamefont {Khalaf}}, \bibinfo {author} {\bibfnamefont {S.}~\bibnamefont {Liu}}, \bibinfo {author} {\bibfnamefont {S.}~\bibnamefont {Chatterjee}}, \bibinfo {author} {\bibfnamefont {A.}~\bibnamefont {Vishwanath}},\ and\ \bibinfo {author} {\bibfnamefont {M.~P.}\ \bibnamefont {Zaletel}},\ }\bibfield  {title} {\bibinfo {title} {Ground {State} and {Hidden} {Symmetry} of {Magic}-{Angle} {Graphene} at {Even} {Integer} {Filling}},\ }\href {https://doi.org/10.1103/PhysRevX.10.031034} {\bibfield  {journal} {\bibinfo  {journal} {Phys. Rev. X}\ }\textbf {\bibinfo {volume} {10}},\ \bibinfo {pages} {031034} (\bibinfo {year} {2020}{\natexlab{b}})},\ \bibinfo {note} {publisher: American Physical Society}\BibitemShut {NoStop}%
\bibitem [{\citenamefont {Lian}\ \emph {et~al.}(2021)\citenamefont {Lian}, \citenamefont {Song}, \citenamefont {Regnault}, \citenamefont {Efetov}, \citenamefont {Yazdani},\ and\ \citenamefont {Bernevig}}]{TBG4}%
  \BibitemOpen
  \bibfield  {author} {\bibinfo {author} {\bibfnamefont {B.}~\bibnamefont {Lian}}, \bibinfo {author} {\bibfnamefont {Z.-D.}\ \bibnamefont {Song}}, \bibinfo {author} {\bibfnamefont {N.}~\bibnamefont {Regnault}}, \bibinfo {author} {\bibfnamefont {D.~K.}\ \bibnamefont {Efetov}}, \bibinfo {author} {\bibfnamefont {A.}~\bibnamefont {Yazdani}},\ and\ \bibinfo {author} {\bibfnamefont {B.~A.}\ \bibnamefont {Bernevig}},\ }\bibfield  {title} {\bibinfo {title} {Twisted bilayer graphene. iv. exact insulator ground states and phase diagram},\ }\href {https://doi.org/10.1103/PhysRevB.103.205414} {\bibfield  {journal} {\bibinfo  {journal} {Phys. Rev. B}\ }\textbf {\bibinfo {volume} {103}},\ \bibinfo {pages} {205414} (\bibinfo {year} {2021})}\BibitemShut {NoStop}%
\bibitem [{\citenamefont {Wang}\ \emph {et~al.}(2022)\citenamefont {Wang}, \citenamefont {Parker}, \citenamefont {Soejima}, \citenamefont {Hauschild}, \citenamefont {Anand}, \citenamefont {Bultinck},\ and\ \citenamefont {Zaletel}}]{wang2022kekule}%
  \BibitemOpen
  \bibfield  {author} {\bibinfo {author} {\bibfnamefont {T.}~\bibnamefont {Wang}}, \bibinfo {author} {\bibfnamefont {D.~E.}\ \bibnamefont {Parker}}, \bibinfo {author} {\bibfnamefont {T.}~\bibnamefont {Soejima}}, \bibinfo {author} {\bibfnamefont {J.}~\bibnamefont {Hauschild}}, \bibinfo {author} {\bibfnamefont {S.}~\bibnamefont {Anand}}, \bibinfo {author} {\bibfnamefont {N.}~\bibnamefont {Bultinck}},\ and\ \bibinfo {author} {\bibfnamefont {M.~P.}\ \bibnamefont {Zaletel}},\ }\href@noop {} {\bibinfo {title} {Kekul\'e spiral order in magic-angle graphene: a density matrix renormalization group study}} (\bibinfo {year} {2022}),\ \Eprint {https://arxiv.org/abs/2211.02693} {arXiv:2211.02693 [cond-mat.str-el]} \BibitemShut {NoStop}%
\bibitem [{\citenamefont {Zhang}\ \emph {et~al.}(2019{\natexlab{c}})\citenamefont {Zhang}, \citenamefont {Mao}, \citenamefont {Cao}, \citenamefont {{Jarillo-Herrero}},\ and\ \citenamefont {Senthil}}]{zhang_nearly_2019}%
  \BibitemOpen
  \bibfield  {author} {\bibinfo {author} {\bibfnamefont {Y.-H.}\ \bibnamefont {Zhang}}, \bibinfo {author} {\bibfnamefont {D.}~\bibnamefont {Mao}}, \bibinfo {author} {\bibfnamefont {Y.}~\bibnamefont {Cao}}, \bibinfo {author} {\bibfnamefont {P.}~\bibnamefont {{Jarillo-Herrero}}},\ and\ \bibinfo {author} {\bibfnamefont {T.}~\bibnamefont {Senthil}},\ }\bibfield  {title} {\bibinfo {title} {Nearly flat {{Chern}} bands in moir{\textbackslash}'e superlattices},\ }\href {https://doi.org/10.1103/PhysRevB.99.075127} {\bibfield  {journal} {\bibinfo  {journal} {Physical Review B}\ }\textbf {\bibinfo {volume} {99}},\ \bibinfo {pages} {075127} (\bibinfo {year} {2019}{\natexlab{c}})}\BibitemShut {NoStop}%
\bibitem [{\citenamefont {Fang}\ \emph {et~al.}(2012)\citenamefont {Fang}, \citenamefont {Gilbert},\ and\ \citenamefont {Bernevig}}]{fang_bulk_2012}%
  \BibitemOpen
  \bibfield  {author} {\bibinfo {author} {\bibfnamefont {C.}~\bibnamefont {Fang}}, \bibinfo {author} {\bibfnamefont {M.~J.}\ \bibnamefont {Gilbert}},\ and\ \bibinfo {author} {\bibfnamefont {B.~A.}\ \bibnamefont {Bernevig}},\ }\bibfield  {title} {\bibinfo {title} {Bulk topological invariants in noninteracting point group symmetric insulators},\ }\href {https://doi.org/10.1103/PhysRevB.86.115112} {\bibfield  {journal} {\bibinfo  {journal} {Physical Review B}\ }\textbf {\bibinfo {volume} {86}},\ \bibinfo {pages} {115112} (\bibinfo {year} {2012})}\BibitemShut {NoStop}%
\bibitem [{\citenamefont {Guerci}\ \emph {et~al.}(2023)\citenamefont {Guerci}, \citenamefont {Mao},\ and\ \citenamefont {Mora}}]{guerci2023chern}%
  \BibitemOpen
  \bibfield  {author} {\bibinfo {author} {\bibfnamefont {D.}~\bibnamefont {Guerci}}, \bibinfo {author} {\bibfnamefont {Y.}~\bibnamefont {Mao}},\ and\ \bibinfo {author} {\bibfnamefont {C.}~\bibnamefont {Mora}},\ }\href@noop {} {\bibinfo {title} {Chern mosaic and ideal flat bands in equal-twist trilayer graphene}} (\bibinfo {year} {2023}),\ \Eprint {https://arxiv.org/abs/2305.03702} {arXiv:2305.03702 [cond-mat.mes-hall]} \BibitemShut {NoStop}%
\bibitem [{\citenamefont {Wu}\ \emph {et~al.}(2019{\natexlab{b}})\citenamefont {Wu}, \citenamefont {Lovorn}, \citenamefont {Tutuc}, \citenamefont {Martin},\ and\ \citenamefont {MacDonald}}]{Wu2019}%
  \BibitemOpen
  \bibfield  {author} {\bibinfo {author} {\bibfnamefont {F.}~\bibnamefont {Wu}}, \bibinfo {author} {\bibfnamefont {T.}~\bibnamefont {Lovorn}}, \bibinfo {author} {\bibfnamefont {E.}~\bibnamefont {Tutuc}}, \bibinfo {author} {\bibfnamefont {I.}~\bibnamefont {Martin}},\ and\ \bibinfo {author} {\bibfnamefont {A.~H.}\ \bibnamefont {MacDonald}},\ }\bibfield  {title} {\bibinfo {title} {Topological insulators in twisted transition metal dichalcogenide homobilayers},\ }\href {https://doi.org/10.1103/PhysRevLett.122.086402} {\bibfield  {journal} {\bibinfo  {journal} {Phys. Rev. Lett.}\ }\textbf {\bibinfo {volume} {122}},\ \bibinfo {pages} {086402} (\bibinfo {year} {2019}{\natexlab{b}})}\BibitemShut {NoStop}%
\bibitem [{\citenamefont {Anderson}\ \emph {et~al.}(2023)\citenamefont {Anderson}, \citenamefont {Fan}, \citenamefont {Cai}, \citenamefont {Holtzmann}, \citenamefont {Taniguchi}, \citenamefont {Watanabe}, \citenamefont {Xiao}, \citenamefont {Yao},\ and\ \citenamefont {Xu}}]{Anderson2023programming}%
  \BibitemOpen
  \bibfield  {author} {\bibinfo {author} {\bibfnamefont {E.}~\bibnamefont {Anderson}}, \bibinfo {author} {\bibfnamefont {F.-R.}\ \bibnamefont {Fan}}, \bibinfo {author} {\bibfnamefont {J.}~\bibnamefont {Cai}}, \bibinfo {author} {\bibfnamefont {W.}~\bibnamefont {Holtzmann}}, \bibinfo {author} {\bibfnamefont {T.}~\bibnamefont {Taniguchi}}, \bibinfo {author} {\bibfnamefont {K.}~\bibnamefont {Watanabe}}, \bibinfo {author} {\bibfnamefont {D.}~\bibnamefont {Xiao}}, \bibinfo {author} {\bibfnamefont {W.}~\bibnamefont {Yao}},\ and\ \bibinfo {author} {\bibfnamefont {X.}~\bibnamefont {Xu}},\ }\bibfield  {title} {\bibinfo {title} {Programming correlated magnetic states with gate-controlled moiré geometry},\ }\href {https://doi.org/10.1126/science.adg4268} {\bibfield  {journal} {\bibinfo  {journal} {Science}\ }\textbf {\bibinfo {volume} {381}},\ \bibinfo {pages} {325} (\bibinfo {year} {2023})},\ \Eprint {https://arxiv.org/abs/https://www.science.org/doi/pdf/10.1126/science.adg4268}
  {https://www.science.org/doi/pdf/10.1126/science.adg4268} \BibitemShut {NoStop}%
\bibitem [{\citenamefont {Kang}\ \emph {et~al.}(2024)\citenamefont {Kang}, \citenamefont {Shen}, \citenamefont {Qiu}, \citenamefont {Watanabe}, \citenamefont {Taniguchi}, \citenamefont {Shan},\ and\ \citenamefont {Mak}}]{kang2024observation}%
  \BibitemOpen
  \bibfield  {author} {\bibinfo {author} {\bibfnamefont {K.}~\bibnamefont {Kang}}, \bibinfo {author} {\bibfnamefont {B.}~\bibnamefont {Shen}}, \bibinfo {author} {\bibfnamefont {Y.}~\bibnamefont {Qiu}}, \bibinfo {author} {\bibfnamefont {K.}~\bibnamefont {Watanabe}}, \bibinfo {author} {\bibfnamefont {T.}~\bibnamefont {Taniguchi}}, \bibinfo {author} {\bibfnamefont {J.}~\bibnamefont {Shan}},\ and\ \bibinfo {author} {\bibfnamefont {K.~F.}\ \bibnamefont {Mak}},\ }\bibfield  {title} {\bibinfo {title} {Observation of the fractional quantum spin hall effect in moir$\backslash$'e mote2},\ }\href@noop {} {\bibfield  {journal} {\bibinfo  {journal} {arXiv preprint arXiv:2402.03294}\ } (\bibinfo {year} {2024})}\BibitemShut {NoStop}%
\bibitem [{\citenamefont {Wang}\ \emph {et~al.}(2023)\citenamefont {Wang}, \citenamefont {Devakul}, \citenamefont {Zaletel},\ and\ \citenamefont {Fu}}]{wang2023topological}%
  \BibitemOpen
  \bibfield  {author} {\bibinfo {author} {\bibfnamefont {T.}~\bibnamefont {Wang}}, \bibinfo {author} {\bibfnamefont {T.}~\bibnamefont {Devakul}}, \bibinfo {author} {\bibfnamefont {M.~P.}\ \bibnamefont {Zaletel}},\ and\ \bibinfo {author} {\bibfnamefont {L.}~\bibnamefont {Fu}},\ }\href@noop {} {\bibinfo {title} {Topological magnets and magnons in twisted bilayer mote$_2$ and wse$_2$}} (\bibinfo {year} {2023}),\ \Eprint {https://arxiv.org/abs/2306.02501} {arXiv:2306.02501 [cond-mat.str-el]} \BibitemShut {NoStop}%
\bibitem [{\citenamefont {Zhang}\ \emph {et~al.}(2024{\natexlab{b}})\citenamefont {Zhang}, \citenamefont {Wang}, \citenamefont {Liu}, \citenamefont {Fan}, \citenamefont {Cao},\ and\ \citenamefont {Xiao}}]{zhang2024polarizationdriven}%
  \BibitemOpen
  \bibfield  {author} {\bibinfo {author} {\bibfnamefont {X.-W.}\ \bibnamefont {Zhang}}, \bibinfo {author} {\bibfnamefont {C.}~\bibnamefont {Wang}}, \bibinfo {author} {\bibfnamefont {X.}~\bibnamefont {Liu}}, \bibinfo {author} {\bibfnamefont {Y.}~\bibnamefont {Fan}}, \bibinfo {author} {\bibfnamefont {T.}~\bibnamefont {Cao}},\ and\ \bibinfo {author} {\bibfnamefont {D.}~\bibnamefont {Xiao}},\ }\href@noop {} {\bibinfo {title} {Polarization-driven band topology evolution in twisted mote$_2$ and wse$_2$}} (\bibinfo {year} {2024}{\natexlab{b}}),\ \Eprint {https://arxiv.org/abs/2311.12776} {arXiv:2311.12776 [cond-mat.mtrl-sci]} \BibitemShut {NoStop}%
\bibitem [{\citenamefont {Li}\ \emph {et~al.}(2021)\citenamefont {Li}, \citenamefont {Jiang}, \citenamefont {Shen}, \citenamefont {Zhang}, \citenamefont {Li}, \citenamefont {Tao}, \citenamefont {Devakul}, \citenamefont {Watanabe}, \citenamefont {Taniguchi}, \citenamefont {Fu} \emph {et~al.}}]{li2021quantum}%
  \BibitemOpen
  \bibfield  {author} {\bibinfo {author} {\bibfnamefont {T.}~\bibnamefont {Li}}, \bibinfo {author} {\bibfnamefont {S.}~\bibnamefont {Jiang}}, \bibinfo {author} {\bibfnamefont {B.}~\bibnamefont {Shen}}, \bibinfo {author} {\bibfnamefont {Y.}~\bibnamefont {Zhang}}, \bibinfo {author} {\bibfnamefont {L.}~\bibnamefont {Li}}, \bibinfo {author} {\bibfnamefont {Z.}~\bibnamefont {Tao}}, \bibinfo {author} {\bibfnamefont {T.}~\bibnamefont {Devakul}}, \bibinfo {author} {\bibfnamefont {K.}~\bibnamefont {Watanabe}}, \bibinfo {author} {\bibfnamefont {T.}~\bibnamefont {Taniguchi}}, \bibinfo {author} {\bibfnamefont {L.}~\bibnamefont {Fu}}, \emph {et~al.},\ }\bibfield  {title} {\bibinfo {title} {Quantum anomalous hall effect from intertwined moir{\'e} bands},\ }\href@noop {} {\bibfield  {journal} {\bibinfo  {journal} {Nature}\ }\textbf {\bibinfo {volume} {600}},\ \bibinfo {pages} {641} (\bibinfo {year} {2021})}\BibitemShut {NoStop}%
\bibitem [{\citenamefont {Wagner}\ \emph {et~al.}(2022)\citenamefont {Wagner}, \citenamefont {Kwan}, \citenamefont {Bultinck}, \citenamefont {Simon},\ and\ \citenamefont {Parameswaran}}]{wagner_global_2021}%
  \BibitemOpen
  \bibfield  {author} {\bibinfo {author} {\bibfnamefont {G.}~\bibnamefont {Wagner}}, \bibinfo {author} {\bibfnamefont {Y.~H.}\ \bibnamefont {Kwan}}, \bibinfo {author} {\bibfnamefont {N.}~\bibnamefont {Bultinck}}, \bibinfo {author} {\bibfnamefont {S.~H.}\ \bibnamefont {Simon}},\ and\ \bibinfo {author} {\bibfnamefont {S.~A.}\ \bibnamefont {Parameswaran}},\ }\bibfield  {title} {\bibinfo {title} {Global phase diagram of the normal state of twisted bilayer graphene},\ }\href {https://doi.org/10.1103/PhysRevLett.128.156401} {\bibfield  {journal} {\bibinfo  {journal} {Phys. Rev. Lett.}\ }\textbf {\bibinfo {volume} {128}},\ \bibinfo {pages} {156401} (\bibinfo {year} {2022})}\BibitemShut {NoStop}%
\bibitem [{\citenamefont {Calugaru}\ \emph {et~al.}(2022)\citenamefont {Calugaru}, \citenamefont {Regnault}, \citenamefont {Oh}, \citenamefont {Nuckolls}, \citenamefont {Wong}, \citenamefont {Lee}, \citenamefont {Yazdani}, \citenamefont {Vafek},\ and\ \citenamefont {Bernevig}}]{Princeton_STM_Theory}%
  \BibitemOpen
  \bibfield  {author} {\bibinfo {author} {\bibfnamefont {D.}~\bibnamefont {Calugaru}}, \bibinfo {author} {\bibfnamefont {N.}~\bibnamefont {Regnault}}, \bibinfo {author} {\bibfnamefont {M.}~\bibnamefont {Oh}}, \bibinfo {author} {\bibfnamefont {K.~P.}\ \bibnamefont {Nuckolls}}, \bibinfo {author} {\bibfnamefont {D.}~\bibnamefont {Wong}}, \bibinfo {author} {\bibfnamefont {R.~L.}\ \bibnamefont {Lee}}, \bibinfo {author} {\bibfnamefont {A.}~\bibnamefont {Yazdani}}, \bibinfo {author} {\bibfnamefont {O.}~\bibnamefont {Vafek}},\ and\ \bibinfo {author} {\bibfnamefont {B.~A.}\ \bibnamefont {Bernevig}},\ }\bibfield  {title} {\bibinfo {title} {Spectroscopy of twisted bilayer graphene correlated insulators},\ }\href {https://doi.org/10.1103/PhysRevLett.129.117602} {\bibfield  {journal} {\bibinfo  {journal} {Phys. Rev. Lett.}\ }\textbf {\bibinfo {volume} {129}},\ \bibinfo {pages} {117602} (\bibinfo {year} {2022})}\BibitemShut {NoStop}%
\bibitem [{\citenamefont {Hong}\ \emph {et~al.}(2022)\citenamefont {Hong}, \citenamefont {Soejima},\ and\ \citenamefont {Zaletel}}]{Berkeley_STM_Theory}%
  \BibitemOpen
  \bibfield  {author} {\bibinfo {author} {\bibfnamefont {J.~P.}\ \bibnamefont {Hong}}, \bibinfo {author} {\bibfnamefont {T.}~\bibnamefont {Soejima}},\ and\ \bibinfo {author} {\bibfnamefont {M.~P.}\ \bibnamefont {Zaletel}},\ }\bibfield  {title} {\bibinfo {title} {Detecting symmetry breaking in magic angle graphene using scanning tunneling microscopy},\ }\href {https://doi.org/10.1103/PhysRevLett.129.147001} {\bibfield  {journal} {\bibinfo  {journal} {Phys. Rev. Lett.}\ }\textbf {\bibinfo {volume} {129}},\ \bibinfo {pages} {147001} (\bibinfo {year} {2022})}\BibitemShut {NoStop}%
\end{thebibliography}%

\setcounter{figure}{0}
\let\oldthefigure\thefigure
\renewcommand{\thefigure}{S\oldthefigure}

\setcounter{table}{0}
\renewcommand{\thetable}{S\arabic{table}}

\newpage
\clearpage
\setcounter{section}{0}
\renewcommand{\thesection}{S\arabic{section}}
\renewcommand{\thesubsection}{\thesection.\arabic{subsection}}
\renewcommand{\thesubsubsection}{\thesubsection.\arabic{subsubsection}}

\begin{appendix}
\onecolumngrid
	\begin{center}\textbf{\large --- Supplementary Material ---}

\end{center}

\section{Additional numerical results}

\subsection{Helical trilayer graphene}\label{secapp:HTG}

\begin{figure*}[h]
    \centering
    \includegraphics[width = \linewidth]{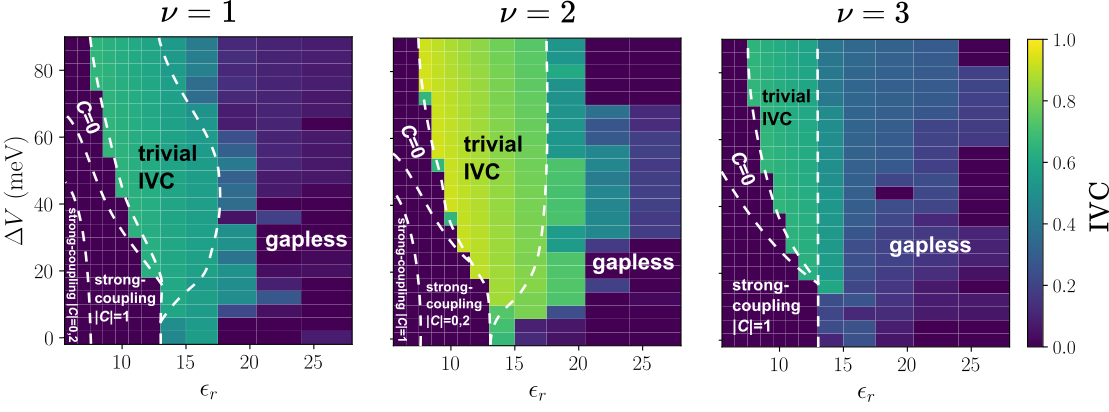}
    \caption{\textbf{Phase diagram of helical trilayer graphene (HTG) at $\theta=1.75^\circ$.} We focus on one h-HTG domain.  Phase diagram calculated using $12\times12$ HF at $\nu=1,2,3$. Momentum-dependent tunneling terms are not included.\label{fig:htg_1.75}}
\end{figure*}

\begin{figure*}[h]
    \centering
    \includegraphics[width = \linewidth]{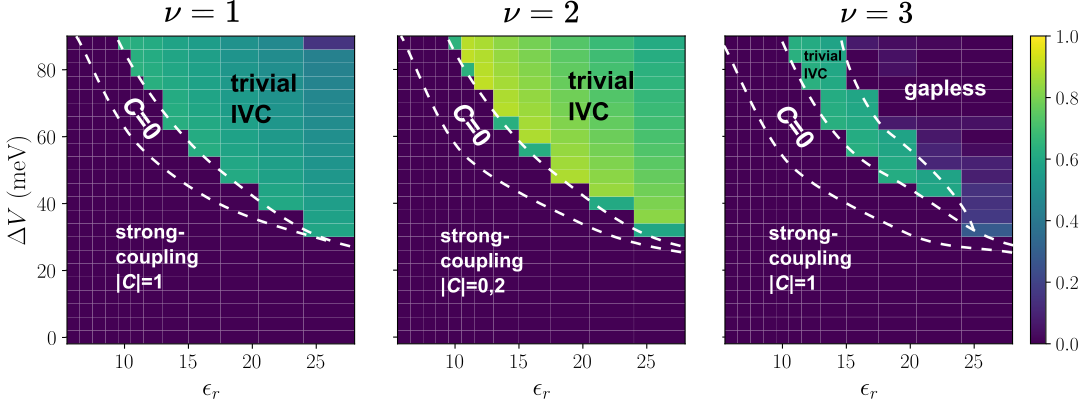}
    \caption{\textbf{Phase diagram of helical trilayer graphene (HTG) at $\theta=1.85^\circ$ with momentum-dependent tunneling.} We focus on one h-HTG domain.  Phase diagram calculated using $12\times12$ HF at $\nu=1,2,3$.\label{fig:htg_kdt}}
\end{figure*}

In Fig.~\ref{fig:htg_1.75}, we show the phase diagram at $\nu=1,2,3$ for the h-HTG domain of HTG at $\theta=1.75^\circ$. While the overall structure of the competition between strong-coupling insulators at small $\epsilon_r$, IVC states at intermediate $\epsilon_r$, and gapless states at weak interactions is similar to that at $\theta=1.85^\circ$ (shown in main text), we do not find any CTIs.

In Fig.~\ref{fig:htg_kdt}, we show the phase diagram at $\nu=1,2,3$ for the h-HTG domain of HTG at $\theta=1.85^\circ$ when momentum-dependent interlayer tunneling is accounted for~\cite{xia2023helical}. This introduces a momentum dependence to the interlayer hopping functions $t_\alpha(\bm{r})$ which varies linearly with the momentum deviation from the Dirac momentum. This enhances the strong-coupling phase at $\nu=1,2,3$. We do not find any CTIs.

\section{Gauge fixing and gauge-invariant velocity}
\subsection{Gauge fixing for intervalley order parameter}\label{secapp:gauge_fixing}
The intervalley order parameter $\Delta_{\bm{q}}(\bm{k})\equiv\langle d^\dagger_{\bm{k},+}d_{\bm{k}+\bm{q},-} \rangle$ is dependent on the choice of gauge for the $d^\dagger_{\bm{k}\tau}$ operators. While many choices are possible, we require that the gauge is smooth within the BZ.  That is, we require $|\braket{u(\bm{k} + \delta \bm{k})|u(\bm{k})} - 1| \rightarrow 0$ when $|\delta \bm{k}| \rightarrow 0$ for all $\bm{k}$, provided no BZ boundary is crossed, where $\ket{u(\bm{k}}$ is the periodic part of the Bloch wavefunction. Numerically, we carry out the procedure described below.

Consider an $N_1 \times N_2$ system. Denote $\bm{g}_{i} = \bm{G}_{i}/N_i$ for $i = 1,2$, where $\bm{G}_i$ is the $i$-th basis reciprocal lattice vector.   Consider first the $\tau = +$ valley. Denote the crystal momentum $\tilde{\bm{k}} = \bm{k} - \tilde{\bm{\gamma}} = m_1\bm{g_1} + m_2\bm{g}_2$ as $(m_1, m_2)$, where the chosen reference point $\tilde{\bm{\gamma}}$ needs not be $\Gamma_M$, and $0 \leq m_i \leq N_i - 1$. Starting from $(0, 0)$, we specify the gauge at $(m, 0)$ by demanding
\begin{equation}
    \arg(\braket{u(m_1 +1, 0)|u(m_1, 0)}) = 0
\end{equation}
for $m_1 = 0, ..., N_1 - 2$. We then fix the gauge at all other points by demanding
\begin{equation}
    \arg(\braket{u(m_1, m_2 + 1)|u(m_1, m_2)}) = 0
\end{equation}
for $m_1 = 0, ..., N_1 - 1$ and $m_2 = 0, ..., N_2 - 2$. This gauge is akin to the `Landau gauge' for an arbitrary Chern band. For the $\tau = -$ valley, we perform the same procedure, except we define $\tilde{\bm{k}} = \bm{k} + \bm{q} - \tilde{\bm{\gamma}} = m_1\bm{g_1} + m_2\bm{g}_2$ to ensure the alignment of the BZ boundaries in both valleys.

\subsection{Gauge-invariant velocity on a grid}
The gauge invariant velocity on a continuous $\bm{k}$-space is defined in Eq.~\ref{eq:invariant_current}. For numerical calculation on a finite grid, we adopt the following definition
\begin{equation}
    \bm{j}(\bm{k}) \cdot \delta\bm{k} = \arg\left(\frac{\Delta_{\bm{q}}(\bm{k} + \delta\bm{k})}{\Delta_{\bm{q}}(\bm{k})}\frac{\braket{u_{\bm{k} + \delta \bm{k}, +}|u_{\bm{k}+}}}{\braket{u_{\bm{k} + \bm{q} + \delta \bm{k}, -}|u_{\bm{k} + \bm{q}, -}}}\right)
\end{equation}
where the principal range is $\arg(x) \in [-\pi, \pi)$, as we expect $|\bm{j}(\bm{k}) \cdot \delta\bm{k}| \ll 1$ for a sufficiently fine momentum grid and not too close to the vortex core, where the velocity diverges. This definition reduces to the correct continuum limit by setting $|\delta \bm{k}| \rightarrow 0$ and has the advantage of being numerically reliable for arbitrarily non-smooth gauge choice. 

\end{appendix}

\end{document}